\documentclass[11pt,a4paper]{article}
\usepackage[utf8]{inputenc}
\usepackage[english]{babel}
\usepackage[top=2.3cm, bottom=3.0cm, left=2.3cm, right=2.3cm]{geometry}
\usepackage{setspace}
\makeatletter
\g@addto@macro\bfseries{\boldmath}
\makeatother
\usepackage{fancyhdr}
\usepackage{amssymb}
\usepackage{amsmath}
\usepackage{amsfonts}
\usepackage{mathrsfs}
\usepackage{bm}
\usepackage{slashed}
\usepackage{mathtools}
\usepackage{amsthm}
\DeclareMathAlphabet{\mathpzc}{OT1}{pzc}{m}{it}
\newcommand{\modj}{\mathpzc{J}}
\numberwithin{equation}{section}
\allowdisplaybreaks
\usepackage{braket}
\usepackage{enumitem}
\usepackage{array}
\renewcommand{\arraystretch}{1.0}
\usepackage{tabularx}

\usepackage[textfont=it]{caption} 
\usepackage{hyperref}
\hypersetup{
    colorlinks,%
    citecolor=blue,%
    filecolor=blue,%
    linkcolor=blue,%
   	urlcolor=blue,
   	linktoc=page
}
\usepackage{cite}
\usepackage[toc,page]{appendix}
\usepackage{graphicx}
\usepackage{float}
\usepackage{subfig}
\usepackage{tikz}
\usetikzlibrary{calc}
\usetikzlibrary{decorations.pathreplacing}
\usetikzlibrary{decorations.pathmorphing}
\usetikzlibrary{decorations.markings}
\usetikzlibrary{arrows.meta}
\usetikzlibrary{positioning}
\usetikzlibrary{math}
\tikzset{every picture/.style={font issue=\footnotesize},
         font issue/.style={execute at begin picture={#1\selectfont}}
        }
\usepackage{xcolor}
\definecolor{DarkRed}{RGB}{192,0,0}
\usepackage{mdframed}
\usepackage[normalem]{ulem} 

\renewcommand{\textbf}[1]{\begingroup\bfseries\mathversion{bold}#1\endgroup}
\newcommand{\D}{d}
\usepackage{pict2e}
\makeatletter
\newcommand{\loplus}{\mathbin{\mathpalette\dog@lsemi{+}}}
\newcommand{\dog@lsemi}[2]{\dog@semi{#1}{#2}{270,90}}
\newcommand{\dog@semi}[3]{%
  \begingroup
  \sbox\z@{$\m@th#1#2$}%
  \setlength{\unitlength}{\dimexpr\ht\z@+\dp\z@\relax}%
  \makebox[\wd\z@]{\raisebox{-\dp\z@}{%
    \begin{picture}(1,1)
    \linethickness{\variable@rule{#1}}
    \roundcap
    \put(0.5,0.5){\makebox(0,0){\raisebox{\dp\z@}{$\m@th#1#2$}}}
    \put(0.5,0.5){\arc[#3]{0.5}}
    \end{picture}%
  }}%
  \endgroup
}
\newcommand{\variable@rule}[1]{%
  \fontdimen8  
  \ifx#1\displaystyle\textfont3\else
    \ifx#1\textstyle\textfont3\else
      \ifx#1\scriptstyle\scriptfont3\else
        \scriptscriptfont3\relax
  \fi\fi\fi
}
\makeatother
\newcommand{\eqholo}{\stackrel{}{\equiv}}
\usepackage{pifont} 
\newcommand{\trefle}{\ding{168}}
\newcommand{\carreau}{{\color{DarkRed}\ding{169}}}
\newcommand{\coeur}{{\color{DarkRed}\ding{170}}}
\newcommand{\pique}{\ding{171}}

\begin{document}

\setstretch{1.1}

\setcounter{tocdepth}{2}

\thispagestyle{empty}
$ $
\begin{center}
\setstretch{2.2}
\vspace{40pt}

{\LARGE{\textbf{Bridging Carrollian and Celestial Holography
}}}
\end{center}

\vspace{15pt}
\begin{center} 
Laura Donnay\textsuperscript{\trefle,\carreau}\footnote{\fontsize{8pt}{10pt}\selectfont \ \href{mailto:laura.donnay@sissa.it}{laura.donnay@sissa.it}}, 
Adrien Fiorucci\textsuperscript{\coeur}\footnote{\fontsize{8pt}{10pt}\selectfont\  \href{mailto:adrien.fiorucci@tuwien.ac.at}{adrien.fiorucci@tuwien.ac.at}},
Yannick Herfray\textsuperscript{\pique}\footnote{\fontsize{8pt}{10pt}\selectfont\  \href{mailto:yannick.herfray@univ-tours.fr}{yannick.herfray@univ-tours.fr}}, Romain Ruzziconi\textsuperscript{\coeur}\footnote{\fontsize{8pt}{10pt}\selectfont\ \href{mailto:romain.ruzziconi@tuwien.ac.at}{romain.ruzziconi@tuwien.ac.at}}

\vspace{30pt}
\normalsize
$^\text{\trefle}$\textit{International School for Advanced Studies (SISSA),\\ Via Bonomea 265, 34136 Trieste, Italy\\
\vspace{2mm}
}

\normalsize
\medskip
$^\text{\carreau}$\textit{Istituto Nazionale di Fisica Nucleare (INFN) --- Sezione di Trieste,\\
Via Valerio 2, 34127 Trieste, Italy\\
\vspace{2mm}
}

\normalsize
\medskip
$^\text{\coeur}$\textit{Institute for Theoretical Physics, Technische Universit\"at Wien,\\
Wiedner Hauptstrasse 8, A-1040 Vienna, Austria\\
\vspace{2mm}
}

\normalsize
\medskip
$^\text{\pique}$\textit{Institut Denis Poisson UMR 7013, Université de Tours,\\
Parc de Grandmont, 37200 Tours, France\\
\vspace{2mm}
}

\vspace{40pt}

\begin{abstract}\noindent
Gravity in $4d$ asymptotically flat spacetime constitutes the archetypal example of a gravitational system with leaky boundary conditions. Pursuing our analysis of \cite{Donnay:2022aba}, we argue that the holographic description of such a system requires the coupling of the dual theory living at null infinity to some external sources encoding the radiation reaching the conformal boundary and responsible for the non-conservation of the charges. In particular, we show that the sourced Ward identities of a conformal Carrollian field theory living at null infinity reproduce the BMS flux-balance laws. We also derive the general form of low-point correlation functions for conformal Carrollian field theories and exhibit a new branch of solutions, which is argued to be the relevant one for holographic purposes. We then relate our Carrollian approach to the celestial holography proposal by mapping the Carrollian Ward identities to those constraining celestial operators through a suitable integral transform.
\end{abstract}

\end{center}

\newpage

\setcounter{page}{2}

\begin{spacing}{0.85}

\tableofcontents

\end{spacing}

\setcounter{footnote}{0} 

\newpage

\section{Introduction}

The program of \textit{flat space holography} consists in building a holographic duality between gravity in asymptotically flat spacetime and a lower-dimensional field theory. The motivations are twofold. Firstly, from a purely theoretical perspective, this program is enshrined in a broader context that aims at understanding how general is the holographic principle. Does it extend beyond the framework of the celebrated AdS/CFT correspondence \cite{Maldacena:1997re,Witten:1998qj,Aharony:1999ti}? Secondly, asymptotically flat spacetimes provide realistic models to describe a huge range of physical processes occurring in our universe, all the way up to astrophysical scales which are below the cosmological scale. 

In spite of repeated early efforts from various points of views \cite{Susskind:1998vk,Polchinski:1999ry,Giddings:1999jq,Arcioni:2003td,Arcioni:2003xx,Mann:2005yr}, it became clear very soon that flat space holography would not merely immediately follow from AdS/CFT via a simple limit procedure where the AdS radius is sent to infinity. But fortunately, the last decades have also taught us that the symmetries of flat spacetime turn out to be way richer and more subtle than originally thought \cite{Bondi:1962px,Sachs:1962zza,Barnich:2010eb,Strominger:2017zoo,Troessaert:2017jcm,Henneaux:2018cst}. Building on the constraints imposed by the BMS symmetries, two bottom-up roads towards flat space holography have emerged over the years: \textit{Carrollian} and \textit{celestial} holography.

The Carrollian approach to flat space holography proposes that the role of the dual theory is played by a \textit{conformal Carrollian field theory} (or Carrollian CFT for short) that lives on the codimension-one boundary of spacetime (null infinity $\mathscr I$).  Alternatively, one could refer to this theory as a BMS-invariant field theory, as it has been known for quite some time that the BMS group is isomorphic to the conformal Carroll group \cite{Duval:2014uva}; but it is fair to say that the ``Carrollian'' name, due to Lévy-Leblond \cite{Levy1965}, has struck physicists' imagination. 

This first approach has proven to be very successful in the context of three-dimensional ($3d$) gravity: among other results, one can find $(i)$ a matching between the entropy of asymptotically flat cosmological solutions and entropy computed with a Cardy-like formula for a Carrollian CFT \cite{Barnich:2012xq,Bagchi:2012xr}; $(ii)$ a computation of the entanglement entropy in the Carrollian CFT from the bulk geometry using some extension of the Ryu-Takayanagi prescription \cite{Li:2010dr,Bagchi:2014iea,Jiang:2017ecm}; $(iii)$ the form of the correlation functions in the dual theory \cite{Bagchi:2009ca,Detournay:2014fva, Bagchi:2015wna,Bagchi:2017cpu,Bagchi:2016geg}; $(iv)$ an effective action for the dual Carrollian CFT \cite{Barnich:2012rz,Barnich:2013yka,Barnich:2017jgw, Merbis:2019wgk}. This construction of effective action has been repeated in $4d$ gravity and shown to describe the dynamics of non-radiative spacetime \cite{Barnich:2022bni}. A complementary approach was adopted earlier in \cite{Adamo:2014yya} where an action encoding the radiative modes at $\mathscr{I}$ was explicitly written. Let us also mention that Carrollian holography is well understood in the fluid/gravity correspondence for both $3d$ and $4d$ spacetimes using a suitable flat limit procedure \cite{Ciambelli:2018wre,Campoleoni:2018ltl,Ciambelli:2020eba,Ciambelli:2020ftk,Campoleoni:2022wmf}. The dual fluid is a Carrollian fluid \cite{deBoer:2017ing,Ciambelli:2018xat, deBoer:2021jej,Petkou:2022bmz,Freidel:2022bai} at $\mathscr{I}$ whose properties are deduced by taking the ultra-relativistic limit of a relativistic fluid. 

The main difficulties inherent to the Carrollian holographic approach are related to two key differences compared to  what one is used to encounter in AdS/CFT. $(i)$ The conformal boundary $\mathscr{I}$ is a \textit{null} hypersurface, implying that the dual theory involves some Carrollian or ``ultralocal'' physics \cite{Bacry:1968zf,Bergshoeff:2014jla,Duval:2014uoa,Hartong:2015xda,Bergshoeff:2017btm,Chen:2022cpx} which intrinsically defines the unconventional features of Carrollian CFTs \cite{Bagchi:2019clu,Bagchi:2019xfx,Gupta:2020dtl,deBoer:2021jej,Pasterski:2022jzc,Bergshoeff:2022eog,Donnay:2022aba,Bagchi:2022emh}, about which very little is known so far. This contrasts with the AdS/CFT correspondence where the boundary is timelike and the dual theory is an honest CFT whose proprieties have been studied for decades. $(ii)$ The gravitational charges at null infinity are generically non-conserved due to the presence of matter or gravitational radiation \cite{Trautman:1958zdi,Bondi:1962px,Sachs:1962wk,Sachs:1962zza,Wald:1999wa,Barnich:2011mi}. This constitutes the plain vanilla example of a gravitational system with \textit{leaky boundary conditions}. Describing the non-conservation of the charges from the point of view of the dual theory requires the development of new techniques such as the coupling with external sources \cite{Donnay:2022aba}. On the contrary, Dirichlet-type of boundary conditions are usually imposed in AdS, preventing leaks through the conformal boundary, which thus acts as a reflective cavity. Let us however mention that leaky boundary conditions can also be imposed in AdS by relaxing the usual boundary conditions and turning on boundary sources \cite{Compere:2019bua,Compere:2020lrt,Ruzziconi:2020cjt,Fiorucci:2020xto,Fiorucci:2021pha}. This relaxation is necessary in order to recover a radiative phase space and BMS symmetries from a large radius limit of AdS. 

In the celestial holography paradigm, the proposed dual theory to gravity in flat space is a \emph{celestial CFT} (or CCFT for short) living on a codimension-two boundary, the celestial sphere. It is anchored in the fact that $\mathcal{S}$-matrix elements in the bulk, once rewritten in boost eigenstate basis, enjoy conformal invariance in a manifest way \cite{deBoer:2003vf,He:2015zea,Pasterski:2016qvg,Cheung:2016iub,Pasterski:2017kqt,Strominger:2017zoo,Pasterski:2017ylz}. The advantages of this approach is that one can use some of the very powerful CFT techniques to study the celestial dual, such as operator product expansions (OPEs) \cite{Fotopoulos:2019tpe,Pate:2019lpp,Fan:2019emx,Banerjee:2020kaa,Fotopoulos:2019vac,Fan:2020xjj,Banerjee:2020zlg,Banerjee:2020vnt,Adamo:2021zpw,Costello:2022upu,Hu:2022bpa,Adamo:2022wjo}, conformal block decomposition \cite{Nandan:2019jas,Atanasov:2021cje,Fan:2021isc,Fan:2021pbp,Guevara:2021abz}, state-operator correspondence \cite{Crawley:2021ivb}, null states and conformal multiplets \cite{Banerjee:2019aoy,Banerjee:2019tam,Pasterski:2021fjn,Pasterski:2021dqe}, crossing symmetry \cite{Mizera:2022sln} or shadow formalism \cite{Pasterski:2017kqt,Kapec:2021eug,Kapec:2022axw,Banerjee:2022wht}. This CFT language has recently allowed to uncover new $w_{1+\infty}$ symmetries in CCFT \cite{Strominger:2021lvk} (see also \cite{Himwich:2021dau,Ball:2021tmb,Adamo:2021lrv}). This remarkable finding suggests that gravity might exhibit much more symmetries than one could have expected. Patterns of these symmetries have been found in the subleading orders of the gravitational solution space in \cite{Freidel:2021dfs,Freidel:2021ytz}. However, there is always a price to pay for the emergence of CFT-like features in flat space, and this is manifested in some of the exotic features that CCFTs exhibit. It is indeed not clear for the moment exactly to which extent they differ from standard (\textit{e.g.} unitary and compact) CFTs or what could be an axiomatic definition of CCFTs. Moreover, the fact that the dual theory is codimension-two with respect to the bulk  makes the link with the AdS/CFT correspondence more nebulous, since it would require more involved steps than simply getting celestial correlators from a flat limit of those in AdS (see however \cite{deBoer:2003vf,Cheung:2016iub,Ball:2019atb,Iacobacci:2022yjo,PipolodeGioia:2022exe,Casali:2022fro,Gonzo:2022tjm} for connections to AdS and \cite{Ogawa:2022fhy} to wedge holography). Finally, because the celestial encoding favors conformal transformations over time translations (which  we recall are not lost but rather reshuffled into shifts in the conformal dimension of celestial operators), the dynamics of the gravitational theory such as the Bondi mass loss formula is not easily interpreted in the celestial CFT.

\begin{figure}[ht!]
\centering
\begin{tikzpicture}
    \tikzset{boxnode/.style={fill=white,draw=black,text width=3.5cm,minimum height=0.8cm,outer sep=10pt,align=center}}
    \def\xx{5};\def\yy{3.2};
    \coordinate (apexL) at (-\xx, \yy);
    \coordinate (apexR) at ( \xx, \yy);
    \coordinate (loweL)	at (-\xx,-\yy);
    \coordinate (loweR) at ( \xx,-\yy);
    \coordinate (midL)  at ($(apexL)!0.5!(loweL)$);
    \coordinate (midR)  at ($(apexR)!0.5!(loweR)$);
    \coordinate (midT)  at ($(apexL)!0.5!(apexR)$);
    \coordinate (midB)  at ($(loweL)!0.5!(loweR)$);
    \def\ofsx{2.1};
    \def\ofsy{0.6};
    \def\dilatx{3};
    \def\dilaty{2.8};
    \draw[white,opacity=0] ($(apexL)+(-\dilatx,\dilaty)$) -- ($(apexR)+(\dilatx,\dilaty)$) --  ($(loweR)+(\dilatx,-0.5*\dilaty)$) -- ($(loweL)+(-\dilatx,-0.5*\dilaty)$) -- cycle;
    \fill[black!15] ($(apexL)+(-\ofsx,\ofsy)$) -- ($(apexR)+(\ofsx,\ofsy)$) -- ($(apexR)+(\ofsx,-\ofsy)$) -- ($(apexL)+(-\ofsx,-\ofsy)$);
    \node[boxnode,dashed] (naL) at (apexL) {Evolution along $\mathscr I$};
    \node[boxnode,dashed] (naR) at (apexR) {Scattering data on $\mathscr I$};
    \node[boxnode] (nlL) at (loweL) {\textbf{3\textit{d} Carrollian CFT}};
    \node[boxnode] (nlR) at (loweR) {\textbf{2\textit{d} Celestial CFT}};
    \draw[Latex-Latex] (naL) -- (nlL);
    \node[rotate=90,align=center,above] at (midL) {\textbf{\textit{Carrollian holography}}};
    \draw[Latex-Latex] (naR) -- (nlR);
    \node[rotate=-90,align=center,above] at (midR) {\textbf{\textit{Celestial holography}}};
    \draw[Latex-Latex] (naL) -- (naR);
    \node[fill=white,draw=black,minimum height=0.8cm,outer sep=10pt,text width=6.5cm,align=center] (topt) at ($(midT)+(0,1.5)$) {\textit{\textbf{Two roads towards flat holography}}};
    \draw[-Latex] (topt) -| (naR);
    \draw[Latex-] (naL) |- (topt);
    \node[fill=black!15,text width=2.5cm,align=center] at (midT) {\textit{Complementary points of view}};
    \draw[Latex-Latex] (nlL) -- (nlR);
    \node[fill=white,text width=0.5cm,align=center] at (midB) {vs};
    \node[above] at ($(midB)+(0,0.1)$) {\textit{Carrollian operators}};
    \node[below] at ($(midB)-(0,0.1)$) {\textit{Conformal operators}};
    \node[align=center,text width=3cm] at (nlL.south) {$\sigma_{(k,\bar k)}(x^a)$};
    \node at (nlR.south) {$\mathcal O_{\Delta,J}(x^A)$};
    \coordinate (cent) at ($(midT)!0.5!(midB)$);
    \node[align=center] at ($(cent)!0.5!(midL)+(0.1,0)$) {
        \begin{tikzpicture}[scale=0.35]
            \scriptsize
        	\coordinate (top) at (0, 5);
        	\coordinate (bot) at (0,-5);
        	\coordinate (lft) at (-5,0);
        	\coordinate (rgt) at ( 5,0);
        	\draw[] (top) -- (lft) -- (bot) -- (rgt) -- cycle;
            \draw[] (lft) node[left,inner sep=2pt]{$i^0$};
            \draw[] (top) node[above,inner sep=2pt]{$i^+$};
            \draw[] (bot) node[below,inner sep=2pt]{$i^-$};
            \draw[] (rgt) node[right,inner sep=2pt]{$i^0$};
            \draw[] ($(lft)!0.5!(top)+(-0.3,0.3)$) node[anchor=south east,inner sep=2pt]{$\mathscr{I}^+$};
            \draw[] ($(lft)!0.5!(bot)+(-0.3,-0.3)$) node[anchor=north east,inner sep=2pt]{$\mathscr{I}^-$};
            \draw[] ($(rgt)!0.5!(top)$) node[above,align=center,inner sep=2pt,outer sep=4pt,rotate=-45]{\color{red}$\frac{dQ}{du}\neq 0$};
            \draw[] ($(rgt)!0.5!(bot)+(0.3,-0.3)$) node[anchor=north west,inner sep=2pt]{$\mathscr{I}^-$};
            \foreach \k in {-3.5,-3.25,...,-2.5} 
            {
            \coordinate (pk) at (\k+4,-\k-2);
	        \coordinate (qk) at ($(pk)+(1.5,1.5)$);
	        \coordinate (ar) at ($(pk)!0.55!(qk)$);
	        \draw[red] (pk) -- (qk);
	        \draw[red, -latex] (pk) -- (ar);
            }
            \def\km{-3.75};\def\kp{-2.25};
            \coordinate (bm) at (\km+3.8,-\km-2.2);
            \coordinate (bp) at (\kp+3.8,-\kp-2.2);
            \coordinate (tm) at ($(bm)+(1.7,1.7)$);
            \coordinate (tp) at ($(bp)+(1.7,1.7)$);
            \draw[blue] ($(tp)+(0.2,0.2)$) -- (bp);
            \draw[blue] ($(tm)+(0.2,0.2)$) -- (bm);
            \draw[-latex,black!30,line width=2pt] ($(bot)!0.5!(rgt)+(-1,1)$) -- ($(lft)!0.5!(top)+(1,-1)$);
            \node[blue,below,outer sep=2pt,rotate=45] at ($(tp)!0.5!(bp)$) {$Q(u_1)$};
            \node[blue,above,outer sep=2pt,rotate=45] at ($(tm)!0.5!(bm)$) {$Q(u_2)$};
        \end{tikzpicture}
        };
    \node[align=center] at  ($(cent)!0.5!(midR)$) {
        \begin{tikzpicture}[scale=0.35]
            \scriptsize
            \coordinate (top) at (0, 5);
            \coordinate (bot) at (0,-5);
            \coordinate (lft) at (-5,0);
            \coordinate (rgt) at ( 5,0);
            \draw[] (top) -- (lft) -- (bot) -- (rgt) -- cycle;
            \draw[] (lft) node[left,inner sep=2pt]{$i^0$};
            \draw[] (top) node[above,inner sep=2pt]{$i^+$};
            \draw[] (bot) node[below,inner sep=2pt]{$i^-$};
            \draw[] (rgt) node[right,inner sep=2pt]{$i^0$};
            \draw[] ($(lft)!0.5!(top)+(-0.3,0.3)$) node[anchor=south east,inner sep=2pt]{$\mathscr{I}^+$};
            \draw[] ($(lft)!0.5!(bot)+(-0.3,-0.3)$) node[anchor=north east,inner sep=2pt]{$\mathscr{I}^-$};
            \draw[] ($(rgt)!0.5!(top)+(0.3,0.3)$) node[anchor=south west,inner sep=2pt]{$\mathscr{I}^+$};
            \draw[] ($(rgt)!0.5!(bot)+(0.3,-0.3)$) node[anchor=north west,inner sep=2pt]{$\mathscr{I}^-$};
            \coordinate (loweranchor) at ($(bot)!0.2!(top)$);
            \coordinate (higheranchor) at ($(bot)!0.8!(top)$);
            \draw[green!50!black] (lft) to [out=-45, in=180] (loweranchor) to [out=0, in=-135] (rgt);
            \draw[green!50!black] (lft) to [out=45, in=180] (higheranchor) to [out=0, in=135] (rgt);
            \node[above] at ($(higheranchor)-(0,0.1)$) {$\langle out|$};
            \draw[-latex,black!30,line width=2pt] ($(loweranchor)+(0,0.8)$) -- ($(higheranchor)+(0,-0.8)$);
            \node[fill=white,draw=black!30,circle,inner sep=1pt,align=center,above] at ($(bot)!0.43!(top)$) {$\mathcal S$};
            \node[below] at ($(loweranchor)+(0,0.1)$) {$|in\rangle$};
        \end{tikzpicture}
        };
    \draw[dashed] ($(apexL)!0.5!(apexR)-(0,1)$) -- ($(loweR)!0.5!(loweL)+(0,0.7)$);
\end{tikzpicture}
\caption{Holographic nature of null infinity: two equivalent visions. $(i)$ On the left picture, $\mathscr{I}^+$ is seen as a boundary along which there is a Carrollian time evolution. This naturally leads to the Carrollian holography proposal where the putative dual theory is a $3d$ Carrollian CFT. This picture is well adapted to describe the dynamics of the system through the flux-balance laws encoding the non-conservation of the charges $Q$. $(ii)$ On the right picture, $\mathscr{I}^+$ is seen as a portion of a Cauchy hypersuface at late time. This naturally leads to the celestial holography proposal where the putative dual theory is a $2d$ CCFT. This second approach is particularly natural when considering a scattering process in flat spacetime. }
\label{fig:equiv}
\end{figure}
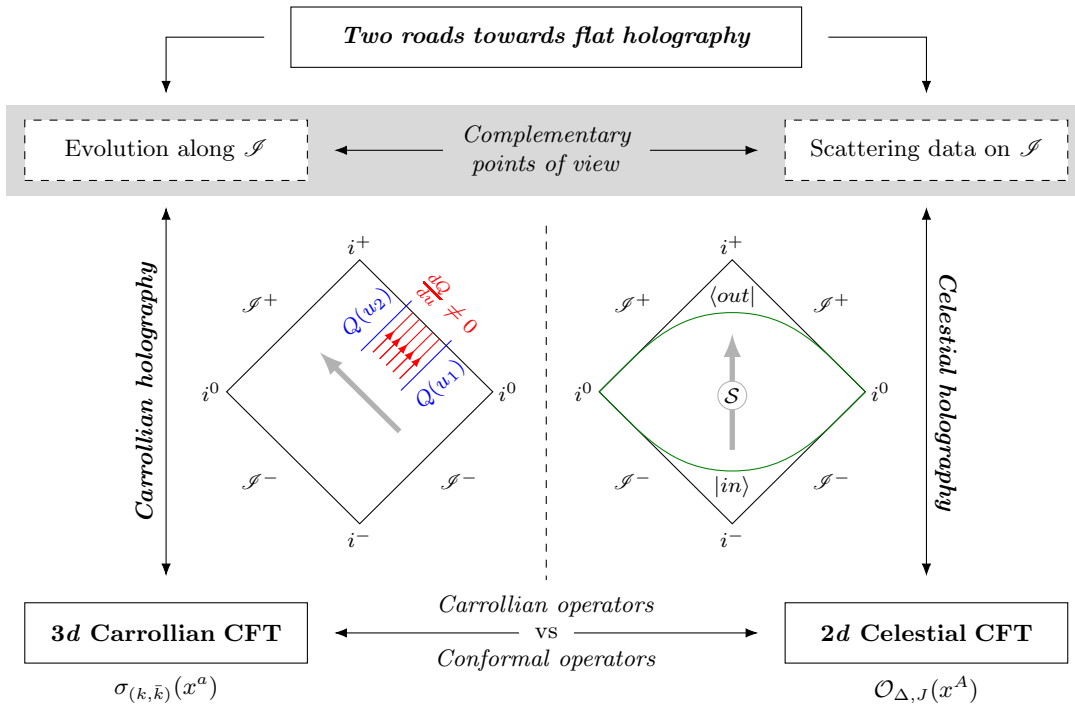

Though these two approaches to flat space holography seem in apparent tension, it has been argued in \cite{Donnay:2022aba} (see also \cite{Bagchi:2022emh}) that they are in fact complementary to each other, as depicted in Fig. \ref{fig:equiv}. Explicit links between them can be established; in particular, Carrollian source operators $\sigma_{(k,\bar k)}$ living at null infinity can be mapped to CCFT operators $\mathcal O_{\Delta, J}$ living on the celestial sphere after using an appropriate integral transform. This allows to relate the correlation functions between the sourced Carrollian CFT and those of the CCFT. In particular, the Ward identities of the sourced Carrollian CFT are then found to be equivalent to those of the CCFT encoding the bulk soft theorem. In this paper, we will pursue our previous analysis of \cite{Donnay:2022aba} and provide more details about the interplay between the Carrollian and the celestial approaches to flat space holography. 

\paragraph{Summary of the paper} The aims of this paper are threefold. The first objective is to review the key ingredients needed to bridge Carrollian and celestial holography. These both build up on the fact that the $\mathcal{S}$-matrix is in a sense holographic by nature, namely scattering amplitudes in the bulk can be identified with correlation functions at the boundary. Depending on the choice of basis for the fields, position space or Mellin space, one ends up with the Carrollian or the celestial approach. We will also review the asymptotic symmetry analysis of both electrodynamics and gravity. These will constitute the two concrete examples that we shall discuss in our set-up. 

After this review part, the second objective is to improve the Carrollian holography proposal. As mentioned above, one of the main obstructions to this approach (obstruction $(ii)$) is the non-conservation of the asymptotic charges $Q$ at null infinity due to the presence of radiation, as illustrated in the left part of Fig. \ref{fig:equiv}. 
We argue that, from the point of view of the dual theory, these dissipative properties can be encoded by coupling the theory to some external sources. We discuss the general framework needed to treat symmetries in that context. This allows us to write flux-balance laws and Ward identities for a sourced field theory. Applying these considerations to a sourced Carrollian CFT, we show that the sourced Ward identities are able to describe the asymptotic dynamics of $4d$ asymptotically flat spacetimes given an appropriate holographic map between the bulk and boundary quantities. In particular, the external sources are argued to encode holographically the radiation in the bulk, which in gravity is contained in the asymptotic shear and the Bondi news tensor.

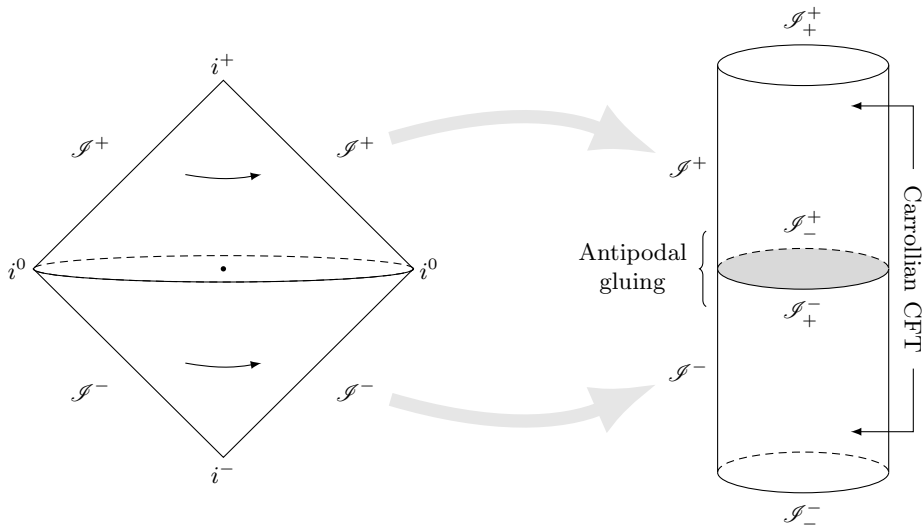
\begin{figure}[ht!]
\begin{center}
\begin{tikzpicture}[scale=0.5]
	\draw[white,opacity=0] (-6,-8) -- (-6,8) -- (19,8) -- (19,-8) -- cycle;
	\coordinate (top) at (0, 5);
	\coordinate (bot) at (0,-5);
	\coordinate (lft) at (-5,0);
	\coordinate (rgt) at ( 5,0);
	\draw[] (top) -- (lft) -- (bot) -- (rgt) -- cycle;
	\draw[densely dashed] (0,0) ellipse (5 and 0.35);
	\draw[] (5,0) arc (0:-180:5 and 0.35);
	\draw[] (lft) node[left,inner sep=2pt]{$i^0$};
	\node[right,inner sep=2pt] (inot) at (rgt) {$i^0$};
	\draw[] (top) node[above,inner sep=2pt]{$i^+$};
	\draw[] (bot) node[below,inner sep=2pt]{$i^-$};
	\draw[] ($(lft)!0.5!(top)+(-0.3,0.3)$) node[anchor=south east,inner sep=2pt]{$\mathscr{I}^+$};
    \draw[] ($(lft)!0.5!(bot)+(-0.3,-0.3)$) node[anchor=north east,inner sep=2pt]{$\mathscr{I}^-$};
    \node[anchor=south west,inner sep=2pt] (scriP) at  ($(rgt)!0.5!(top)+(0.3,0.3)$) {$\mathscr{I}^+\ $};
    \node[anchor=north west,inner sep=2pt] (scriM) at ($(rgt)!0.5!(bot)+(0.3,-0.3)$) {$\mathscr{I}^-\ $};
	\coordinate (midb) at ($(bot)!0.25!(top)$);
	\coordinate (midt) at ($(bot)!0.75!(top)$);
	\draw[-latex] ($(midb)-(1,0)$) to [bend right=10] ($(midb)+(1,0)$);
	\draw[-latex] ($(midt)-(1,0)$) to [bend right=10] ($(midt)+(1,0)$);
	\draw[] (0,0) node[circle, fill=black, minimum size=2pt,inner sep=0pt]{};
	%
    	\def\cx{4.5}; \def\cy{5.4}; \def\arq{0.54};
    	\def\decx{13}; \def\decy{-5.4};
    	\coordinate (smL) at ($(0,0)+(\decx,\decy)$);
    	\coordinate (smR) at ($(\cx,0)+(\decx,\decy)$);
    	\coordinate (szL) at ($(0,\cy)+(\decx,\decy)$);
    	\coordinate (szR) at ($(\cx,\cy)+(\decx,\decy)$);
    	\coordinate (spL) at ($(0,2*\cy)+(\decx,\decy)$);
    	\coordinate (spR) at ($(\cx,2*\cy)+(\decx,\decy)$);
    	\draw[] (smL) -- (spL);
    	\draw[] (smR) -- (spR);
    	\draw[densely dashed] (smL) arc[x radius=\cx/2, y radius=\arq, start angle=180, end angle=0];
    	\draw[] (smL) arc[x radius=\cx/2, y radius=\arq, start angle=-180, end angle=0];
    	\fill[black!15] (szR) arc[x radius=\cx/2, y radius=\arq, start angle=0, end angle=360];
    	\draw[densely dashed] (szL) arc[x radius=\cx/2, y radius=\arq, start angle=180, end angle=0];
    	\draw[] (szL) arc[x radius=\cx/2, y radius=\arq, start angle=-180, end angle=0];
    	\draw[] (spL) arc[x radius=\cx/2, y radius=\arq, start angle=180, end angle=0];
    	\draw[] (spL) arc[x radius=\cx/2, y radius=\arq, start angle=-180, end angle=0];
    	\node[left] (scriMc) at ($(smL)!0.5!(szL)$) {$\mathscr I^-$};
    	\node[left] (scriPc) at ($(szL)!0.5!(spL)$) {$\mathscr I^+$};
    	\node[above] at ($(spL)!0.5!(spR)+(0,\arq)$){$\mathscr I^+_+$};
    	\node[below] at ($(smL)!0.5!(smR)-(0,\arq)$){$\mathscr I^-_-$};
    	\node[above] at ($(szL)!0.5!(szR)+(0,\arq)$){$\mathscr I^+_-$};
    	\node[below] at ($(szL)!0.5!(szR)-(0,\arq)$){$\mathscr I^-_+$};
    	\node[left,align=center,text width=2cm] (antipod) at (szL) {Antipodal gluing};
    	\draw[decorate,decoration={brace}] ($(szL)+(-0.3,-1)$) -- ($(szL)+(-0.3,1)$);
    	\node[rotate=-90,above,outer sep=3pt,align=center] (cft) at (szR) {Carrollian CFT};
    	\draw[-latex] (cft.west) |- ($(szR)!0.8!(spR)-(1,0)$);
    	\draw[-latex] (cft.east) |- ($(szR)!0.8!(smR)-(1,0)$);
    	\draw[-Latex,black!10,line width=5pt] (scriP) to [bend left=20] (scriPc);
    	\draw[-Latex,black!10,line width=5pt] (scriM) to [bend right=20] (scriMc);
\end{tikzpicture}
\end{center}
\caption{Geometric implementation of the antipodal matching.}
\label{fig:antipodal_schematic}
\end{figure}

While the insertion of external sources locally spoils the symmetries of the theory, leading to non-conservation of the currents and dissipation, one can still extract some useful constraints implied by the symmetries on the correlators by using some holographic inputs. A crucial ingredient to implement this constraint is to consider that the Carrollian CFT is not living on $\mathscr{I}^+$ and $\mathscr{I}^-$ separately, but on the whole conformal boundary $\hat{\mathscr{I}} = \mathscr{I}^+ \sqcup \mathscr{I}^-$ obtained by gluing antipodally $\mathscr{I}^+$ and $\mathscr{I}^-$ along $\mathscr{I}^+_-$ and $\mathscr{I}^-_+$, which provides a geometric implementation of the antipodal matching from the Carrollian point of view, see Fig. \ref{fig:antipodal_schematic}.

After providing details of this construction, we show that the sourced Ward identities of the Carrollian CFT integrated over $\hat{\mathscr{I}}$ determine the general form of low-point ($1$-, $2$- and $3$-point) correlation functions of this theory. In particular, for the $2$-point function, we exhibit a new branch of solutions that does not seem to have appeared in previous literature. We then argue that this new branch is precisely the one that is relevant for holographic Carrollian CFT. We relate it to the usual bulk propagator via Fourier transform.

The third objective is to provide further details on the relation between the Carrollian and the celestial approaches to flat space holography. We show that the Carrollian source operators are mapped on celestial operators through an integral transform, coined as the $\mathcal{B}$-transform, which is the combination of a Mellin and a Fourier transform. This allows to relate the correlation functions in the two theories. After providing the properties of the $\mathcal{B}$-transform, we demonstrate the equivalence between the sourced Ward identities in the Carrollian CFT that holographically encode the asymptotic bulk dynamics and the Ward identities in the CCFT that encode the leading and subleading soft theorems. For each of these two points, we consider both electrodynamics and gravity in the bulk.

\paragraph{Organization of the paper} The remaining of the paper is organized as follows. In Section \ref{sec:Massless scattering in flat spacetime}, we review the scattering of massless particles in flat spacetime, which serves to introduce important notations and conventions for the remaining of the article. In Section \ref{sec:Asymptotic symmetry analysis}, we review the asymptotic symmetry analysis of electrodynamics and gravity in $4d$ asymptotically flat spacetime. In Section \ref{sec:Sourced quantum field theory} we develop a framework to treat the symmetries and their associated flux-balance laws / Ward identities in presence of external sources. We then specify this formalism to sourced theories exhibiting $U(1)$ or conformal Carrollian symmetries. In Section \ref{sec:Holographic conformal Carrollian field theory} we provide more insights about the nature of the proposed $3d$ holographic sourced conformal Carrollian field theory. We derive the low-point functions for this theory using the sourced Ward identities. We also propose a holographic correspondence between bulk metric and boundary Carrollian stress tensor, which allows to show that the sourced Ward identities of the Carrollian CFT encode the BMS flux-balance laws. In Section \ref{sec:Relation with celestial holography}, we relate the sourced conformal Carrollian field theory with the celestial CFT. We provide the integral transform that maps Carrollian source operators to celestial operators and show that the BMS Ward identities in both theories are equivalent. The low-point functions are also related with each other. Finally, in Section \ref{sec:Discussion}, we conclude our analysis with some comments and future directions. This paper is also complemented with some appendices: Appendix \ref{app:Bondi adv and ret} describes our coordinate conventions; Appendix \ref{sec:isomorphism Poincare Carroll} reviews the isomorphism between global conformal Carroll algebra in three dimensions and the Poincar\'e algebra in four dimensions; finally, the last Appendix \ref{sec:Constrains on the Carrollian stress tensor} provides details on the derivation of the classical constraints on the Carrollian stress tensor.

\section{Massless scattering in flat spacetime}
\label{sec:Massless scattering in flat spacetime}

This section is a review of salient features of scattering of massless fields in flat spacetime. We consider a scattering of massless bosonic spin-$s$ fields through the $\mathcal{S}$-matrix approach via perturbation theory in Minkowski spacetime. We describe the asymptotic free states in terms of plane wave and conformal primary wave function bases. The boundary operators are obtained by taking the large radius expansion of the bulk fields as they approach null infinity. Reciprocally, the bulk fields can be reconstructed from the boundary operators using the Kirchhoff-d'Adh\'emar formula that defines a boundary-to-bulk propagator. Finally, we discuss various integral transforms that relate position space, momentum space and Mellin space. This allows us to formulate the scattering problem in position and Mellin spaces, which will be suitable for the subsequent discussion on flat space holography. To write this section, we abundantly used references \cite{Pasterski:2016qvg,Pasterski:2017kqt,Strominger:2017zoo, Donnay:2018neh, Pasterski:2021rjz, Pasterski:2020pdk, Donnay:2022sdg} where complementary material can be found.

\subsection{Bulk operators}
\label{sec:Bulk operators}

\subsubsection{Massless fields in Minkowski spacetime and plane wave expansion}
Let us consider a free massless bosonic spin-$s$ Fronsdal field $\phi^{(s)}_I$ in Minkowski spacetime \cite{osti_4283250,Chang:1967zzc,Singh:1974qz} ($s=0,1,2,\dots$). $I = (\mu_1\mu_2\dots\mu_s)$ is a symmetrized multi-index notation, $\mu_i\in \{0,\dots, 3\}$, and spacetime is covered by standard Cartesian coordinates $X^\mu =\{t,\vec x\}$. The massless spin-$s$ field $\phi^{(s)}_I$ can be conveniently put in the De Donder traceless gauge
\begin{equation}
    \partial^\nu \phi^{(s)}_{\nu\mu_2\dots\mu_s}(X) =0 , \qquad \eta^{\mu\nu}\phi^{(s)}_{\mu\nu\mu_3\dots\mu_s}(X)=0 , \label{eq:de donder}
\end{equation}
in which the equations of motion reduce to
\begin{equation}
    \partial^\mu \partial_\mu \phi^{(s)}_I(X)=0 \label{Alembert}
\end{equation}
and the residual gauge transformations are
\begin{equation}
    \begin{aligned}
    &\delta_\lambda \phi^{(s)}_{\mu_1\dots\mu_s}(X) = \partial_{(\mu_1}\lambda_{\mu_2\dots\mu_s)}(X)\\
    &\text{such that } \partial^\nu \lambda_{\nu\mu_3\dots\mu_s}(X) = 0\ \text{and}\ \eta^{\mu\nu}\lambda_{\mu\nu\mu_4\dots\mu_s}(X)=0.
    \end{aligned}\label{Residual gauge transformations}
\end{equation}
The field can be quantized in the Heisenberg representation and expanded in Fourier modes. Each mode is labelled by an on-shell null $4$-momentum vector (see \textit{e.g.} \cite{Strominger:2017zoo})
\begin{equation}
    p^\mu(\omega,w,\bar w) = \omega\, q^\mu(w,\bar w) ,\quad   q^\mu(w,\bar w) = \frac{1}{\sqrt{2}} \Big(1+w\bar w,w+\bar w,-i(w-\bar w),1-w\bar w\Big) \label{q in terms of w bar w future}
\end{equation}
parametrized by the light-cone energy $\omega>0$ and coordinates $(w,\bar w)$ on the complex plane. Let $\varepsilon^{\pm}_{\mu}(\vec q)$ be the polarization co-vectors,
 \begin{equation}
    \begin{split}
        \varepsilon^{+}_{\mu}(\vec q) &= \partial_w q_\mu = \frac{1}{\sqrt{2}}\big(-\bar w,1,-i,-\bar w\big), \\
        \varepsilon^{-}_{\mu}(\vec q) &= [\varepsilon^{+}_{\mu}(\vec q)]^* = \partial_{\bar w} q_\mu = \frac{1}{\sqrt{2}}\big(-w,1,i,-w\big).
    \end{split}
    \label{epsilon pola}
\end{equation}
We can complete these into a null co-tetrad $\mathcal N = \{q_\mu,  n_\mu, \varepsilon^+_\mu, \varepsilon^-_\mu\}$ satisfying
\begin{equation}
    q^\mu n_\mu = -1,\quad \varepsilon_+^\mu \varepsilon^-_\mu = 1, \quad q^\mu \varepsilon^\pm_\mu = 0 = n^\mu \varepsilon^\pm_\mu,\label{properties of polarization vectors}
\end{equation}
by setting $n_{\mu} \equiv \partial_{w}\partial_{\bar w}q_{\mu}$. The spin-$s$ field in De Donder gauge \eqref{eq:de donder} can be expanded in Fourier modes as
\begin{equation}
    \phi^{(s)}_I(X) = K_{(s)} \sum_{\alpha=\pm} \int \frac{\D^3 p}{(2\pi)^3\, 2p^0} \Big[\varepsilon_I^{*\alpha}(\vec q)\, a^{(s)}_\alpha(\vec p) \, e^{i p^\mu X_\mu} + \varepsilon_I^{\alpha}(\vec q)\, a^{(s)}_\alpha(\vec p)^\dagger \, e^{-i p^\mu X_\mu}\Big] \label{pre-fourier}
\end{equation}
after choosing a Lorentz-invariant measure in momentum space and defining the polarization tensors
\begin{equation}
    \varepsilon^\pm_{\mu_1\dots \mu_s}(\vec q) = \varepsilon^\pm_{\mu_1}(\vec q)\varepsilon^\pm_{\mu_2}(\vec q)\dots \varepsilon^\pm_{\mu_s}(\vec q) , \label{polarisation tensor}
\end{equation}
which are fully symmetric and transverse tensors, \textit{i.e.} $q^{\mu_i}\varepsilon^\pm_{\mu_1\dots\mu_i\dots\mu_s}(\vec q)$ for any $i=1,\dots,s$. The overall constant $K_{(s)}\in\mathbb R^+_0$ may depend on the coupling constant for the relevant spin (\textit{e.g.} the elementary charge $e$ for the spin-$1$ field or the Newton-Cavendish constant $G$ for the spin-$2$ field). One usually takes
\begin{equation}
    K_{(1)} = e\sqrt{\hbar},\quad K_{(2)} = \sqrt{32\pi G\hbar} . \label{values of Ks}
\end{equation}
 Ladder operators obey the usual commutation relations
\begin{equation}
   \left[ a^{(s)}_\alpha(\vec p),a^{(s)}_{\alpha'}(\vec p')^\dagger \right] = (2\pi)^3\,2p^0\,\delta^{(3)}(\vec p-\vec p') \label{a comm adagger in p}\,\delta_{\alpha,\alpha'}
\end{equation}
as induced by the canonical commutation relations of the field \eqref{pre-fourier}.

Importantly, a residual gauge transformation \eqref{Residual gauge transformations} driven by $\lambda_{\mu_2 ... \mu_s}(X)$ admitting a well defined Fourier transform acts on the field \eqref{pre-fourier} by addition of a term of the form
\begin{equation}
    \partial_{(\mu_1}\lambda_{\mu_2 \dots \mu_s)}(X) = \int \frac{\D^3 p}{(2\pi)^3\, 2p^0} \Big[ p_{(\mu_1}\hat \lambda_{\mu_2 \dots \mu_s)}(\vec p) \, e^{i p^\mu X_\mu} + p_{(\mu_1}\hat \lambda^*_{\mu_2 \dots \mu_s)}(\vec p) \, e^{-i p^\mu X_\mu}\Big]. \label{gauge in fourier lambda}
\end{equation}
This impacts the polarization tensors as
\begin{equation}
\delta_{\lambda}\varepsilon^\alpha_{\mu_1\dots\mu_s}(\vec q) = q_{(\mu_1} \tilde\lambda^\alpha_{\mu_2\dots\mu_s)}(\vec p) \label{gauge in fourier}
\end{equation}
writing $\hat\lambda_{\mu_2\dots\mu_s}(\vec p) \equiv \frac{K_{(s)}}{\omega}\tilde\lambda^\alpha_{\mu_2\dots\mu_s}(\vec p)a_\alpha^{(s)}(\vec p)$ and $\eta^{\mu\nu}\tilde\lambda^\alpha_{\mu\nu\mu_4\dots\mu_s}=0$, but the ladder operators $a^{(s)}_{\alpha}$ are left invariant by these gauge transformations. However, as we will review later, asymptotic gauge symmetries (whose generating parameters do not have finite energy) do act non trivially on the operators.

Transforming the integration measure for the parametrization \eqref{q in terms of w bar w future}, we can write \begin{equation}
    \phi^{(s)}_I (X) 
    = \frac{K_{(s)}}{16\pi^3} \sum_{\alpha=\pm}  \int \omega\,\D\omega\,\D^2 w \Big[a^{(s)}_\alpha(\omega, w,\bar w) \varphi_I^{*\alpha}(\omega,w,\bar w|X) + a^{(s)}_\alpha(\omega,w,\bar w)^\dagger \varphi_I^{\alpha}(\omega,w,\bar w|X) \Big]
 \label{fourier decomposition bulk}
\end{equation}
where $\D^2 w = i\,\D w\D \bar w$ denotes the integration measure on the complex plane with local holomorphic coordinates $(w,\bar w)$ and
\begin{equation}
    \varphi_I^{*\alpha}(\omega,w,\bar w|X) \equiv \varepsilon_I^{*\alpha}(w,\bar w)\, e^{i \omega q^\mu X_\mu} \label{plane wave basis}
\end{equation}
are the basis vectors of plane waves. In this parametrization, the canonical commutation relations \eqref{a comm adagger in p} become
\begin{equation}
    \left[ a^{(s)}_\alpha(\omega,w,\bar w),a^{(s)}_{\alpha'} (\omega',w',\bar w')^\dagger\right] = 16\pi^3 \, \omega^{-1}\delta(\omega-\omega')\,\delta^{(2)}(w-w')\,\delta_{\alpha,\alpha'} . \label{commu a adagger}
\end{equation}
At the quantum level, $a^{(s)}_+(\omega, w,\bar w)^\dagger$ (resp. $a^{(s)}_-(\omega, w,\bar w)^\dagger$) creates a massless particle of spin $s$ and helicity $J = +s$ (resp. $J=-s$), with energy $\omega$ and a null momentum pointing towards the direction $q^{\mu}(w,\bar w)$.

Poincaré transformations $X'^\mu = {\Lambda^\mu}_\nu X^\nu + t^\mu$ act on the gauge field as
\begin{equation}
    \phi^{(s)}_{\mu_1\mu_2\dots\mu_s}(X)\mapsto \phi'^{(s)}_{\mu_1\mu_2\dots\mu_s}(X') = {\Lambda_{\mu_1}}^{\nu_1}{\Lambda_{\mu_2}}^{\nu_2}\dots {\Lambda_{\mu_s}}^{\nu_s}\phi^{(s)}_{\nu_1\nu_2\dots \nu_s}(X) \label{lorentz rep}.
\end{equation}
Lorentz transformations induce a $SL(2,\mathbb C)$ Möbius transformation
\begin{equation}
w\mapsto w'(w) = \frac{aw+b}{cw+d} \label{mobius}
\end{equation}
with $ad-bc = 1$ on the complex coordinates determining the direction of the null momentum $q^\mu$ with the embedding \eqref{q in terms of w bar w future} of the Riemann sphere into the light cone. The expression of the matrix ${\Lambda^\mu}_\nu$ in terms of the Möbius parameters $(a,b,c,d)$ can be found \textit{e.g.} in \cite{Oblak:2015qia}. Since $p^\mu$ is a Lorentz vector, one has
\begin{equation}
\omega' = \left|\frac{\partial w'}{\partial w}\right|^{-1}\omega,\quad q^\mu(w',\bar w') = \left|\frac{\partial w'}{\partial w}\right|{\Lambda^\mu}_\nu q^\nu(w,\bar w). \label{transfo omega et q}
\end{equation}
The second equation states the fact that a Lorentz transformation on the light cone induces the corresponding M\"obius transformation on the Riemann sphere via the embedding \eqref{q in terms of w bar w future}. Owing to \eqref{epsilon pola} and \eqref{transfo omega et q}, one can show that $\varepsilon_\mu^\pm(w,\bar w)$ do not transform homogeneously under the action of \eqref{mobius} but the supplementary terms are part of the residual gauge freedom \eqref{gauge in fourier}, \textit{i.e.}
\begin{equation}
\varepsilon'^\pm_\mu(w',\bar w') = \left(\frac{\partial w'}{\partial w}\right)^{\mp\frac{1}{2}}\left(\frac{\partial \bar w'}{\partial \bar w}\right)^{\pm\frac{1}{2}}
{\Lambda_\mu}^\nu \varepsilon_\nu^\pm(w,\bar w) + \mathscr A(w,\bar w) {\Lambda_\mu}^\nu q_\nu(w,\bar w), \label{transfo pola conformal}
\end{equation}
where $\mathscr A$ is a fixed function of $(w,\bar w)$. For the expansion \eqref{fourier decomposition bulk}--\eqref{plane wave basis} recalling that the integration measure and the plane wave are Lorentz-invariant, we deduce from \eqref{transfo pola conformal} that the transformation of ladder operators under the Poincaré group is
\begin{equation}
    a_\pm'^{(s)}(\omega',w',\bar w') = \left(\frac{\partial w'}{\partial w}\right)^{-\frac{J}{2}}\left(\frac{\partial \bar w'}{\partial \bar w}\right)^{\frac{J}{2}}\,e^{-i\omega q^{\mu}(w, \bar w) {\Lambda^{\nu}}_{\mu} t_{\nu}}\,a_\pm^{(s)}(\omega,w,\bar w). \label{transfo a poincare}
\end{equation}
In particular, one recovers that the ladder operators are eigenvectors of translations, which is expected since they are assumed to create/annihilate energy eigenstates. The infinitesimal version of \eqref{transfo a poincare} can be obtained by setting $X'^\mu = X^\mu - \epsilon\xi^\mu$, with $\xi^\mu = {\varpi^\mu}_\nu X^\nu +\tau^\mu$ ($\varpi_{\mu\nu} = \varpi_{[\mu\nu]}$) and $w'(w) = w-\epsilon \mathcal Y^w(w)$, and retaining only the linear terms in $\epsilon$. One concludes that
\begin{equation}
    \begin{split}
        &\delta_{\xi(\mathcal T,\mathcal Y)}a_\pm^{(s)}(\omega,w,\bar w) \\
        &= \left[ -i\omega\mathcal T + \mathcal Y^w\partial_w + \mathcal Y^{\bar w}\partial_{\bar w} + \frac{J}{2}\partial_w \mathcal Y^w - \frac{J}{2}\partial_{\bar w} \mathcal Y^{\bar w} -\frac{\omega}{2} (\partial_w \mathcal Y^w + \partial_{\bar w} \mathcal Y^{\bar w})\partial_\omega  \right] a_\pm^{(s)}(\omega,w,\bar w)
    \end{split}
    \label{transfo a poincare infinitesimal}
\end{equation}
where $\xi(\mathcal T,\mathcal Y)$ is now parametrized by the function $\mathcal T(w,\bar w) = -q^{\mu}(w,\bar w)\tau_\mu$ and the vector $\mathcal Y = \mathcal Y^w(w)\partial_w + \mathcal Y^{\bar w}(\bar w)\partial_{\bar w}$ on the Riemann sphere, while acting in momentum space covered by coordinates $(\omega,w,\bar w)$.

\subsubsection{Mellin transform and conformal primary basis}
\label{sec:Mellin transform section}

In \eqref{fourier decomposition bulk}, we expanded the massless spin-$s$ field in the plane wave basis \eqref{plane wave basis} whose elements are energy eigenstates. Another convenient choice is the conformal primary wave function basis \cite{deBoer:2003vf,Cheung:2016iub,Pasterski:2016qvg,Pasterski:2017kqt} whose elements are boost eigenstates. This basis trades the energy parameter $\omega$ for the eigenvalue $\Delta$ of the Lorentz boost along the direction fixed by the null momentum $p^\mu$. 

The map between the two bases is given by the Mellin transform defined as
\begin{equation}
    F(\Delta) = \mathcal M[f(\omega),\Delta] \equiv \int_0^{+\infty} \D\omega\,\omega^{\Delta-1}\, f(\omega)
    \label{definition mellin}
\end{equation}
for $f : \mathbb R^+ \to \mathbb C$ where $\Delta = c+i\nu$ is generically complex (see \textit{e.g.} \cite{titchmarsh1948introduction} for the precise assumptions that make the integral \eqref{definition mellin} well defined). 
According to the Mellin inversion theorem, if $F(\Delta)$ is analytic in the complex strip defined by $c\in \ ]a,b[\ \subset\mathbb R$ and the integral
\begin{equation}
f(\omega) = \frac{1}{2\pi }\int_{-\infty}^{+\infty}\D \nu \, \omega^{-(c+i\nu)}F(c+i\nu) \label{inverse mellin}
\end{equation}
converges for any $c \in ]a,b[$, then it defines the inverse Mellin transform $f(\omega)\equiv \mathcal M^{-1}[F(\Delta),\omega]$ acting on $F(\Delta)$.

The Mellin transform of the plane waves \eqref{plane wave basis}, for $\omega>0$, yields (see \textit{e.g.} \cite{deBoer:2003vf,Pasterski:2016qvg,Pasterski:2017kqt})
\begin{equation}
    \begin{split}
        V_{I}^{*\alpha}(\Delta,w,\bar w|X) &= \lim_{\epsilon \to 0^+} \varepsilon^{*\alpha}_I(w,\bar w) \int_0^{+\infty}\D\omega\, \omega^{\Delta-1}\, e^{i \omega q^{\mu} X_{\mu} -\epsilon\omega} \\
        &= \lim_{\epsilon \to 0^+} \varepsilon^{*\alpha}_I(w,\bar w) \frac{ i^\Delta \Gamma[\Delta]}{(q^{\mu} X_{\mu} + i \epsilon)^\Delta} 
    \end{split} \label{mellin of plane wave}
\end{equation}
where $\epsilon>0$ is a regulator that can be arbitrarily close to zero. The computation of the integral \eqref{mellin of plane wave} is made particularly easy if one uses Ramanujan's master theorem, stating that if $f(\omega)$ is a complex-valued function on the positive real axis admitting a power-series expansion of the form $f(\omega) = \sum_{k=0}^{+\infty} \frac{\varphi(k)}{k!}(-\omega)^k$ in some neighborhood of the origin, then its Mellin transform is simply given by $\mathcal M[f(\omega),\Delta] = \Gamma[\Delta]\varphi(-\Delta)$ in a certain range of validity $\Delta\in\mathcal R\subset\mathbb C$. For the plane wave, the regulation of the integral \eqref{mellin of plane wave} by a decaying exponential function $\sim\, e^{-\epsilon\omega}$ allows it to converge in the whole half complex plane $\mathcal R = \{ c+i\nu\,|\, c>0,\nu\in\mathbb R\}$.
 Using the identity
\begin{equation}
    \int_0^{+\infty}\D\omega\, \omega^{i\nu-1} = 2\pi\,\delta(\nu), \label{delta represented in mellin}
\end{equation}
one can show \cite{Pasterski:2017kqt} that the statement that plane waves form a delta-function normalizable basis     
for the Klein-Gordon inner product translates, after Mellin transform, into requiring that $\Delta$ lays on the principal continuous series of the irreducible unitary representations of the Lorentz group, \textit{i.e.} $\Delta = 1+i\nu$.

Let us first consider the expansion of the bulk field \eqref{fourier decomposition bulk}, where we recall that the polarization is taken as products of \eqref{epsilon pola}. In terms of the Mellin representatives \eqref{mellin of plane wave}, such an expansion reads \cite{Donnay:2018neh,Donnay:2020guq}
\begin{equation}
    \phi^{(s)}_I (X) 
    = \frac{K_{(s)}}{32\pi^4} \sum_{\alpha=\pm}\int \D \nu\, \D^2 w \left[a^{(s)}_{2-\Delta,\alpha}(w,\bar w) V_I^{*\alpha}(\Delta,w,\bar w|X) + a^{(s)}_{2-\Delta,\alpha}(w,\bar w)^\dagger V^{\alpha}_I (\Delta,w,\bar w|X) \right]
    \label{Phi expanded in Mellin bulk}
\end{equation} 
where we defined the ladder operators in the Mellin basis as in \cite{Pasterski:2021dqe} 
\begin{equation}
\begin{aligned} 
a_{\Delta, \alpha}^{(s)}(w,\bar w) &\equiv \mathcal M\big[a^{(s)}_\alpha(\omega,w,\bar w),\Delta\big] = \int_0^{+\infty} \D\omega\,\omega^{\Delta-1}\,a^{(s)}_\alpha(\omega,w,\bar w), \\
    a_{\Delta, \alpha}^{(s)}(w,\bar w)^\dagger &\equiv \mathcal M\big[a^{(s)}_\alpha(\omega,w,\bar w)^\dagger,\Delta\big] = \int_0^{+\infty} \D\omega\,\omega^{\Delta-1}\,a^{(s)}_\alpha(\omega,w,\bar w)^\dagger. 
     \label{mellin O}
\end{aligned}
\end{equation}
These last relations can be inverted as
\begin{equation}
    \begin{split}
        a^{(s)}_\alpha(\omega, w,\bar w) &= \frac{1}{2\pi} \int_{-\infty}^{+\infty}\D\nu \, \omega^{-\Delta} \, a^{(s)}_{\Delta, \alpha}(w,\bar w) , \\ a^{(s)}_\alpha(\omega, w,\bar w)^\dagger &= \frac{1}{2\pi} \int_{-\infty}^{+\infty}\D\nu \, \omega^{-\Delta} \, a^{(s)}_{\Delta, \alpha}(w,\bar w)^\dagger .
    \end{split}
\label{ladder op inverse mellin}
\end{equation}
Crucially, applying the Mellin transforms \eqref{mellin O} to \eqref{transfo a poincare}, it has been observed that these operators transform as
\begin{equation}
a'^{(s)}_{\Delta,\alpha}(w',\bar w') = \left(\frac{\partial w'}{\partial w}\right)^{ -\frac{\Delta +J}{2}}\left(\frac{\partial \bar w'}{\partial \bar w}\right)^{-\frac{\Delta -J}{2}} a^{(s)}_{\Delta,\alpha}(w,\bar w).
\label{a delta transfo lorentz finite}
\end{equation}
under the action of the Lorentz group, which is precisely the definition of a conformal primary field of conformal dimension $\Delta$ and $2d$ spin $J=\alpha s$ (see \textit{e.g.} \cite{DiFrancesco:1997nk}). The action of the Poincaré translations in the Mellin basis is given by \cite{Stieberger:2018onx}
\begin{equation}
a_{\Delta,\alpha}'^{(s)}(w',\bar w') = e^{-i q^\mu(w,\bar w)t_\mu \hat\partial_{\Delta}}a^{(s)}_{\Delta,\alpha}(w,\bar w) = \sum_{n=0}^{+\infty} \frac{\left(-i q^\mu(w,\bar w)t_\mu\right)^n}{n!} a^{(s)}_{\Delta+n,\alpha}(w,\bar w) \label{transfo a mellin translations}
\end{equation}
defining the discrete derivative operator $\hat\partial_\Delta F(\Delta) \equiv F(\Delta+1)$. The infinitesimal action of Poincaré transformations in the Mellin basis reads as
\begin{equation}
\begin{split}
    &\delta_{\xi(\mathcal T,\mathcal Y)} a^{(s)}_{\Delta,\alpha}(w,\bar w) = \left(-i\mathcal T\hat\partial_{\Delta} + \mathcal{Y}^w \partial_{w} + \mathcal Y^{\bar w} \partial_{\bar w} + \frac{\Delta+ J}{2} \partial_{w} \mathcal{Y}^w  + \frac{\Delta- J}{2} \partial_{\bar w} \mathcal Y^{\bar w}\right) a^{(s)}_{\Delta,\alpha}(w,\bar w).
\end{split}
\label{a delta transfo poincare inf}
\end{equation}

\paragraph{Remark} The functions $V_{I}^\alpha(\Delta,w,\bar w|X)$ defined in \eqref{mellin of plane wave} are not strictly speaking boost eigenstates since they do not transform covariantly under the action of the Lorentz group: this is due to the transformation law of the polarization vector \eqref{transfo pola conformal}. Instead, one can define the conformal primary wave functions \cite{Pasterski:2017kqt} by rescaling the functions $V_{I}^\alpha(\Delta,w,\bar w|X)$ and performing a gauge transformation
\begin{equation}
    A_I^\alpha (\Delta,w,\bar w|X) = c^{(s)}(\Delta) V_{I}^\alpha(\Delta,w,\bar w|X) + \partial_{(\mu_1} f^{(s)\alpha}_{\mu_2 \dots \mu_s)} (\Delta,w,\bar w|X) . \label{def A in terms of V}
\end{equation} 
The precise expressions of scaling factor $c^{(s)}(\Delta)$ and gauge parameter $f^{(s)\alpha}_{\mu_2 \dots \mu_s} (\Delta,w,\bar w|X)$ will depend on the spin $s$ (explicit expressions for $s=1,2$ can be found \textit{e.g.} in \cite{Pasterski:2017kqt,Donnay:2020guq}). It is convenient to use the compact form \cite{Pasterski:2020pdk}
\begin{equation}
    A_I^\pm(\Delta,w,\bar w|X) =  m^\pm_{\mu_1}(w,\bar w|X)\dots m^\pm_{\mu_s}(w,\bar w|X) \frac{1}{(q_\nu X^\nu)^\Delta},
    \label{conformal primaries true}
\end{equation}
for any (integer) spin $s$, where $m_\mu^\pm(w,\bar w|X)$ represent the modified polarization co-vectors
\begin{equation}
    m^\pm_{\mu}(w,\bar w|X)\equiv \varepsilon^\pm_\mu(w,\bar w) - \frac{\varepsilon^\pm_\nu X^\nu}{q_\rho X^\rho} q_\mu (w,\bar w). \label{def m}
\end{equation}
One can show that the set of functions \eqref{conformal primaries true} are actual boost eigenstates, because now \eqref{def m} transforms homogeneously under $SL(2,\mathbb C)$ transformations \eqref{mobius}, \textit{i.e.}
\begin{equation}
m'^\pm_\mu(w',\bar w'|X') = \left(\frac{\partial w'}{\partial w}\right)^{\mp\frac{1}{2}}\left(\frac{\partial \bar w'}{\partial \bar w}\right)^{\pm\frac{1}{2}}
{\Lambda_\mu}^\nu m^\pm_\nu(w,\bar w|X).
\end{equation}
While the basis elements $V_I^\alpha(\Delta,w,\bar w|X)$ only satisfy the De Donder gauge fixing conditions \eqref{eq:de donder} because $q^\mu \varepsilon^\pm_\mu = 0$, the new basis functions $A_I^\alpha(\Delta,w,\bar w|X)$ also satisfy the radial (or Fock-Schwinger) gauge condition $X^\mu A^\alpha_{\mu\mu_2\dots\mu_s}(\Delta,w,\bar w|X) = 0$ because $X^\mu m_\mu^\pm = 0$ \cite{Pasterski:2017kqt}. The price to pay for trading $\varepsilon^\pm_\mu$ for $m_\mu^\pm$ is thus the breaking of manifest translation invariance, because the supplementary gauge fixing is sensitive to the choice of an origin. Notice that the fixation of the radial gauge is compatible with the De Donder gauge, as argued in \cite{Magliaro:2007qr}. The so-called celestial operators are defined as \cite{Donnay:2020guq}
\begin{equation}
\mathscr{O}^{(s)}_{\Delta, \alpha}(w,\bar w) \equiv \left\langle A_{I}^{\alpha}(\Delta,w,\bar w|X) , \phi_{I}^{(s)}(X) \right\rangle \label{KG}
\end{equation}
where $\langle \cdot,\cdot \rangle$ denotes the relevant spin-$s$ inner-product. For scalar fields ($s=0$), it will be the usual Klein-Gordon inner product; more involved expressions are needed for $s=1,2$ which can be constructed from symplectic structures available in the literature (see \cite{Donnay:2020guq} and references therein).
Notice that the operators $\mathscr{O}^{(s)}_{\Delta, \alpha}(w,\bar w)$ and $a^{(s)}_{\Delta, \alpha}(w,\bar w)$ are proportional to each other \cite{Donnay:2020guq,Pasterski:2021dqe} (see also \cite{Donnay:2022ijr} for more details). Hence we will persist in using consistently the Mellin ladder operators $a^{(s)}_{\Delta, \alpha}(w,\bar w)$ in the rest of the paper.

\subsection{Boundary operators}
\label{sec: Boundary operators}

\subsubsection{From bulk to boundary: the large-\textit{r} expansion}

In this section, we review how to construct boundary operators by performing a large-$r$ expansion around $\mathscr{I}^+$ and $\mathscr{I}^-$ \cite{Strominger:2017zoo}.

We first analyze the late-time behavior of free massless fields around $\mathscr{I}^+$: these will ultimately approximate outgoing free fields of the $\mathcal{S}$-matrix. We start from the Fourier expansion 
\begin{equation}
    \phi^{(s)}_I (X) 
    = \frac{K_{(s)}}{16\pi^3} \int \omega\,\D\omega\,\D^2 w \,\Big[\varepsilon_I^{*\alpha}(w,\bar w) \, a^{(s)}_\alpha(\omega, w,\bar w) \, e^{i\omega q^\mu X_\mu} + \varepsilon_I^{\alpha}(w,\bar w)\, a^{(s)}_\alpha(\omega,w,\bar w)^\dagger \, e^{-i\omega q^\mu X_\mu} \Big]\label{Fourier expansion 3}
\end{equation}
and use retarded Bondi coordinates $\{u,r,z,\bar z\}$ with flat boundary representatives, as defined in Appendix \ref{app:Bondi adv and ret}, to perform an asymptotic expansion in $r$ near $\mathscr I^+=\{r\to+\infty\}$. Rewriting $X^\mu$ as \eqref{Retarded flat BMS coordinates (appendix)} in terms of retarded Bondi coordinates one has $q^\mu X_\mu = -u-r|z-w|^2$. We now introduce polar coordinates in the complex plane as $z-w=\rho\, e^{i\vartheta}$. The integration measure becomes $\D^2 w\to 2\,\rho\,\D\rho\,\D\vartheta$ and altogether \eqref{Fourier expansion 3} can be rewritten as
\begin{equation}
\phi^{(s)}_I (X) = \frac{K_{(s)}}{8\pi^3} \int_{0}^{+\infty}\D \omega\,\omega \int_0^{+\infty}\D \rho \, \rho \int_0^{2\pi}\D\vartheta \left[\varepsilon^{*\alpha}_I(w,\bar w) a^{(s)}_\alpha(\omega,w,\bar w) e^{ -i\omega u - i\omega r \rho^2} + \text{h.c.} \right], \label{phi to compute}
\end{equation}
where ``$\text{h.c.}$'' stands for hermitian conjugation. The expansion of \eqref{phi to compute} in the limit $r\to+\infty$ (as approaching $\mathscr I^+$) is given by the stationary phase approximation of the $\rho$-integral around the point $\rho = 0$ corresponding to the situation where $q$ and $x$ are collinear. Evaluating the integral in $\rho$ for large-$r$ gives the property \cite{Donnay:2022sdg}
\begin{equation}
    \rho\,e^{-i\omega r\rho^2} = -\frac{i}{2r\omega}\delta(\rho) + \mathcal O(r^{-2}). \label{saddle point}
\end{equation}
In Bondi retarded coordinates, the expressions of the elements of the co-tetrad $\mathcal N$ are
\begin{equation}
    \begin{split}
        q_\mu\D X^\mu &= -\D u - |z-w|^2\D r - r(\bar z-\bar w)\D z - r( z- w)\D \bar z ,\quad n_\mu\D X^\mu = -\D r,\\
        \varepsilon_\mu^+\D X^\mu &= (\bar z-\bar w)\,\D r+r\,\D \bar z , \quad \varepsilon_\mu^-\D X^\mu = ( z- w)\,\D r+r\,\D z,
    \end{split} \label{Polarisation in BMS coordinates}
\end{equation}
and, in the collinear limit $w = z$, we have $\varepsilon_\mu^+\D X^\mu = r\,\D\bar z$ and $\varepsilon_\mu^-\D X^\mu = r\,\D z$.
Using \eqref{saddle point}, the leading components of $\phi^{(s)}_I(X)$ near $\mathscr I^+$ are thus
\begin{equation}
\phi^{(s)}_{z\dots z} (X) = -\frac{i K_{(s)}}{8\pi^2}r^{s-1} \int_{0}^{+\infty}\D \omega  \left[ a_+^{(s)}(\omega,z,\bar z) e^{ -i\omega u} - a_-^{(s)}(\omega,z,\bar z)^\dagger e^{+i\omega u}\right] + \mathcal O(r^{s-2}), \label{Phi s final}
\end{equation} 
and its complex conjugated component $\phi^{(s)}_{\bar z\dots\bar z}(X)$ as well as components of the form $\phi^{(s)}_{ r z \dots z}(X)$, $\phi^{(s)}_{r rz\dots z}(X)$ \textit{etc.}, all generically of order $\mathcal{O}(r^{s-1})$ as well. Indeed, the subleading $\mathcal O(r^{-2})$ terms in \eqref{saddle point} contribute in the radial component of in \eqref{Polarisation in BMS coordinates}, recalling that $\D r = -r^2 \D (r^{-1})$,  producing a contribution at leading order. One can extract the \emph{boundary value} of the field by
\begin{equation}
    \begin{split}
        &\bar{\phi}^{(s)}_{z\dots z} (u, z, \bar z)\, \D z \otimes \cdots \otimes \D z + \bar{\phi}^{(s)}_{\bar z\dots \bar z} (u, z, \bar z)\, \D \bar z \otimes \cdots \otimes \D \bar z \\
        &\equiv \lim_{r\to+\infty} \iota^*\left(r^{1-s}\, \phi^{(s)}_{\mu_1\dots\mu_s}\D X^{\mu_1}\otimes \dots \otimes \D X^{\mu_s}\right)  ,
    \end{split}
\label{Boundary value}
\end{equation} 
where $\iota^*$ denotes the pull-back on the constant $r$ hypersurfaces, hence deleting the leading radial components. Importantly, the resulting fields are independent of the residual gauge ambiguity \eqref{gauge in fourier lambda}--\eqref{gauge in fourier} as can be seen from the fact that $q_{\mu}\D X^{\mu} = \mathcal{O}(1)$ in the collinear limit while $\varepsilon_\mu^\pm\D X^\mu = \mathcal{O}(r)$ (for a discussion on this phenomenon, see \cite{Ashtekar:2017wgq}).

To summarize, the asymptotic behavior of the field $\phi^{(s)}_{I} (X)$ is encoded by the boundary value \eqref{Boundary value}: its  Fourier expansion can be directly read off from \eqref{Phi s final} as \cite{He:2014laa,Strominger:2017zoo}
\begin{equation}
\bar\phi^{(s)}_{z\dots z} (u,z,\bar z) = -\frac{i K_{(s)}}{8\pi^2} \int_{0}^{+\infty}\D \omega  \left[ a_+^{(s)}(\omega,z,\bar z) e^{ -i\omega u} - a_-^{(s)}(\omega,z,\bar z)^\dagger e^{+i\omega u}\right], \label{outgoing spin s}
\end{equation}
while the Fourier modes for $\bar \phi^{(s)}_{\bar z \dots \bar z} = (\bar \phi^{(s)}_{z \dots z})^{\dagger}$ are obtained from  \eqref{outgoing spin s} by exchanging $a^{(s)}_\pm \to a^{(s)}_\mp$. 

\paragraph{Remark} As stated below equation \eqref{Boundary value}, boundary values \eqref{outgoing spin s} of the fields are left invariant under residual gauge transformations  \eqref{Residual gauge transformations} whose parameters $\lambda_{\mu_2\dots\mu_s}(X)$ admit a Fourier decomposition \eqref{gauge in fourier lambda}. Notice that there exist gauge transformations which do not admit such Fourier decomposition and do modify the boundary value of the fields. For instance, picking $s=1$, the action of a gauge transformation $\delta_\lambda \phi^{(1)}_\mu = \partial_\mu \lambda$ of parameter $\lambda$ satisfying $\partial^\mu\partial_\mu\lambda=0$, which is solved asymptotically by $\lambda(u,r,z,\bar z) = \lambda^{(0)}(z,\bar z)+u\, \partial_z\partial_{\bar z} \lambda^{(0)}\, r^{-1}\ln r+\mathcal O (r^{-1})$, would shift the boundary gauge field by 
\begin{equation}
\delta_\lambda \bar \phi^{(1)}_z(u,z,\bar z)=\partial_z \lambda^{(0)}(z,\bar z).  
\label{shift}
\end{equation}
Another point of view is to re-interpret this shift as the following change in the zero mode of the ladder operators
\begin{equation}
    \delta_\lambda^S a^{(s)}_+(\omega,z,\bar z) = -\delta_\lambda^S a^{(s)}_-(\omega,z,\bar z)^\dagger \equiv \frac{8 \pi^2 i}{K_{(1)}}\,\partial_z \lambda^{(0)}(z,\bar z)\,\delta(\omega). \label{add goldstone}
\end{equation}
Hence the operator $\delta_\lambda^S$ (where $S$ stands for ``soft'') only affects the ``Goldstone mode'' (\textit{i.e.} the zero-energy mode transforming in a pure inhomogeneous way) but leaves the bulk field \eqref{Fourier expansion 3} unchanged since $\omega\,\delta(\omega)\simeq 0$ in the sense of distributions. Notice that this mode has to be distinguished from the leading soft photon mode \cite{He:2014cra}
\begin{equation}
   \mathcal N^{(0)}_{z}(z,\bar{z}) \equiv \int_{-\infty}^{+\infty}\D u\,\partial_{u} \bar\phi^{(1)}_z(u,z,\bar z) = \bar{\phi}^{(1)}_{z} (u,z,\bar{z})\big|_{u\to+\infty} - \bar{\phi}^{(1)}_{z}(u,z,\bar z) \big|_{u\to-\infty}
\end{equation}
whose expression in terms of ladder operators is obtained 
using \eqref{outgoing spin s} and $\int_{-\infty}^{+\infty}\D u\,e^{\pm i\omega u} = 2\pi\delta(\omega)$ and gives
\begin{equation}
    \mathcal N^{(0)}_{z}(z,\bar{z}) = -\frac{K_{(1)}}{8\pi}\lim_{\omega\to 0^+}\left[ \omega\,a_+(\omega,z,\bar z) + \omega\,a_-(\omega,z,\bar z)^\dagger \right].
\end{equation}
The non-triviality of the result corresponds to the presence of a pole in the ladder operators in Fourier space. The discussion is easily extended to any spin $s\geq 1$.

The commutation relation \eqref{commu a adagger} implies (see \textit{e.g.} \cite{Ashtekar:1987tt,He:2014laa,He:2014cra})
\begin{equation}
\begin{split}
    \big[ \bar{\phi}^{(s)}_{z\dots z}(u_1,z_1,\bar z_1),\bar{\phi}^{(s)}_{\bar z\dots \bar z} (u_2,z_2,\bar z_2)\big] &= -\frac{i}{4} K_{(s)}^2\, \text{sign}(u_1-u_2)\,\delta^{(2)}(z_1-z_2) , \\
    \big[ \partial_{u_1}\bar{\phi}^{(s)}_{z\dots z}(u_1,z_1,\bar z_1),\bar{\phi}^{(s)}_{\bar z\dots  \bar z} (u_2,z_2,\bar z_2)\big] &= -\frac{i}{2} K_{(s)}^2\, \delta(u_1-u_2)\,\delta^{(2)}(z_1-z_2) .    
\end{split}
\label{commu boundary fields}
\end{equation}
In our notations, $\text{sign}(x)$ is the sign distribution related to the Heaviside and Dirac distributions $\Theta(x)$, $\delta(x)$ as
\begin{equation}
    \text{sign}(x) = 2\,\Theta(x) - 1,\quad \text{sign}'(x) = 2\,\delta(x),
\end{equation}
for the particular choice $\Theta(0) = \frac{1}{2}$. Owing to \eqref{transfo omega et q}, the Fourier transform of \eqref{transfo a poincare} gives
\begin{equation}
\bar\phi'^{(s)}_{z\dots z}(u',z',\bar z') = \left(\frac{\partial z'}{\partial z}\right)^{-\frac{1+J}{2}}\left(\frac{\partial \bar z'}{\partial \bar z}\right)^{-\frac{1-J}{2}} \bar\phi^{(s)}_{z\dots z}(u,z,\bar z) \label{bar phi transfo lorentz}
\end{equation}
in the collinear limit $w=z$, where $u' = \left|\frac{\partial z'}{\partial z}\right|\big(u-q^\mu(z,\bar z){\Lambda_\mu}^\nu t_\nu\big)$ and $J=s$. The complex conjugated component $\bar\phi^{(s)}_{\bar z\dots \bar z}$ transforms in the same way but with the flipped helicity $J=-s$. This translates infinitesimally into
\begin{equation}
\begin{split}
    &\delta_{\xi(\mathcal T,\mathcal Y)} \bar\phi_{z...z}^{(s)}(u, z, \bar z) = \left(\mathcal T + \frac{u}{2}\left(\partial_z \mathcal{Y}^z + \partial_{\bar z} \mathcal{Y}^{\bar z}  \right) \right)\partial_u \bar\phi_{z...z}^{(s)}(u,z,\bar z) \\
    &\quad + \left(\mathcal{Y}^z \partial_{z} + \mathcal{Y}^{\bar z} \partial_{\bar z} + \frac{1+ J}{2} \partial_{z} \mathcal{Y}^z  + \frac{1- J}{2} \partial_{\bar z} \mathcal{Y}^{\bar z}\right) \bar\phi_{z...z}^{(s)}(u, z, \bar z ).
\end{split} \label{Poincare on boundary value}
\end{equation}

The same analysis can be performed for operators in the vicinity of past null infinity $\mathscr{I}^-$. The first step consists in trading the retarded Bondi coordinates $\{u,r,z,\bar z\}$ for the advanced Bondi coordinates $\{v,r',z',\bar z'\}$, also with flat boundary representative (see Appendix \ref{app:Bondi adv and ret}) in order to perform the large-$r'$ expansion around $\mathscr I^- = \{r'=+\infty\}$. According to the change of coordinates \eqref{u to v}, this amounts effectively to make the permutation $\{u\mapsto v, r\mapsto -r', z\mapsto z' \}$ in all expressions. This induces a sign flip in the polarization co-vectors $\varepsilon_\mu^\pm$ in \eqref{Polarisation in BMS coordinates}. As now $q^\mu X_\mu = -v+r'|z'-w|^2$, another global minus sign is induced by the evaluation of the $(w,\bar w)$ integral in \eqref{fourier decomposition bulk} in the stationary phase approximation around the collinear configuration $z'=w$, see \eqref{saddle point}. Therefore, the large-$r'$ limit of the field goes in a parallel way as before and provides the following expansion of operators near $\mathscr I^-$:
\begin{equation}
\phi^{(s)}_{z\dots z} (X) = (-1)^{s}\frac{i K_{(s)}}{8\pi^2}r'^{s-1} \int_{0}^{+\infty}\D \omega  \left[ a_+^{(s)}(\omega,z',\bar z') e^{ -i\omega v} - a_-^{(s)}(\omega,z',\bar z')^\dagger e^{+i\omega v}\right] + \mathcal O(r'^{s-2}). 
\end{equation} 
Because the considered coordinate systems $\{u,r,z,\bar z\}$ and $\{v,r',z',\bar z'\}$ both interpolate between $\mathscr I^-$ and $\mathscr I^+$, the most natural choice to extract the boundary value at $\mathscr I^-$ is
\begin{equation}
    \begin{split}
        &\bar{\phi}^{(s)}_{z\dots z} (v, z, \bar z)\, \D z \otimes \cdots \otimes \D z + \bar{\phi}^{(s)}_{\bar z\dots \bar z} (v, z, \bar z)\, \D \bar z \otimes \cdots \otimes \D \bar z \\
        &\equiv \lim_{r\to-\infty} \iota^*\left(r^{1-s}\, \phi^{(s)}_{\mu_1\dots\mu_s}\D X^{\mu_1}\otimes \dots \otimes \D X^{\mu_s}\right)  
    \end{split}
\label{Boundary value PAST}
\end{equation} 
as before, where the primes on the $z$'s have been dropped because of \eqref{u to v}. This implies 
\begin{equation}
\bar\phi^{(s)}_{z\dots z} (v,z,\bar z) = - \frac{i K_{(s)}}{8\pi^2} \int_{0}^{+\infty}\D \omega  \Big[ a^{(s)}_+(\omega,z,\bar z) e^{- i\omega v} - a^{(s)}_-(\omega,z,\bar z)^\dagger e^{i\omega v}\Big],
 \label{incoming spin s}
\end{equation}
and in particular $\bar\phi^{(s)}_{z\dots z}(u,z,\bar z) \equiv \bar\phi^{(s)}_{z\dots z}(v,z,\bar z)$, which is also consistent with the change of coordinates \eqref{u to v}. This finally shows that the boundary value at $\mathscr I^-$ also transforms as \eqref{bar phi transfo lorentz}--\eqref{Poincare on boundary value} up to the mere replacement $u\mapsto v$.

\subsubsection{From boundary to bulk: the Kirchhoff-d'Adh\'emar formula}

Up to this stage, we have reviewed how a bulk free field induces a boundary field in the asymptotic region. We now make the link with the Kirchhoff-d'Adh\'emar formula \cite{Penrose1980GoldenON,Penrose:1985bww} that describes how to reconstruct the bulk free field from its boundary value via a boundary-to-bulk propagator.

Inverting the Fourier transform \eqref{outgoing spin s}, we get
\begin{equation}
    \begin{split}
        a_{+}^{(s)}(\omega,z,\bar z) & = \frac{4\pi i}{K_{(s)}} \int_{-\infty}^{+\infty} \D u\, e^{i \omega u}\, \bar\phi^{(s)}_{z\dots z}(u, z, \bar z) ,\\
        a_{-}^{(s)}(\omega,z,\bar z)^{\dagger} & = - \frac{4\pi i}{K_{(s)}} \int_{-\infty}^{+\infty} \D u\, e^{-i \omega u}\, \bar\phi^{(s)}_{z\dots z}(u, z, \bar z),
    \end{split} \label{fourier transform formula at scri}
\end{equation}
for $\omega >0$. Rewriting the expression for the field \eqref{fourier decomposition bulk} as
\begin{equation}
\begin{split}
    \phi_I^{(s)}(X) &= \frac{K_{(s)}}{16\pi^3}\int \omega\,\D\omega\, \D^2 w\, \varepsilon_I^{*+}(w,\bar w) \left[a^{(s)}_+(\omega,w,\bar w)e^{i\omega q^\mu X_\mu} + a^{(s)}_-(\omega,w,\bar w)^\dagger e^{-i\omega q^\mu X_\mu}\right] +\text{h.c.}
\end{split}
\end{equation}
and inserting \eqref{fourier transform formula at scri} gives 
\begin{align}
    \phi_I^{(s)}(X) &= \frac{i}{4\pi^2}\int \omega\, \D\omega\, \D^2 w\, \varepsilon^{*+}_I(w,\bar w)\int_{-\infty}^{+\infty} \D\tilde u\left[e^{i\omega(q^\mu X_\mu+\tilde u)} - e^{-i\omega(q^\mu X_\mu+\tilde u)}\right]\bar\phi_{z\dots z}^{(s)}(\tilde u,w,\bar w) + \text{h.c.} \nonumber \\
    &= \frac{i}{4\pi^2}\int \D^2 w\, \varepsilon^{*+}_I(w,\bar w) \int_{-\infty}^{+\infty} \D\tilde u \int_{-\infty}^{+\infty}\D\omega\,\omega \, e^{i\omega(q^\mu X_\mu+\tilde u)}\, \bar\phi_{z\dots z}^{(s)}(\tilde u,w,\bar w) + \text{h.c.} 
\end{align}
One concludes, making use of the identity 
$
    \int_{-\infty}^{+\infty} dx\, x\, e^{ ipx} = - 2\pi i\,  \partial_{p}\delta(p),
$
that
\begin{align}
    \phi^{(s)}_I (X) &=  \frac{1}{2\pi} \int\,\D^2 w\, \D \tilde u   \;  \varepsilon^{*+}_I(w,\bar w) \Big[   \partial_{\tilde u}\delta\left(\tilde u+q^\mu X_\mu\right) \bar\phi^{(s)}_{z\dots z}( \tilde u, w,\bar w) \Big] + \text{h.c.}
\end{align}
In this sense, the boundary-to-bulk propagator is $\mathcal P(\tilde u,w,\bar w | X ) =\partial_{\tilde u}\delta(q^\mu X_\mu + \tilde u)$. Integrating out the $\delta$-distribution, we obtain the Kirchhoff-d'Adh\'emar formula \cite{Penrose:1985bww,Penrose1980GoldenON}
\begin{equation}\label{KA}
    \phi^{(s)}_I (X) =  -\frac{1}{2\pi} \int\,\D^2 w  \; \varepsilon^{*+}_I(w,\bar w)   \;  \partial_{\tilde u}\bar\phi_{z\dots z}^{(s)} (\tilde u = -q^\mu X_\mu, w,\bar  w)  + \text{h.c.}
\end{equation} 
which allows to reconstruct the bulk field $\phi^{(s)}_I (X)$ from its boundary value at $\mathscr{I}^+$. A similar relation holds for the boundary value at $\mathscr{I}^-$. 

\paragraph{Remark} Consistently with the remark of the previous section, we recover from \eqref{KA} that a shift of the form \eqref{shift} does not alter the bulk field. Notice that, by contrast, changing the value of the pole in the Fourier transform $a_{\pm}^{(s)}(\omega,z,\bar{z}) \mapsto a_{\pm}^{(s)}(\omega,z,\bar{z}) + \Delta_{\eta} a_{\pm}^{(s)}(\omega,z,\bar{z})$ with
\begin{equation}
    \Delta_{\eta} a_+^{(s)}(\omega,z,\bar{z})= - \Delta_{\eta} a^{(s)}_-(\omega,z,\bar{z})^\dagger \equiv -\frac{4 \pi}{K_{(s)}}  \eta(z,\bar{z})\frac{1}{\omega}, \label{pole change}
\end{equation}
which corresponds to shifting the boundary field by $\Delta_{\eta} \bar{\phi}^{(s)}_{z\dots z} (u,z,\bar{z})= \frac{1}{2}\eta(z,\bar{z})\text{sign}(u)$, does affect the bulk field as
\begin{equation}
\Delta_{\eta} \phi^{(s)}_{I} (X) = -\frac{1}{2\pi} \int\,\D^2 w  \; \varepsilon^{*+}_I(w,\bar w) \eta(w,\bar{w})\delta(q^{\mu}X_{\mu}) + \text{h.c.} \label{temp1}
\end{equation}

\subsection{From null infinity to the celestial sphere}
\label{sec:From null infinity to the celestial sphere}

\subsubsection{From position space to Mellin basis}

Combining the results of the previous sections, we can construct a dictionary between boundary operators $\bar\phi^{(s)}_{z\dots z}(u,z,\bar z)$ (resp. $\bar\phi^{(s)}_{z\dots z}(v,z,\bar z)$) evolving in retarded (resp. advanced) time and the ladder operators in the Mellin basis $a^{(s)}_{\Delta,\alpha}(z,\bar z)$ and $a^{(s)}_{\Delta,\alpha}(z,\bar z)^\dagger$ living on the celestial sphere. Injecting \eqref{fourier transform formula at scri} into \eqref{mellin O}, and using \eqref{mellin of plane wave} to compute the integral on $\omega$, yields
\begin{equation}
\begin{split}
a^{(s)}_{\Delta,+}(z,\bar z) &= \frac{4\pi i} {K_{(s)}} \, \lim_{\epsilon\to 0^+} \int_0^{+\infty} \D\omega\,  \omega^{\Delta-1} \int_{-\infty}^{+\infty} \D u\, e^{i\omega u-\omega\epsilon}\, \bar\phi^{(s)}_{z\dots z}(u,z,\bar z)\\
&= \frac{4\pi}{K_{(s)}} \, i^{\Delta+1} \Gamma[\Delta] \, \lim_{\epsilon\to 0^+}  \int_{-\infty}^{+\infty} \D u\, (u +  i\epsilon)^{-\Delta}\, \bar\phi^{(s)}_{z\dots z}(u,z,\bar z) ,
\end{split} \label{Btransform in two steps}
\end{equation}
where $\Delta = c+i\nu$, $c>0$. The integral transform \eqref{Btransform in two steps} coincides with the one discussed in the extrapolate-style dictionary presented in \cite{Pasterski:2021dqe}. It essentially trades the time dependence $u$ of the operators in position space for the conformal dimension $\Delta$ of the operators in Mellin space. This motivates introducing
\begin{equation}
\boxed{
    \mathcal B_\pm\big[f(u),\Delta\big] \equiv \kappa_{\Delta}^{\pm}\lim_{\epsilon\to 0^+} \int_{-\infty}^{+\infty} \D u\, (u\pm i\epsilon)^{-\Delta}\, f(u)
} \label{Btransform}
\end{equation}
for any smooth functions $f$ defined on $\mathscr I^+$, with $\kappa_{\Delta}^{\pm} = 4\pi \, (\pm i)^{\Delta+1}\Gamma[\Delta]$. The integral \eqref{Btransform}, here referred to as the $\mathcal B$-transform, is merely the composition of a Fourier transform (which maps $u\mapsto \omega$) and a Mellin transform (which maps $\omega\mapsto \Delta$), see \eqref{Btransform in two steps}. One should stress that the $\mathcal B_+$-transform (resp. $\mathcal B_-$) alone is \textit{not} invertible since it projects out the positive (resp. negative) frequency modes of $f$, \textit{i.e.} it annihilates the second (resp. first) term in the decomposition
\begin{equation}
    f(u) = \int_0^{+\infty}\D\omega\big[ f_+(\omega)\,e^{-i\omega u} + f_-(\omega)\,e^{i\omega u} \big]. \label{frequency decomposition of f}
\end{equation}
This is easily seen from the first line of \eqref{Btransform in two steps} where the integral in $u$ effectively gets rid of all positive frequency modes of $\bar\phi^{(s)}_{z\dots z}(u,z,\bar z)$. However $\mathcal B_+$ (resp. $\mathcal B_-$) is invertible when restricted to functions with negative (resp. positive) frequency only. This comes from the fact that composing a Fourier and a Mellin transform is only possible if the integral over frequencies is taken over $\mathbb R^+$, which amounts to use ``unilateral'' (hence improper) Fourier transforms that are also not invertible. The physical meaning of this choice of integration domain is again that the asymptotic excitations of $\phi^{(s)}_I$ have strictly positive energy, $\omega>0$.

Keeping this in mind, we have
\begin{equation}
a^{(s)}_{\Delta,+}(z,\bar z) = \frac{1}{K_{(s)}}\, \mathcal{B}_{+}\Big[\bar\phi_{z\dots z}^{(s)}(u,z,\bar z),\Delta\Big],
\quad a^{(s)}_{\Delta,-}(z,\bar z)^{\dagger} = \frac{1}{K_{(s)}}\, \mathcal{B}_{-}\Big[\bar\phi_{z\dots z}^{(s)}(u,z,\bar z),\Delta\Big],
\label{a(Delta) from Phi(u)}
\end{equation}
and
\begin{equation}
a^{(s)}_{\Delta,-}(z,\bar z) = \frac{1}{K_{(s)}}\, \mathcal{B}_{+}\Big[\bar\phi_{\bar z\dots \bar z}^{(s)}(u,z,\bar z),\Delta\Big],
\quad a^{(s)}_{\Delta,+}(z,\bar z)^{\dagger} = \frac{1}{K_{(s)}}\, \mathcal{B}_{-}\Big[\bar\phi_{\bar z\dots \bar z}^{(s)}(u,z,\bar z),\Delta\Big].
\end{equation}
One can write similar expressions at $\mathscr I^-$ for $\bar\phi_{z...z}^{(s)}(v,z,\bar z)$ expanded as \eqref{incoming spin s} by making the replacement $u\mapsto v$.

\subsubsection{From Mellin basis to position space}

We now want to discuss how to reconstruct the boundary field $\bar \phi_{z...z}^{(s)}(u,z,\bar z)$ from the celestial operators $a^{(s)}_{\Delta,\pm}(z,\bar z)$. To achieve this, let us first define
\begin{equation}
\boxed{
    \mathcal B^{-1}_\pm \big[F(\Delta),u\big] \equiv -\frac{1}{16\pi^3}\lim_{\epsilon\to 0^+} \int_{-\infty}^{+\infty}\D\nu\, \frac{(\pm i)^\Delta \Gamma[1-\Delta]}{(u\mp i\epsilon)^{1-\Delta}}\, F(\Delta).
}
\label{inverse B transform}
\end{equation}
These operators are effectively obtained by the successive action of two integral transforms
\begin{equation}
 \mathcal B^{-1}_{\pm} [F(\Delta),u]  = \frac{\mp i}{16\pi^3} \int_0^{+\infty}\D \omega\, e^{\mp i\omega u} \int_{-\infty}^{+\infty} \D \nu\, \omega^{-\Delta} F(\Delta). \label{inverse B transform decomposition}
\end{equation}
The second integral in \eqref{inverse B transform decomposition} is the inverse Mellin transform \eqref{inverse mellin}, while the first integral means that functions in the image of $\mathcal B_{+}^{-1}$ (resp. $\mathcal B_{+}^{-1}$) have negative (resp. positive) frequencies only. Now, one always has
\begin{align}
\mathcal B_{\pm} \circ \mathcal B^{-1}_{\pm}\big[F(\Delta)\big] =  F(\Delta).
\label{B o B^-1}
\end{align}
This relation can checked explicitly by providing a realization of the $\delta$-distribution on conformal weights as
\begin{equation}
\lim_{\epsilon\to 0^+}  \int_{-\infty}^{+\infty}\D u \, \frac{i^{\Delta} \Gamma[\Delta] }{(u+i\epsilon)^{\Delta} } \frac{(-i)^{1-\Delta'} \Gamma[1-\Delta']}{(u-i\epsilon)^{1-\Delta'}} = 4\pi^2\delta(\nu-\nu')\label{dirac(delta) identity}
\end{equation}
where $\Delta = c+i\nu$ and $\Delta' = c+i\nu'$. This identity is the analog of \eqref{delta represented in mellin} but adapted for the $\mathcal B$-transform instead of the Mellin transform: it can be derived by first rewriting each of the fractions in the integrand of \eqref{dirac(delta) identity} as Mellin transforms of plane waves, then performing the integral in $u$ to obtain a $\delta$-function and concluding using \eqref{delta represented in mellin}.

However, as we already emphasized, the transforms $\mathcal B_{\pm}$ are not invertible. Rather, if we decompose $f(u)$ in positive/negative frequency modes as in \eqref{frequency decomposition of f}, then
\begin{equation}
\mathcal B_{\pm}^{-1} \circ \mathcal B_{\pm}\big[f(u)\big] =  \int_{0}^{+\infty}\D\omega\, e^{\mp i\omega u} f_{\pm}(\omega), \label{B^-1 o B}
\end{equation}
\textit{i.e.} $\mathcal B_{\pm}^{-1} \circ \mathcal B_{\pm}$ are the projectors on negative/positive frequency modes. This follows from
\begin{equation}
\lim_{\epsilon \to 0^+}  \int_{-\infty}^{+\infty}\D \nu \, \frac{i^{\Delta} \Gamma[\Delta] }{(u+i\epsilon)^{\Delta} } \frac{(-i)^{1-\Delta} \Gamma[1-\Delta]}{(u'-i\epsilon)^{1-\Delta}} = 2\pi \int_{0}^{+\infty} \D \omega\,  e^{i\omega(u-u')}, \label{proj(u) identity}
\end{equation}
which is not $\delta(u-u')$ because of the restricted domain of integration. This is responsible for the fact that $\mathcal B_{\pm}^{-1} \circ \mathcal B_{\pm}$ is a projection rather than the identity map. Equation \eqref{proj(u) identity} can be derived by rewriting the second fraction in the integrand as the Mellin transform of a plane wave and recognizing the resulting integral over $\nu$ as an inverse Mellin transform acting on the first fraction.

From all these remarks, we conclude that the relations \eqref{a(Delta) from Phi(u)} are inverted as follows:
\begin{equation}
   \bar\phi_{z\dots z}^{(s)}(u,z,\bar z) = K_{(s)}\, \mathcal B^{-1}_+\Big[a^{(s)}_{\Delta,+}(z,\bar z),u\Big] + K_{(s)}\,\mathcal B^{-1}_-\Big[a^{(s)}_{\Delta,-}(z,\bar z)^{\dagger},u\Big]. \label{full inverse B transform}
\end{equation}
with a similar expression for $\bar\phi_{z\dots z}^{(s)}(v,z,\bar z)$ obtained by making the replacement $u \mapsto v$.

\subsection{\texorpdfstring{$\mathcal{S}$-matrix}{S-matrix} in the three scattering bases}

We now turn to the scattering of interacting massless spin-$s$ fields in flat spacetime. This discussion will be repeated in the three scattering bases \cite{Donnay:2022sdg}: momentum space, Mellin space and position space. Even though the $\mathcal{S}$-matrix is known to be trivial for higher spins $s>2$ \cite{PhysRev.135.B1049,Grisaru:1976vm,Aragone:1979hx,Weinberg:1980kq,Porrati:2008rm} (see \textit{e.g.} \cite{Bekaert:2022poo} for a recent review), it will be valuable to keep the spin arbitrary throughout to highlight some general structure. In fact, some patterns of an infrared triangle \cite{Strominger:2017zoo} have been shown to persist for higher spin theories \cite{Campoleoni:2017mbt,Campoleoni:2017qot,Heissenberg:2019fbn,Campoleoni:2020ejn}, which makes the analysis for $s>2$ interesting by itself. Moreover, there exist non-trivial (interacting) chiral higher-spin theories \cite{Metsaev:1991mt,Metsaev:1991nb,Ponomarev:2016lrm,Skvortsov:2018jea,Skvortsov:2020wtf,Krasnov:2021nsq,Herfray:2022prf} (see \cite{Ren:2022sws,Monteiro:2022xwq} for the relevance of these theories in the context of celestial OPEs). The presence of interactions in those theories does not necessarily mean that the $\mathcal{S}$-matrix is non-trivial: typically, the usual theorems apply and, as a result of the higher-spin symmetries, scattering amplitudes vanish. However there are examples of chiral higher-spin theories with a non-trivial $\mathcal{S}$-matrix \cite{Adamo:2022lah,Tran:2022amg}.

\subsubsection{Asymptotically free fields for null scattering processes}
As usual in the $\mathcal S$-matrix picture, the interacting theory is taken to be asymptotically free. More explicitly, in retarded coordinates \eqref{Retarded flat BMS coordinates (appendix)}, we suppose that the bulk operator $\phi_I^{(s)}$ of the full interacting theory can be expanded around $\mathscr{I}^+$ as
\begin{equation}
    \phi_I^{(s)}(X) = \phi_I^{(s)}(X)^{\text{out}} + \mathcal{O}(r^{s-2}),
\end{equation}
and in advanced coordinates \eqref{Advanced flat BMS coordinates (appendix)} around $\mathscr{I}^-$ as
\begin{equation}
    \phi_I^{(s)} (X) = \phi_I^{(s)} (X)^{\text{in}} + \mathcal{O}(r^{s-2}).
\end{equation}
Here $\phi_I^{(s)}(X)^{\text{out/in}}$ are the free fields discussed in the previous sections and $\mathcal{O}(r^{s-2})$ stands for the contributions of the interactions in the bulk that are not seen near the boundary.

Under these assumptions, the boundary values \eqref{outgoing spin s} and \eqref{incoming spin s} of the interacting field at $\mathscr{I}^+$ and $\mathscr{I}^-$ are finite and coincide with the boundary values of $\phi_I^{(s)}(X)^{\text{out/in}}$,
 \begin{align}
\bar\phi^{(s)}_{z\dots z} (u,z,\bar z)^{\text{out}} &= -\frac{i K_{(s)}}{8\pi^2} \int_{0}^{+\infty}\D \omega  \Big[ a^{(s)\,\text{out}}_+(\omega,z,\bar z) e^{ -i\omega u} - a_-^{(s)\,\text{out}}(\omega,z,\bar z)^\dagger e^{i\omega u}\Big]
, \label{phi a out}\\
\bar\phi^{(s)}_{z\dots z} (v,z,\bar z)^{\text{in}} &= -\frac{i K_{(s)}}{8\pi^2} \int_{0}^{+\infty}\D \omega  \Big[ a^{(s)\,\text{in}}_+(\omega,z,\bar z) e^{- i\omega v} - a_-^{(s)\,\text{in}}(\omega,z,\bar z)^\dagger e^{i\omega v}\Big]. \label{phi a in} 
\end{align} 
We therefore asymptotically end up with two free theories related by a non-trivial interaction process. In the free theory, our conventions \eqref{Boundary value}--\eqref{Boundary value PAST} mean that incoming and outgoing boundary values are equal, up to the mere replacement $u\mapsto v$.  In presence of interactions, the antipodal matching conditions of \cite{Strominger:2013jfa,He:2014cra} read
\begin{equation}
    \bar\phi^{(s)}_{z\dots z}(u,z,\bar z)^{\text{out}}\big|_{\mathscr I^+_-} = \bar\phi^{(s)}_{z\dots z}(v,z,\bar z)^{\text{in}}\big|_{\mathscr I^-_+}, \label{antipodal on fields}
\end{equation}
where $\mathscr I^+_- = \{X\in\mathscr I^+ \,|\, u\to-\infty\}$ and $\mathscr I^-_+ = \{X\in\mathscr I^- \,|\, v\to +\infty\}$. This is because the coordinates $z=z'$ define antipodal directions on the celestial sphere between $\mathscr I^-$ and $\mathscr I^+$. 

In Section \ref{sec:From null infinity to the celestial sphere}, three descriptions of these fields have been presented: the direct interpretation in position space and their decomposition in terms of ladder operators both in Fourier and Mellin spaces. In the following, we relate the asymptotic states in these three equivalent descriptions by integral transforms.

\subsubsection{Scattering amplitudes in momentum space}

In momentum space, we can construct an incoming or outgoing state, respectively denoted as $\ket{\omega,z,\bar z,s,\alpha}$ and $\bra{\omega,z,\bar z,s,\alpha}$, representing a particle of light-cone energy $\omega$, spin $s$ and helicity $J = \alpha s=\pm s$ coming from or heading to the point $(z,\bar z)$ on the celestial sphere, by acting on the respective vacua with the ladder operators $a^{(s)\,\text{in/out}}_\alpha(\omega,z,\bar z)$, \textit{i.e.} 
\begin{equation}
\begin{aligned}
\bra{\omega,z,\bar z,s,\alpha} &= -  \frac{i K_{(s)}}{4\pi} \bra{0} a_\alpha^{(s)\,\text{out}} (\omega, z, \bar{z} ) , \\ \ket{\omega,z,\bar z,s,\alpha} &=  \frac{iK_{(s)}}{4\pi} a_\alpha^{(s)\,\text{in}}(\omega,z,\bar z)^\dagger |0\rangle .
\end{aligned} \label{energy eigenstates}
\end{equation}
The perhaps unusual choice of normalization of energy eigenstates in \eqref{energy eigenstates} is inspired by the relations \eqref{fourier transform formula at scri} and will be motivated in Section \ref{sec:asymptotic states in direct basis}. The scattering amplitudes in momentum space involving $N$ massless particles -- $n$ of which are outgoing -- are given by the $\mathcal{S}$-matrix elements $\mathcal A_N(p_1;\dots; p_N) = \braket{\text{out}|\text{in}}_{\text{mom}}$ with
\begin{equation}
    \bra{\text{out}} = \bra{\omega_{1},z_{1},\bar z_{1},s_{1},\alpha_{1}}\otimes \dots\otimes \bra{\omega_n,z_n,\bar z_n,s_n,\alpha_n}
\end{equation}
and
\begin{equation}
    \ket{\text{in}} = \ket{\omega_{n+1},z_{n+1},\bar z_{n+1},s_{n+1},\alpha_{n+1}}\otimes \dots\otimes |\omega_N,z_N,\bar z_N,s_N,\alpha_N\rangle.
\end{equation}

\subsubsection{Scattering amplitudes in Mellin space}
\label{sec:asymptotic states in Mellin basis}

Alternatively, as it was pointed out in \cite{Pasterski:2016qvg,Pasterski:2017kqt} (see \cite{Pasterski:2021rjz} for a review), one can express the $\mathcal{S}$-matrix elements in Mellin space using the change of representation discussed in Section \ref{sec:Mellin transform section}. The Mellin transform \eqref{mellin O} turns the in/out asymptotic ladder operators in momentum space to asymptotic operators in the Mellin basis. Correspondingly, we obtain the asymptotic states on the celestial sphere from the usual momentum eigenstates by 
\begin{equation}
\bra{\Delta,z,\bar z,s,\alpha}= -\frac{iK_{(s)}}{4\pi} \bra{0}a_{\Delta,\alpha}^{(s)\,\text{out}}(z,\bar z)  = \int_0^{+\infty} \D\omega \,\omega^{\Delta-1} \bra{\omega,z,\bar z,s,\alpha}  \label{outgoing celestial}
\end{equation}
for outgoing particles, and
\begin{equation}
\ket{\Delta,z,\bar z,s,\alpha} =   \frac{iK_{(s)}}{4\pi} a_{\Delta,\alpha}^{(s)\,\text{in}}(z,\bar z)^\dagger \ket{0} = \int_0^{+\infty} \D\omega \,\omega^{\Delta-1} \ket{\omega,z,\bar z,s,\alpha} \label{incoming celestial}
\end{equation}
for incoming particles. Because of \eqref{a delta transfo lorentz finite}, states defined as \eqref{outgoing celestial}--\eqref{incoming celestial} are boost eigenstates. The so-called ``celestial amplitudes'' $\mathcal M_N = \braket{\text{out}|\text{in}}_{\text{boost}}$ involving $N$ inserted particles on the celestial sphere are now obtained as (dropping the spin indices)
\begin{equation}
\begin{split}
    &\mathcal M_N (\Delta_1,z_1,\bar z_1;\dots; \Delta_N,z_N,\bar z_N) = \int_0^{+\infty} \D\omega_1 \, \omega_1^{\Delta_1-1}  \dots \int_0^{+\infty} \D\omega_N \, \omega_N^{\Delta_N-1} \mathcal A_N (p_1;\dots; p_N).
\end{split}
    \label{Celestial S-matrix}
\end{equation}

\subsubsection{Scattering amplitudes in position space}
\label{sec:asymptotic states in direct basis}

Finally, we can define the asymptotic quantum states directly in position space. Outgoing states at $\mathscr{I}^+$ are naturally defined as
\begin{equation}
\begin{aligned}
\bra{u,z,\bar z,s,+} &= \bra{0}\bar\phi^{(s)}_{z\dots z}(u,z,\bar z) = \frac{1}{2\pi} \int_{0}^{+\infty} \D \omega\, e^{-i\omega u} \langle \omega,z,\bar z,s,+| ,\\
\bra{u,z,\bar z,s,-}  &= \bra{0}\bar\phi^{(s)}_{\bar z\dots \bar z}(u,z,\bar z)= \frac{1}{2\pi} \int_{0}^{+\infty} \D \omega\, e^{-i\omega u} \langle \omega,z,\bar z,s,-|,
\end{aligned} \label{direct to fourier boundary}
\end{equation} 
because of \eqref{phi a out}--\eqref{phi a in}. The boundary field $\bar\phi^{(s)}_{z\dots z}(u,z,\bar z)$ creates outgoing spin-$s$ particles with positive helicity and destroys outgoing spin-$s$ particles with negative helicity, while $\bar\phi^{(s)}_{\bar z\dots\bar z}(u,z,\bar z)$ $=$ $\bar\phi^{(s)}_{z\dots z}(u,z,\bar z)^{\dagger}$ acts in the opposite way. The normalization of \eqref{energy eigenstates} allows to represent the creation and annihilation operators in position space \eqref{direct to fourier boundary} without extra normalization factors. Playing the same game at $\mathscr{I}^-$, we construct the incoming states in position space as
\begin{equation}
\begin{aligned}
\ket{v,z,\bar z,s,+} &=  \bar\phi^{(s)}_{z\dots z}(v,z,\bar z)^{\dagger}\ket{0} = \frac{1}{2\pi}  \int_{0}^{+\infty} \D \omega\, e^{i\omega v} | \omega,z,\bar z,s,+\rangle ,\\
\ket{v,z,\bar z,s,-} &=  \bar\phi^{(s)}_{\bar z\dots \bar z}(v,z,\bar z)^\dagger\ket{0} = \frac{1}{2\pi}  \int_{0}^{+\infty} \D \omega\, e^{i\omega v} |\omega,z,\bar z,s,-\rangle .
\end{aligned} \label{direct to fourier boundary PAST}
\end{equation}
We then introduce ``position space amplitudes''
$\mathcal C_N = \braket{\text{out}|\text{in}}_{\text{pos}}$, which are obtained from the usual momentum representation of the $\mathcal S$-matrix as
\begin{equation}
\boxed{
    \begin{aligned}
        &\mathcal C_N( u_{1},z_{1},\bar z_{1};\dots ;u_n, z_n,\bar z_n;v_{n+1}, z_{n+1},\bar z_{n+1};\dots; v_N,z_N,\bar z_N) \\
        &= \frac{1}{(2\pi)^N}\prod_{k=1}^n \int_0^{+\infty} \D\omega_k \, e^{-i\omega_k u_k}  \prod_{\ell=n+1}^{N} \int_0^{+\infty} \D\omega_{\ell} \, e^{i\omega_{\ell} v_\ell}  \mathcal A_N ( p_1;\dots; p_N). \label{Carrollian S-matrix}
    \end{aligned}
}
\end{equation}
These amplitudes can be translated in the Mellin representation thanks to the $\mathcal B$-transform \eqref{Btransform} by noticing that
\begin{equation}
    \begin{split}
        \bra{\Delta,z,\bar z,s,\alpha} &= \frac{1}{4\pi i}\mathcal B_+\big[\bra{u,z,\bar z,s,\alpha},\Delta\big], \quad \ket{\Delta,z,\bar z,s,\alpha} = -\frac{1}{4\pi i} \mathcal B_-\big[\ket{v,z,\bar z,s,\alpha},\Delta\big],
    \end{split} 
    \label{ket bra Delta in terms of bra u}
\end{equation}
owing to the definitions \eqref{outgoing celestial}--\eqref{incoming celestial} and \eqref{a(Delta) from Phi(u)}. Hence
\begin{equation}
    \begin{split}
        &\mathcal M_N (\Delta_1,z_1,\bar z_1;\dots; \Delta_N,z_N,\bar z_N)  \\
        & = \frac{(-1)^{N-n}}{(4\pi i)^N} \mathcal B_-^{(N-n)}\Big[ \mathcal B_+^{(n)}\big[ \mathcal C_N(u_1,z_1,\bar z_1;\dots; v_N,z_N,\bar z_N),\{\Delta_{1},\dots, \Delta_n\}\big],\{\Delta_{n+1},\dots,\Delta_N\}\Big],
    \end{split} \label{B transform of position space amplitudes}
\end{equation}
where $\mathcal B^{(k)}_+$ represents $k$ successive applications of the $\mathcal B$-transform.

\subsubsection{Low-point amplitudes}
\label{sec:Low-point amplitudes}
Let us make more concrete the action of the various integral transforms introduced so far by considering propagation of one particle, ignoring quantum loop effects. In momentum basis, the tree-level amplitude $\mathcal A_2$ is 
\begin{equation}
    \mathcal A_2(\omega_1,z_1,\bar z_1;\omega_2,z_2,\bar z_2) = K_{(s)}^2\,\pi\,\frac{\delta(\omega_1-\omega_2)}{\omega_1}\,\delta^{(2)}(z_1-z_2)\,\delta_{\alpha_1,\alpha_2},
\label{two point amplitude in fourier}
\end{equation}
for $\omega_i>0$. This simply describes the travel of a particle of helicity $J=\alpha s$ inserted at $\mathscr I^-$ crossing Minkowski spacetime towards $\mathscr I^+$.

Recalling the definition \eqref{outgoing celestial}--\eqref{incoming celestial} and using the transformation \eqref{Celestial S-matrix}, the two-point amplitude in Mellin space can be deduced from \eqref{two point amplitude in fourier} and is given by \cite{Pasterski:2017ylz}
\begin{equation}
    \begin{split}
        &\mathcal M_2(\Delta_1,z_1,\bar z_1;\Delta_2,z_2,\bar z_2) = \int_0^{+\infty}\D\omega_1\,\omega_1^{\Delta_1-1}\int_0^{+\infty}\D\omega_2\,\omega_2^{\Delta_2-1}\, \mathcal A_2(\omega_1,z_1,\bar z_1;\omega_2,z_2,\bar z_2) \\
        &= K^2_{(s)}\,\pi\int_0^{+\infty}\D\omega\,\omega^{\Delta_1+\Delta_2-3}\,\delta^{(2)}(z_1-z_2)\,\delta_{\alpha_1,\alpha_2} = 2\pi^2\,K_{(s)}^2\,\delta(\nu_1+\nu_2)\,\delta^{(2)}(z_1-z_2)\,\delta_{\alpha_1,\alpha_2},
    \end{split}
\label{2pt celestial M2}
\end{equation}
where $\Delta_i = 1+i\nu_i$ is assumed for the integral over $\omega$ to be converging, as prescribed by \eqref{delta represented in mellin}.

The position space amplitude $\mathcal C_2$ can also be computed from \eqref{two point amplitude in fourier} as
\begin{equation}
    \begin{split}
        \mathcal C_2(u_1,z_1,\bar z_1;u_2,z_2,\bar z_2) &= \frac{1}{4\pi^2} \int_0^{+\infty}\D\omega_1\int_0^{+\infty}\D\omega_2\,e^{-i\omega_1 u_1}e^{i\omega_2 u_2} \mathcal A_2(\omega_1,z_1,\bar z_1;\omega_2,z_2,\bar z_2) \\
        &= \frac{K^2_{(s)}}{4\pi} \underbrace{\int_0^{+\infty}\frac{\D\omega}{\omega}\,e^{-i\omega(u_1-u_2)}}_{\equiv\,\mathcal I_0(u_1-u_2)}\,\delta^{(2)}(z_1-z_2)\,\delta_{\alpha_1,\alpha_2},
    \end{split} \label{W12 def}
\end{equation}
owing to \eqref{Carrollian S-matrix} and using \eqref{u to v} to trade $v$ for $u$. The integral $\mathcal I_0(u_1-u_2)$ is divergent but can nevertheless be regulated noticing that $\displaystyle\mathcal I_0(u_1-u_2) = \lim_{\beta\to 0^+} \mathcal I_\beta(u_1-u_2)$ with the definition
\begin{equation}
    \begin{split}
        \mathcal I_\beta(x) &= \lim_{\epsilon\to 0^+}\int_0^{+\infty} \D\omega\,\omega^{\beta-1}\,e^{-i\omega x-\omega\epsilon} = \lim_{\epsilon\to 0^+}\frac{\Gamma[\beta](-i)^\beta}{(x-i\epsilon)^\beta}.
    \end{split} \label{I beta}
\end{equation}
In the limit $\beta\to 0^+$, we obtain 
\begin{equation}
        \mathcal I_\beta(x) = \frac{1}{\beta}-\left[\gamma+\ln|x| + \frac{i\pi}{2}\text{sign}(x) \right] + \mathcal O(\beta), \label{curly I beta}
\end{equation}
where $\gamma$ is the Euler-Mascheroni constant. So \eqref{W12 def} evaluates to
\begin{equation}
\boxed{
    \mathcal C_2(u_1;u_2) = \frac{K^2_{(s)}}{4\pi} \left[\frac{1}{\beta} -\left(\gamma + \ln|u_1-u_2| + \frac{i\pi}{2}\text{sign}(u_1-u_2) \right) \right]\delta^{(2)}(z_1-z_2)\,\delta_{\alpha_1,\alpha_2} + \mathcal O(\beta).
} \label{Wuup}
\end{equation}
 We show in Section \ref{sec:Conformal Carrollian invariant correlation functions} that \eqref{Wuup} is a Poincaré invariant object as a consequence of \eqref{bar phi transfo lorentz}: importantly, the pole in $1/\beta$ shall be kept because it is essential to ensure boost invariance. Let us stress that, fundamentally, this divergence can be interpreted as an infrared pole. Indeed, if we choose to regularize the integral $\mathcal I_0(x)$ as
\begin{equation}
    \mathcal I_0(x) = \lim_{\epsilon\to 0^+} \int_\epsilon^{+\infty} \frac{\D\omega}{\omega}\,e^{-i\omega x} \equiv\lim_{\epsilon\to 0^+} E_1(ix\epsilon)
\end{equation}
where $E_1(z)$ represents the principal branch $\text{Arg}(z)\in\,\,]-\pi,\pi[$ of the complex exponential integral, for any $z\notin \mathbb R_-$, we have the following series \cite{abramowitz+stegun}
\begin{equation}
    E_1(z) = -\gamma -\ln z-\sum_{k=1}^{+\infty}\frac{1}{k!}\frac{(-z)^k}{k}\quad \Rightarrow \quad \mathcal I_0(x) = \ln\frac{1}{\epsilon} -\gamma - \ln|x| - \text{sign}(x)\,\frac{i\pi}{2} + \mathcal O(\epsilon)
\end{equation}
in the limit $\epsilon\to 0^+$. Boost invariance requires that $\ln(\epsilon^{-1})$ behaves around zero as the pole of the Euler Gamma function, hence $\beta^{-1} \simeq \ln(\epsilon^{-1})$ and both treatments of the divergence agree on the final result.

Let us now observe thanks to \eqref{direct to fourier boundary}--\eqref{direct to fourier boundary PAST} that
\begin{equation}
    \begin{split}
        \mathcal C_2(u_1,z_1,\bar z_1;u_2,z_2,\bar z_2) \equiv \langle 0 | \bar\phi^{(s)}_{z\dots z}(u_1,z_1,\bar z_1) \bar\phi^{(s)}_{z\dots z}(u_2,z_2,\bar z_2)^\dagger | 0 \rangle \label{2pt amplitude position}
    \end{split}
\end{equation}
picking $\alpha = +$ for definiteness (similar considerations apply, \textit{mutatis mutandis}, for $\alpha=-$). In Section \ref{sec:Holographic conformal Carrollian field theory}, we will interpret this object as a (holographic) boundary two-point function: it results from the double limit of the bulk two-point function where one point is sent to $\mathscr I^+$ and the other one to $\mathscr I^-$. Although \eqref{2pt amplitude position} has an infrared divergence, we directly observe from \eqref{curly I beta} that the difference $\mathcal I_\beta(u_1-u_2) - \mathcal I_\beta(u_2-u_1)$ is finite in the limit $\beta\to 0^+$ and reproduces the commutation relation \eqref{commu boundary fields}, \textit{i.e.}
\begin{equation}
    \begin{split}
        &\langle 0 | \big[\bar\phi^{(s)}_{z\dots z}(u_1,z_1,\bar z_1),\bar\phi^{(s)}_{z\dots z}(u_2,z_2,\bar z_2)^\dagger\big] | 0 \rangle \\
        &= \lim_{\beta\to 0^+}\frac{K^2_{(s)}}{4\pi}\left[\mathcal I_\beta(u_1-u_2) - \mathcal I_\beta(u_2-u_1)\right]\,\delta^{(2)}(z_1-z_2) \\
        &= -\frac{iK_{(s)}^2}{4}\,\text{sign}(u_1-u_2)\,\delta^{(2)}(z_1-z_2). 
    \end{split} \label{commu phi phibar} 
\end{equation}
The behavior \eqref{Wuup} of the boundary two-point function \eqref{2pt amplitude position} has also been established independently in \cite{Liu:2022mne}.

Finally, let us close the loop by showing that applying twice the $\mathcal B$-transform as prescribed in \eqref{B transform of position space amplitudes} gives us back the celestial amplitude. Keeping \eqref{ket bra Delta in terms of bra u} in mind, we first compute
\begin{equation}
    \begin{split}
        &\mathcal B_+\big[\mathcal I_0(u_1-u_2),\Delta_1\big] = 4\pi \, i^{\Delta_1+1}\Gamma[\Delta_1]\lim_{\epsilon\to 0^+}\int_{-\infty}^{+\infty} \frac{\D u_1}{(u_1+i\epsilon)^{\Delta_1}}\mathcal I_0(u_1-u_2) \\
        &= 4\pi i \lim_{\epsilon\to 0^+}\int_{-\infty}^{+\infty}\D u_1\int_0^{+\infty}\omega^{\Delta_1-1}\,e^{i\omega u_1-\omega\epsilon}\int_0^{+\infty}\frac{\D\omega'}{\omega'}e^{-i\omega'(u_1-u_2)} \\
        &=8\pi^2 i \lim_{\epsilon\to 0^+}\int_0^{+\infty} \D\omega\,\omega^{\Delta_1-2}\,e^{i\omega u_2-\omega \epsilon} = 8\pi^2 i \lim_{\epsilon\to 0^+} \frac{i^{\Delta_1-1}\Gamma[\Delta_1-1]}{(u_2+i\epsilon)^{\Delta_1-1}}.
    \end{split}
\end{equation}
Computing the second $\mathcal B$-transform leads to
\begin{equation}
    \begin{split}
        &\mathcal B_-\Big[\mathcal B_+\big[\mathcal I_0(u_1-u_2),\Delta_1\big],\Delta_2\Big] \\
        &= 32\pi^3 \lim_{\epsilon\to 0^+} \int_{-\infty}^{+\infty}\D u_2\, \frac{i^{\Delta_1-1}\Gamma[\Delta_1-1]}{(u_2+i\epsilon)^{\Delta_1-1}} \frac{(-i)^{\Delta_2}\Gamma[\Delta_2]}{(u_2-i\epsilon)^{\Delta_2}} = 128\pi^5 \delta(\nu_1+\nu_2),
    \end{split}
\end{equation}
as a corollary of \eqref{dirac(delta) identity} assuming $\Delta_i=1+i\nu_i$. Therefore, using the dictionary \eqref{B transform of position space amplitudes}, 
\begin{equation}
    \begin{split}
        \mathcal M_2(\Delta_1,z_1,\bar z_1;\Delta_2,z_2,\bar z_2) &=\frac{1}{(4\pi)^2}\mathcal B_-\Big[\mathcal B_+\big[\mathcal C_2(u_1,z_1,\bar z_1;u_2,z_2,\bar z_2),\Delta_1\big],\Delta_2\Big]\\
        &= 2\pi^2\, K^2_{(s)}\,\delta(\nu_1+\nu_2)\,\delta^{(2)}(z_1-z_2)\,\delta_{\alpha_1,\alpha_2},
    \end{split}
\end{equation}
we recover \eqref{2pt celestial M2} as it should.

A similar treatment could be applied to three-point amplitudes but kinematic constraints imply that the latter have to vanish in Lorentzian signature $(-,+,+,+)$. This issue can be circumvented by working in split signature $(-,+,-,+)$ in which $z$ and $\bar z$ are no longer related by complex conjugation (see \textit{e.g.} \cite{Henn:2014yza}) and the description of amplitudes in position space can be adapted to this framework. These considerations, as well as the computation of the higher-point amplitudes ($N\geq 4$) will be discussed elsewhere. 

\subsubsection{The three bases for scattering amplitudes in flat spacetime}
\label{sec:The three bases for scattering amplitudes in flat spacetime}

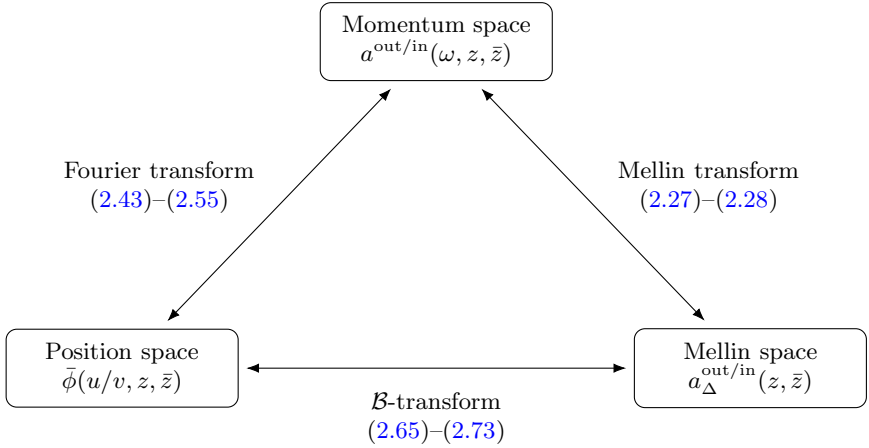
\begin{figure}[ht!]
    \centering
    \begin{tikzpicture}
        \tikzmath{\a = 2.5; \b = sqrt(3)*\a;} 
        \tikzstyle{nb} = [rectangle,draw,fill=white,rounded corners,outer sep=3pt,text width=2.7cm,align=center,inner sep=5pt];
        \coordinate (A) at (-\a, 0);
        \coordinate (B) at ( \a, 0);
        \coordinate (C) at ( 0, \b);
        \draw[white,opacity=0] ($(A)+(-4,-1.5)$) -- ($(B)+(+4,-1.5)$) -- ($(B)+(0,\b)+(+4,1)$) -- ($(A)+(0,\b)+(-4,1)$) -- cycle;
        \node[nb,left] (An) at (A) {Position space $\bar\phi(u/v,z,\bar z)$};
        \node[nb,right] (Bn) at (B) {Mellin space $a^{\text{out/in}}_{\Delta}(z,\bar z)$};
        \node[nb] (Cn) at (C) {Momentum space $a^{\text{out/in}}(\omega,z,\bar z)$};
        \draw[Latex-Latex] (An) -- (Bn);
        \draw[Latex-Latex] (Bn) -- (Cn);
        \draw[Latex-Latex] (An) -- (Cn);
        \node[below,align=center,text width=2cm] at ($(A)!0.5!(B)-(0,0.2)$) {$\mathcal B$-transform \eqref{a(Delta) from Phi(u)}--\eqref{full inverse B transform}};
        \node[align=center,anchor=south east] at ($(An)!0.45!(Cn)$) {Fourier transform\\ \eqref{outgoing spin s}--\eqref{fourier transform formula at scri}};
        \node[align=center,anchor=south west] at ($(Bn)!0.45!(Cn)$) {Mellin transform\\ \eqref{mellin O}--\eqref{ladder op inverse mellin}};
    \end{tikzpicture}
    \caption{Interplay between the three bases of scattering in flat spacetime.}
    \label{fig:FBM}
\end{figure}

From the above discussion, it follows that one can work in three different spaces to discuss the boundary fields:
\begin{enumerate}[label={$(\roman*)$}]
    \item The position space involving the boundary field $\bar \phi^{(s)} (u/v,z,\bar z)$,
    \item The Fourier space providing the usual description by ladder operators $a_{\pm}^{(s)\,\text{out/in}}(\omega, z, \bar{z})$,
    \item The Mellin space with the celestial ladder operators $a^{(s)\,\text{out/in}}_{\Delta, \pm}(z, \bar{z})$,
\end{enumerate}
see Figure \ref{fig:FBM} for a summary. This observation was already pointed out in \cite{Donnay:2022sdg} where it was argued that the three spaces, or ``bases'', are useful, depending on the question of interest. For instance, to discuss infrared issues, Fourier space will be suited to write the soft theorems, Mellin space will be more convenient to discuss symmetries through some Ward identities, and position space will be appropriate to highlight the memory effects. 

We observe that the $\mathcal{S}$-matrix is holographic by nature, in the sense that its elements can be re-interpreted as correlations between boundary operators. Interestingly, as we will argue in Section \ref{sec:Holographic conformal Carrollian field theory}, the boundary operators in position space can be interpreted as operators sourcing a dual Carrollian CFT. The $\mathcal{S}$-matrix elements are then just seen as correlation functions for these operators. This points towards the Carrollian approach of flat space holography. Similarly, the boundary operators in Mellin space can be interpreted as operators in the CCFT, pointing towards the celestial holography proposal. Therefore, one can already deduce that the $\mathcal{B}$-transform discussed in Section \ref{sec:From null infinity to the celestial sphere} relates the Carrollian CFT and the CCFT. It trades the null time dependence of the Carrollian operators for the conformal dimension of the CCFT operators. We will come back to this observation in Section \ref{sec:Relation with celestial holography} when discussing the link between Carrollian and celestial holographies.

\section{Asymptotic symmetries in flat spacetime}
\label{sec:Asymptotic symmetry analysis}

This section reviews the asymptotic symmetry analysis of electrodynamics and gravity. This allows us to fix the notations and conventions that will be useful in order to discuss the holographic Carrollian correspondence in Section \ref{sec:Holographic conformal Carrollian field theory}. The Bondi-Metzner-Sachs (BMS) asymptotic symmetries of gravity will play the role of global spacetime symmetries in the dual Carrollian CFT, while the $U(1)$ asymptotic symmetries of electrodynamics will provide an example of global internal symmetry.

\subsection{Scalar electrodynamics}
\label{sec:electro}

We start by reviewing the asymptotic symmetry structure of electrodynamics on a flat background (see \cite{He:2014cra,Lysov:2014csa} for the original references and \cite{Strominger:2017zoo} for a review). We denote the electromagnetic potential as $A_\mu(X)$ and the matter current as $\modj_\mu(X)$. For definiteness, we focus on complex scalar matter field $\phi(X)$ carrying a charge $Qe$, where $Q\in\mathbb R$ and $e$ is the elementary charge. The electromagnetic current is thus
\begin{equation}
    \modj_\mu = iQe\big(\phi^*\partial_\mu \phi - \phi\partial_\mu \phi^*\big) .
\end{equation} In the retarded coordinates $\{u, r, z,\bar z\}$, radiative boundary conditions for the potential $A_\mu$ near future null infinity are
\begin{equation}
    \begin{split}
        A_z(u,r,z,\bar z) &= A_z^{(0)}(u,z,\bar z) + \mathcal O(r^{-1}), \\
        A_r(u,r,z,\bar z) &= \mathcal{O}(r^{-2}) , \quad A_u(u,r,z,\bar z) = \mathcal{O }(r^{-1}) ,
    \end{split} \label{falloff A}
\end{equation}
consistently with Section \ref{sec:Massless scattering in flat spacetime}. Comparing with \eqref{outgoing spin s}, the role of the boundary field is played here by $A_z^{(0)}(u,z,\bar z)$. These boundary conditions are preserved by gauge transformations with $\lambda(u,r,z,\bar z) = \lambda^{(0)}(z,\bar z)+\mathcal O(r^{-1})$ such that $\delta_\lambda A^{(0)}_z = \partial_z \lambda^{(0)}$. On the matter side, the massless scalar field decays as
\begin{equation}
    \phi(u,r,z,\bar z) = \frac{\phi^{(0)}(u,z,\bar z)}{r} + \mathcal O(r^{-2})
\end{equation}
at future null infinity. The boundary field $\phi^{(0)}(u,z,\bar z)$ is left unconstrained by the massless Klein-Gordon equation and transforms as 
\begin{equation}
    \delta_\lambda \phi^{(0)}(u,z,\bar z) = -i\lambda^{(0)}(z,\bar z)Qe\phi^{(0)} (u,z,\bar z) \label{delta lambda matter}
\end{equation}
under the gauge symmetry. 

Expanding the field strength $F_{\mu\nu} = \partial_\mu A_\nu - \partial_\nu A_\mu$ and current $\modj_\mu$ in $1/r$ near $\mathscr I^+$ taking \eqref{falloff A}  into account, we have
\begin{equation}
    \begin{split}
        F_{uz}(u,r,z,\bar z) &= F_{uz}^{(0)}(u,z,\bar z) + \mathcal O(r^{-1})  , \\
        F_{ru}(u,r,z,\bar z) &= F_{ru}^{(2)}(u,z,\bar z)r^{-2} + \mathcal O(r^{-3}) ,\\
        F_{z\bar{z}}(u,r,z,\bar z)  &=  F_{z\bar{z}}^{(0)}(u,z,\bar z) + \mathcal O(r^{-1}), \\
        \modj_u(u,r,z,\bar z) &= \modj_u^{(2)}(u,z,\bar z)r^{-2} + \mathcal O(r^{-3}) .
    \end{split}
\end{equation} In addition, one requires that the magnetic field is vanishing at $\mathscr I^+_\pm$ (see \cite{Strominger:2015bla} for a consideration of magnetic contributions at $\mathscr{I}^+_\pm$), \textit{i.e.}
\begin{equation}
    \begin{split}
        F_{z\bar z}^{(0)}\Big|_{\mathscr I^+_\pm} = (\partial_z A^{(0)}_{\bar z} - \partial_{\bar z} A^{(0)}_{z})\Big|_{\mathscr I^+_\pm} = 0 . \label{no magnetic}
    \end{split}
\end{equation}
The leading equation of motion is
\begin{equation}
   \partial_u F_{ru}^{(2)} + \partial_{\bar z} F_{uz}^{(0)} + \partial_z F_{u\bar{z}}^{(0)} + e^2 \modj^{(2)}_u = 0 . \label{maxwell on the bndry}
\end{equation}
The surface charges associated with the residual gauge symmetry driven by $\lambda^{(0)}$ can be derived as \cite{He:2014cra,Lysov:2014csa,Strominger:2017zoo}
\begin{equation}
    Q_\lambda[A] = -\frac{1}{e^2}\int_{\Sigma} \D^2 z\, \lambda^{(0)}(z,\bar z)\, F^{(2)}_{ru}(u,z,\bar z) , \label{charge maxwell future}
\end{equation} where $\Sigma$ is a $\{u=\text{const.}\}$ cut of $\mathscr{I}^+$. As retarded time evolves along $\mathscr I^+$, these charges are not conserved because of the presence of electromagnetic radiation, encoded in the leading of piece of the Maxwell field $A^{(0)}_z$, and the null matter current $\modj^{(2)}_u$. Indeed, using \eqref{maxwell on the bndry}, one has the following flux-balance law:
\begin{equation}
    \frac{\D Q_{\lambda}}{\D u} = \frac{1}{e^2} \int_{\Sigma} \D^2 z\,\lambda^{(0)} \left( \partial_{\bar z} F_{uz}^{(0)} + \partial_z F_{u\bar{z}}^{(0)} + e^2\,\modj^{(2)}_u \right) \equiv \int_{\Sigma} \D^2 z\, F_\lambda. \label{flux balance maxwell future}
\end{equation}

In order to distinguish between hard (finite energy) and soft (zero energy) radiative excitations of the gauge field, one can split it as \cite{He:2014cra}
\begin{equation}
    A^{(0)}_z(u,z,\bar z) = \tilde A_z^{(0)}(u,z,\bar z) + \partial_z \chi(z,\bar z) \label{split hard soft QED}
\end{equation}
where $\chi(z,\bar z)$ is related to the early and late-time values of the boundary field as
\begin{equation}
    \partial_z \chi(z,\bar z) = \frac{1}{2}\left(A_z^{(0)}\Big|_{\mathscr I^+_+} + A_z^{(0)}\Big|_{\mathscr I^+_-}\right).
\end{equation}
Because $\chi(z,\bar z)$ transforms by an inhomogeneous shift ($\delta_\lambda \chi = \lambda^{(0)}$), it is interpreted as the Goldstone mode of asymptotic symmetry breaking while $\tilde A_z^{(0)}(u,z,\bar z)$ is taken to be gauge-invariant (\textit{i.e.} $\delta_\lambda \tilde A_z^{(0)} = 0$). The latter encodes the hard modes of the boundary value $A^{(0)}_z$ of the gauge field $A_\mu$, which are captured by the Fourier expansion \eqref{outgoing spin s} for $s=1$. In particular, it transforms as \eqref{Poincare on boundary value} with $J=1$ under Poincaré symmetries. Defining $\mathcal N^{(0)}_z\equiv \int_{-\infty}^{+\infty}\D u\, F_{uz}^{(0)}$ and recalling the boundary condition \eqref{no magnetic}, we have
\begin{equation}
    \begin{split}
        \partial_z \mathcal N^{(0)}_{\bar z} - \partial_{\bar z} \mathcal N^{(0)}_z = \int_{-\infty}^{+\infty} \D u \, \partial_u F_{z\bar z}^{(0)} = F_{z\bar z}^{(0)}\Big|_{\mathscr I^+_+} - F_{z\bar z}^{(0)}\Big|_{\mathscr I^+_-} = 0, \label{electricity condition for QED}
    \end{split}
\end{equation}
where we used the Bianchi identity $\partial_{[\mu} F_{\nu\rho]} = 0$ expressed for the leading components of the Faraday tensor. This equation is solved if there exists a scalar field $N(z,\bar z)$ such that
\begin{equation}
    \partial_z N(z,\bar z) \equiv \frac{1}{e^2}\int_{-\infty}^{+\infty}\D u\, F^{(0)}_{uz}(u,z,\bar z) = \frac{1}{e^2}\left(A_z^{(0)}\Big|_{\mathscr I^+_+} - A_z^{(0)}\Big|_{\mathscr I^+_-}\right) . \label{def of N QED}
\end{equation}
Notice that the so-defined $N(z,\bar z)$, referred to as the memory mode, is gauge-invariant. Hence the radiative variables are organized in hard and soft fields as
\begin{equation}
    \Gamma^H = \{\tilde A^{(0)}_z,\partial_u \tilde A^{(0)}_z,\tilde A^{(0)}_{\bar z},\partial_u\tilde A^{(0)}_{\bar z}\},\quad \Gamma^S = \{\partial_z\chi,\partial_z N,\partial_{\bar z}\chi,\partial_{\bar z}N\}.
\end{equation}
The evaluation of the symplectic structure at the boundary $\mathscr I^+$ has been detailed \textit{e.g.} in \cite{He:2014cra,Strominger:2017zoo} and gives rise to the following Poisson brackets: 
\begin{align}
        &\left\{ \partial_{u_1} \tilde A_{z_1}^{(0)}(u_1,z_1,\bar z_1),\tilde A_{\bar z_2}^{(0)}(u_2,z_2,\bar z_2) \right\} = -\frac{e^2}{2}\,\delta(u_1-u_2)\,\delta^{(2)}(z_1-z_2) , \label{bracket hard maxwell} \\
        &\left\{ \partial_{z_1} N(z_1,\bar z_1), \partial_{\bar z_2} \chi(z_2,\bar z_2)\right\} = -\frac{1}{2}\,\delta^{(2)}(z_1-z_2)
\label{bracket soft maxwell}
\end{align}
for the hard and soft symplectic pairs $(\tilde A^{(0)}_z,\partial_u \tilde A^{(0)}_{\bar z})$ and $(\partial_z\chi,\partial_{\bar z} N)$. The quantum version of the bracket \eqref{bracket hard maxwell} is nothing but \eqref{commu boundary fields} for the choice of normalization \eqref{values of Ks}. Because the theory is linear, the hard flux only encompasses matter contributions while the flux terms depending on the gauge field are soft, \textit{i.e.}
\begin{align}
    \mathcal F^H_\lambda &= \int_{\mathscr I^+} \D u\,\D^2 z\,\lambda^{(0)}\,\modj_u^{(2)}, \label{flux hard maxwell} \\
    \mathcal F^S_\lambda &= \frac{1}{e^2}\int_{\mathscr I^+} \D u\,\D^2 z\,\lambda^{(0)}\left(\partial_{\bar z}F_{uz}^{(0)} + \partial_z F_{u\bar z}^{(0)}\right) = 2\int_\Sigma \D^2 z\,\lambda^{(0)}\,\partial_z\partial_{\bar z}N, \label{flux soft maxwell} 
\end{align}
where the second equality of \eqref{flux soft maxwell} holds because of \eqref{electricity condition for QED} and \eqref{def of N QED}. From \eqref{bracket hard maxwell}--\eqref{bracket soft maxwell} together with the canonical commutation relation $\{\phi^{(0)}(u,z,\bar z),\partial_{u'}\phi^{(0)}(u',z',\bar z')\} = \delta(u-u')\,\delta^{(2)}(z-z')$ for the matter field, one deduces the very important properties
\begin{equation}
\begin{split}
    &\{ \mathcal F_\lambda^H,\Gamma^S \} = 0 = \{ \mathcal F_\lambda^S,\Gamma^H \}, \\
    &\{ \mathcal F_\lambda^H,\tilde A_z^{(0)}\} = 0,\quad \{ \mathcal F_\lambda^S,\partial_z\chi\} = \partial_z\lambda^{(0)}, \\
    &\{ \mathcal F_\lambda^H,\phi^{(0)}\} = \delta_\lambda\phi^{(0)},\quad \{ \mathcal F_\lambda^S,\phi^{(0)}\} = 0.
\end{split} \label{symmetry generated by fluxes maxwell}
\end{equation}
Because $N$ and $\modj_\mu$ are gauge-invariant, we also have that hard and soft fluxes represent separately the $U(1)$ algebra, \textit{i.e.} $\delta_{\lambda_1} \mathcal F_{\lambda_2}^{H/S} = \{\mathcal F_{\lambda_1}^{H/S},\mathcal F_{\lambda_2}^{H/S}\} = 0$.

A similar analysis can be performed at $\mathscr{I}^-$, choosing the advanced Bondi coordinates $\{v,r,z,\bar z\}$ as defined in Appendix \ref{app:Bondi adv and ret}. The equation of motion \eqref{maxwell on the bndry} becomes
\begin{equation}
    -\partial_v F_{rv}^{(2)} + \partial_{\bar z} F_{vz}^{(0)}+ \partial_z F_{v\bar z}^{(0)}+ e^2 \modj_v^{(2)} = 0 .
    \label{maxwell EOM past}
\end{equation}
The induced surface charges are
\begin{equation}
    Q_\lambda[A] = -\frac{1}{e^2}\int_\Sigma \D^2 z\, \lambda^{(0)}(z,\bar z)\, F_{rv}^{(2)}(v,z,\bar z) ,
\end{equation}
with no sign change with respect to \eqref{charge maxwell future}. The fluxes get a sign flip because of the corresponding sign flip in the equation of motion \eqref{maxwell EOM past}, \textit{i.e.}
\begin{equation}
    \frac{\D Q_{\lambda}}{\D v} = - \frac{1}{e^2} \int_{\Sigma} \D^2 z\,\lambda^{(0)} \left( \partial_{\bar z} F_{vz}^{(0)} + \partial_z F_{v\bar{z}}^{(0)} + e^2\,\modj^{(2)}_v \right)=  \int_\Sigma \D^2 z\, F_\lambda. \label{flux balance maxwell past}
\end{equation}
Using an homologous split between hard and soft modes in the boundary value $A^{(0)}_z(v,z,\bar z)$ of the gauge field, it can be shown that the property \eqref{symmetry generated by fluxes maxwell} also holds at past null infinity without any relative change of sign. Notice finally that the convention \eqref{antipodal on fields} implies $A^{(0)}_z(u\to-\infty,z,\bar z) = A^{(0)}_z(v\to+\infty,z,\bar z)$, \textit{i.e.} the antipodal matching condition around spatial infinity as stated in \cite{Kapec:2015ena}.

\subsection{Gravity}
\label{sec:gravity asymptotics}

We now switch to gravity in $4d$ asymptotically flat spacetimes. We follow the same notations and conventions than \cite{Donnay:2021wrk,Donnay:2022aba,Donnay:2022hkf}. In retarded Bondi coordinates $\{u, r, x^A\}$, $x^A = (z, \bar{z})$, the solution space of four-dimensional asymptotically flat metrics reads as 
\begin{equation}
    \begin{split}
        ds^2 = \, &\left(\frac{2M}{r}+\mathcal O(r^{-2})\right) \D u^2 - 2 \left( 1+\mathcal O(r^{-2})\right) \D u \D r\\
        &+ \left(r^2 \mathring{q}_{AB} + r\, C_{AB} + \mathcal{O}(r^{0})\right) \D x^A \D x^B \\
    &+ \left(\frac{1}{2}\partial^B C_{AB} + \frac{2}{3r}(N_A + \frac{1}{4}C_{AB} \partial_C C^{BC}) + \mathcal O(r^{-2}) \right)\D u \D x^A ,
    \end{split} \label{Bondi gauge metric}
\end{equation}
where $\mathring q_{AB} \D x^A \D x^B = 2\D z\D\bar z$ is the flat metric on the punctured complex plane $\mathbb{C}^*$. The topology of future null infinity is taken to be $\mathscr I^+\simeq \mathbb R\times \mathbb C^*$ in order to allow meromorphic superrotations among the set of asymptotic symmetries (see \textit{e.g.} \cite{Barnich:2021dta,Barnich:2017ubf,Donnay:2021wrk} for details on the precise set-up). Indices $A$, $B$ are lowered and raised by $\mathring q_{AB}$ and its inverse. The asymptotic shear $C_{AB}$ is a symmetric trace-free tensor ($\mathring{q}^{AB} C_{AB} = 0$). The Bondi mass aspect $M(u,z,\bar z)$ and the angular momentum aspect $N_A (u,z,\bar z)$ satisfy the time evolution/constraint equations
\begin{equation}
\begin{split}
\partial_u M &= - \frac{1}{8} N_{AB} N^{AB} + \frac{1}{4} \partial_A \partial_B N^{AB} - 4\pi G\, T_{uu}^{m(2)} , \\ 
\partial_u N_A &= \partial_A M + \frac{1}{16} \partial_A (N_{BC} C^{BC}) - \frac{1}{4} N^{BC} \partial_A C_{BC} - 8\pi G\, T_{uA}^{m(2)}\\
&\quad  -\frac{1}{4} \partial_B (C^{BC} N_{AC} - N^{BC} C_{AC}) - \frac{1}{4} \partial_B \partial^B \partial^C C_{AC}+ \frac{1}{4} \partial_B \partial_A \partial_C C^{BC} ,
\end{split}\label{EOM1} 
\end{equation} 
with $N_{AB} = \partial_u C_{AB}$ the Bondi news tensor encoding the outgoing radiation and $T_{\mu\nu}^m$ the null matter stress tensor whose expansion near null infinity is taken to be
\begin{equation}
    \begin{split}
        T_{uu}^m(u,r,z,\bar z) &= T_{uu}^{m(2)}(u,z,\bar z)\frac{1}{r^2}+\mathcal O(r^{-3}) , \\ 
        T_{uA}^m(u,r,z,\bar z) &= T_{uA}^{m(2)}(u,z,\bar z)\frac{1}{r^2}+\mathcal O(r^{-3})  .
    \end{split}
\end{equation}

The diffeomorphisms preserving the solution space \eqref{Bondi gauge metric} are generated by vectors fields $\xi = \xi^u \partial_u + \xi^z \partial_z + \xi^{\bar{z}} \partial_{\bar z} + \xi^r \partial_r$ whose leading order components read 
\begin{equation}
\begin{split}
    &\xi^u =\mathcal{T} + \frac{u}{2} (\partial_z \mathcal{Y}^z +  \partial_{\bar z} \mathcal{Y}^{\bar z}) , \\
    &\xi^z = \mathcal{Y}^z + \mathcal{O}(r^{-1}) , \quad \xi^{\bar{z}} = \mathcal{Y}^{\bar z} + \mathcal{O}(r^{-1}) , \\ 
    &\xi^r = - \frac{r}{2} (\partial_z \mathcal{Y}^z + \partial_{\bar z} \mathcal{Y}^{\bar z}) + \mathcal{O}(r^0) ,
    \end{split}
    \label{AKV Bondi}
\end{equation} where $\mathcal{T}= \mathcal{T}(z, \bar{z})$ is the supertranslation parameter and $\mathcal{Y}^z = \mathcal{Y}^z(z)$, $\mathcal{Y}^{\bar z} = \mathcal{Y}^{\bar z}(\bar{z})$ are the superrotation parameters satisfying the conformal Killing equation
\begin{equation}
    \partial_{\bar z} \mathcal{Y}^z = 0 , \quad \partial_z \mathcal{Y}^{\bar z} = 0   .\label{CKV Y}
\end{equation} 
Using the modified Lie bracket $[\xi_1, \xi_2]_\star \equiv [\xi_1, \xi_2] - \delta_{\xi_1}\xi_2 + \delta_{\xi_2} \xi_1$, where the last two terms take into account the field-dependence of \eqref{AKV Bondi} at subleading orders in $r$ \cite{Barnich:2010eb}, these asymptotic Killing vectors satisfy the commutation relations
\begin{equation}
    \big[\xi (\mathcal{T}_1, \mathcal{Y}^z_1, \mathcal{Y}^{\bar z}_1), \xi (\mathcal{T}_2, \mathcal{Y}^z_2, \mathcal{Y}^{\bar z}_2) \big]_\star = \xi (\mathcal{T}_{12}, \mathcal{Y}^z_{12}, \mathcal{Y}^{\bar z}_{12}) ,
    \label{commutation relations 1}
\end{equation} with
\begin{equation}
\begin{split}
    \mathcal{T}_{12} &= \mathcal{Y}^z_1 \partial_z \mathcal{T}_2 - \frac{1}{2} \partial_z \mathcal{Y}^z_1 \mathcal{T}_2 + \text{c.c.} - (1 \leftrightarrow 2)   \, , \\
    \mathcal{Y}^z_{12} &= \mathcal{Y}^z_1 \partial_z \mathcal{Y}^z_2 - (1 \leftrightarrow 2)\,, \quad \mathcal{Y}^{\bar z}_{12} = \mathcal{Y}^{\bar z}_1 \partial_{\bar z} \mathcal{Y}^{\bar z}_2 - (1 \leftrightarrow 2) .
\end{split}
    \label{commutation relations 2}
\end{equation} 
This corresponds to the (extended) BMS algebra \cite{Barnich:2009se, Barnich:2010eb, Barnich:2011ct}. The six globally well-defined solutions of \eqref{CKV Y} and the four linearly-independent solutions of $(\partial_A\partial_B \mathcal T)^{TF} = 0$ generate the Poincaré subgroup. The latter condition, where $TF$ means the tracefree part with respect to the boundary metric $\mathring q_{AB}$, is generally solved as $\mathcal T(z,\bar z) = b_\mu q^\mu(z,\bar z)$ where $b^\mu$ are the constant parameters of an infinitesimal bulk translation in Cartesian coordinates.

The infinitesimal transformation of the solution space is induced by the Lie derivative of the bulk metric along BMS generators \eqref{AKV Bondi}. The most crucial transformation for the purpose of this paper is the variation of the asymptotic shear, which reads as
\begin{equation}
    \begin{split}
        &\delta_\xi C_{zz} = \delta^H_\xi C_{zz} + \delta^S_\xi C_{zz} , \\
        &\delta^H_\xi C_{zz} = \left[\left(\mathcal{T} + \frac{u}{2}(\partial_z\mathcal{Y}^z + \partial_{\bar z} \mathcal Y^{\bar z})\right) \partial_u + \mathcal Y^z \partial_z + \mathcal Y^{\bar z}\partial_{\bar z} + \frac{3}{2} \partial_z \mathcal Y^z - \frac{1}{2} \partial_{\bar z} \mathcal Y^{\bar z}\right] C_{zz} , \\
        &\delta^S_\xi C_{zz} = -2 \partial_z^2 \mathcal{T} - u\, \partial_z^3 \mathcal Y^z ,
    \end{split}\label{transfo CAB}
\end{equation}
together with the complex conjugate relations for $C_{\bar{z}\bar{z}}$. The hard and soft parts of the transformation are denoted by $\delta_\xi^H C_{AB}$ and $\delta_\xi^S C_{AB}$, respectively, and are built up from the respective homogeneous and inhomogeneous terms appearing in the induced variation.

Following the discussion of \cite{Compere:2018ylh,Campiglia:2020qvc,Compere:2020lrt,Fiorucci:2021pha,Donnay:2021wrk,Freidel:2021qpz,Freidel:2021dfs,Freidel:2021ytz,Donnay:2022hkf}, the BMS charges are taken to be
\begin{equation}
    Q_\xi[g] = \frac{1}{16\pi G} \int_{\Sigma} \D^2 z\, \big(4 \mathcal{T} \bar M + 2 \mathcal Y^A \bar N_A\big) , \label{grav charge}
\end{equation}
where the redefined mass $\bar M$ and angular momentum aspect $\bar N_A$ are given by
\begin{equation}
\begin{split}
    \bar M = M &+ \frac{1}{8}N_{AB}C^{AB} , \\
    \bar N_A = N_A &- u\, \partial_A \bar M + \frac{1}{4}C_{AB} \partial_C C^{BC} +\frac{3}{32}\partial_A (C_{BC}C^{BC}) \\
    &+ \frac{u}{4}  \partial^B \big[(\partial_B \partial_C - \frac{1}{2}N_{BC}) {C_A}^C \big] - \frac{u}{4} \partial^B \big[(\partial_A \partial_C - \frac{1}{2}N_{AC}) {C_B}^C\big] . \label{BMS momenta}
    \end{split}
\end{equation}
The latter can be rewritten in a more elegant way in terms of Newman-Penrose coefficients \cite{Newman:1961qr,Newman:1962cia} as the compact expressions $\bar M = -\frac{1}{2}(\Psi_2^0+\bar\Psi_2^0)$, $\bar N_z = -\Psi_1^0+u\,\partial_z \Psi_2^0$ and $\bar N_{\bar z} = \bar N_z^*$; see \cite{Donnay:2022hkf}. Using the time evolution/constraint equations \eqref{EOM1}, these charges satisfy the flux-balance laws
\begin{equation}
    \frac{\D Q_\xi}{\D u} = \int_{\Sigma} \D^2 z\, F_{\xi(\mathcal T,\mathcal Y)} , \label{flux balance BMS 1}
\end{equation}
where the local fluxes (respectively associated with supertranslations and superrotations) read as 
\begin{align}
    F_{\xi(\mathcal T,0)} &= \frac{1}{16\pi G}\,\mathcal T\left[\partial_z^2 N_{\bar z\bar z} + \frac{1}{2} C_{\bar z\bar z}\partial_u N_{zz}+\text{c.c.} \right] - \mathcal T\,T_{uu}^{m(2)}, \label{FT} \\
    F_{\xi(0,\mathcal Y)} &= \frac{1}{16\pi G}\,\mathcal Y^z \left[-u\partial^3_z N_{\bar z\bar z} + C_{zz}\partial_z N_{\bar z\bar z} - \frac{u}{2}\partial_z C_{zz}\partial_u N_{\bar z\bar z} - \frac{u}{2}C_{zz}\partial_z\partial_u N_{\bar z\bar z}\right] \nonumber \\
    &\quad -\mathcal Y^z T^{m(2)}_{uz} + \frac{u}{2}\mathcal Y^z\partial_z T_{uu}^{m(2)} +\text{c.c.} \label{FY}
\end{align}

As in electrodynamics, a careful split between hard and soft boundary degrees of freedom can be performed, taking into account the factorization of the radiative phase space between hard and soft sectors \cite{Campiglia:2021bap,Donnay:2022hkf,Campiglia:2020qvc}. At quantum level, this split allows for the BMS Ward identities to correctly reproduce the leading and subleading soft theorems \cite{He:2014laa,Kapec:2014opa} at all orders of perturbation \cite{Donnay:2022hkf,Pasterski:2022djr}. For completeness, we provide here a mere summary of the construction and refer to \cite{Donnay:2022hkf} for details. 

For the analysis of the radiative phase spaces including superrotations, the following early and late time behavior of the radiative fields are imposed \cite{Campiglia:2021bap,Campiglia:2020qvc,Compere:2020lrt,Donnay:2021wrk,Fiorucci:2021pha}:
\begin{equation}
    C_{AB}\Big|_{\mathscr I^+_\pm} = (u+C_\pm)N^{vac}_{AB} - 2 (\partial_A\partial_B C_\pm)^{TF} + o(u^{-1}),\quad N_{AB} = N_{AB}^{vac} + o(u^{-2}), \label{falloff in u for SR}
\end{equation}
where $C_\pm(z,\bar z)$ correspond to the values of the supertranslation field for $u\to\pm \infty$ and whose difference encodes the displacement memory effect, and $N_{AB}^{vac}(z,\bar z)$ is the vacuum news tensor \cite{Compere:2016jwb,Compere:2018ylh}, identified with the tracefree part of the Geroch tensor \cite{1977asst.conf....1G, Campiglia:2020qvc, Nguyen:2022zgs}. The latter can be expressed as
\begin{equation}
    N_{zz}^{vac} = \frac{1}{2}(\partial_z\varphi)^2-\partial^2_z\varphi, \quad N_{\bar z\bar z}^{vac} = \frac{1}{2}(\partial_{\bar z}\bar\varphi)^2-\partial^2_{\bar z}\bar\varphi,
\end{equation}
where $\varphi(z)$, $\bar\varphi(\bar z)$ are the holomorphic superboost fields encoding the velocity kick (or refraction) memory effect \cite{Compere:2018ylh}. For conformal primary fields $\psi_{k,\bar k}(z,\bar z)$ of weights $(k,\bar k)$ that transform as
\begin{equation}
    \delta_{\mathcal Y}\psi_{(k,\bar k)}(z,\bar z) = \left(\mathcal Y^z\partial_z+\mathcal Y^{\bar z}\partial_{\bar z}+k\,\partial_z\mathcal Y^z+\bar k\,\partial_{\bar z}\mathcal Y^{\bar z}\right)\psi_{(k,\bar k)}(z,\bar z)
\end{equation}
under meromorphic superrotations, it is convenient to introduce the conformally covariant derivative operators \cite{Barnich:2021dta,Campiglia:2020qvc}
\begin{equation}
    \begin{split}
        &\mathscr D_z : (k,\bar k)\to (k+1,\bar k) : \psi_{(k,\bar k)}\mapsto \mathscr D_z \psi_{(k,\bar k)} = [\partial_z - k\,\partial_z\varphi]\psi_{(k,\bar k)}\\
        &\mathscr D_{\bar z} : (k,\bar k)\to (k,\bar k+1) : \psi_{(k,\bar k)}\mapsto \mathscr D_{\bar z} \psi_{(k,\bar k)} = [\partial_{\bar z} - \bar k\,\partial_{\bar z}\bar\varphi]\psi_{(k,\bar k)}
    \end{split} \label{mathscr D op}
\end{equation}
satisfying $[\mathscr D_z,\mathscr D_{\bar z}]\psi_{(k,\bar k)} = 0$. With these definition, the split between hard and soft variables works as follows. Defining
\begin{equation}
    C_{zz}\equiv u N_{zz}^{vac} + C^{(\circ)}_{zz}+\tilde C_{zz},\quad N_{zz} = N_{zz}^{vac}+\tilde N_{zz} \label{split CAB hard soft}
\end{equation}
where $C_{zz}^{(\circ)}(z,\bar z) = -2\mathscr D_z^2 C^{(\circ)}(z,\bar z)$ for $C^{(\circ)} = \frac{1}{2}(C_+ + C_-)$, the Goldstone mode of supertranslation. The falloff conditions \eqref{falloff in u for SR} imply that $\tilde N_{zz} = \partial_u \tilde C_{zz} = o(u^{-2})$ and that the asymptotic shear is ``purely electric'' at early and late times
\begin{equation}
    (\mathscr D_z^2 C_{\bar z\bar z} - \mathscr D_{\bar z}^2 C_{zz})\big|_{\mathscr I^+_\pm} = 0 \quad \Rightarrow \quad (\mathscr D_z^2 \tilde C_{\bar z\bar z} - \mathscr D_{\bar z}^2 \tilde C_{zz})\big|_{\mathscr I^+_\pm} = 0. \label{electricity condition}
\end{equation} 
The implication comes from the fact that $\mathscr D_z$ and $\mathscr D_{\bar z}$ commute, hence $\mathscr D_z C^{(\circ)}_{\bar z\bar z} - \mathscr D_{\bar z} C^{(\circ)}_{zz} = 0$ and by definition $\mathscr D_z N^{vac}_{\bar z\bar z} = 0 = \mathscr D_{\bar z} N^{vac}_{zz}$. The condition \eqref{electricity condition} is the analog of \eqref{electricity condition for QED} for gravity and is solved by
\begin{equation}
    \tilde C_{zz}\big|_{\mathscr I^+_\pm} = \mp 2\mathscr D^2_z N^{(\circ)}, \quad N^{(\circ)}(z,\bar z) = \frac{1}{2}(C_+ - C_-). \label{tilde C corner}
\end{equation}
The hard variables of the radiative phase space are collectively denoted as
\begin{equation}
    \Gamma^{H} = \{\tilde C_{zz},\tilde N_{zz},\tilde C_{\bar z\bar z},\tilde N_{\bar z\bar z}\}. \label{Gamma hard}
\end{equation}
On the other hand, soft variables are identified as
\begin{equation}
    \Gamma^{S} = \{C_{zz}^{(\circ)},C_{\bar z\bar z}^{(\circ)},\mathcal N^{(0)}_{zz},\mathcal N^{(0)}_{\bar z\bar z},\Pi_{zz},\Pi_{\bar z\bar z},N^{vac}_{zz},N^{vac}_{\bar z\bar z}\},\quad \Pi_{zz}\equiv 2\mathcal N_{zz}^{(1)}+C^{(0)}\mathcal N_{zz}^{(0)}, \label{Gamma soft}
\end{equation}
where $\mathcal N^{(0)}_{zz}$ and $\mathcal N^{(1)}_{zz}$ are respectively the leading and subleading soft news \cite{He:2014laa,Kapec:2014opa} given by
\begin{equation}
    \mathcal N^{(0)}_{zz}(z,\bar z) \equiv \int_{-\infty}^{+\infty} \D u\, \tilde N_{zz}(u,z,\bar z),\quad \mathcal N^{(1)}_{zz}(z,\bar z) \equiv \int_{-\infty}^{+\infty} \D u\, u\,\tilde N_{zz}(u,z,\bar z).
\end{equation}
The boundary condition $\tilde N_{AB} = o(u^{-2})$ implies that both integrals converge and in particular $\mathcal N^{(0)}_{zz}(z,\bar z) = -4\mathscr D^2_{z}N^{(\circ)}(z,\bar z)$ because of \eqref{tilde C corner}. The transformations of all these objects, which can be worked out from \eqref{transfo CAB}, are reproduced in Equation (3.6) of \cite{Donnay:2022hkf}. The fields in \eqref{Gamma hard} are functions of $(u,z,\bar z)$ and transform homogeneously under the action of extended BMS symmetries. For instance,
\begin{equation}
    \delta_{\xi(\mathcal T,\mathcal Y)}\tilde C_{zz} = \left(\mathcal T+\frac{u}{2}(\partial_z\mathcal Y^z+\partial_{\bar z}\mathcal Y^{\bar z}\right)\tilde N_{zz} + \left(\mathcal Y^z\partial_z +\mathcal Y^{\bar z}\partial_{\bar z} + \frac{3}{2}\partial_z\mathcal Y^z - \frac{1}{2}\partial_{\bar z}\mathcal Y^{\bar z}\right)\tilde C_{zz}.
\end{equation}
The quantized modes of $\tilde C_{zz}(u,z,\bar z)$ are simply captured by the expansion \eqref{outgoing spin s} for $s=2$. The fields in \eqref{Gamma soft}, on the other hand, are functions of $(z,\bar z)$ only, defined at the corners of $\mathscr I^+$ and their transformation laws
\begin{align}
    \delta_{\xi(\mathcal T,\mathcal Y)}C_{zz}^{(\circ)} &= \left(\mathcal Y^z\partial_z + \mathcal Y^{\bar z}\partial_{\bar z} + \frac{3}{2}\partial_z\mathcal Y^z - \frac{1}{2}\partial_{\bar z}\mathcal Y^{\bar z}\right)C_{zz}^{(\circ)} - 2\mathscr D_z^2 \mathcal T,\\
    \delta_{\xi(\mathcal T,\mathcal Y)} N^{vac}_{zz} &= \left(\mathcal Y^z\partial_z + 2\partial_z\mathcal Y^z \right)N_{zz}^{vac} - \partial_z^3 \mathcal Y^z,
\end{align}
account for the inhomogeneous pieces in \eqref{transfo CAB}. As shown in \cite{Donnay:2022hkf,Campiglia:2021bap}, \eqref{Gamma hard}--\eqref{Gamma soft} constitute Darboux variables parametrizing the radiative phase space of asymptotically flat gravity at $\mathscr I^+$. Using them, the suitable split of the integrated BMS fluxes $\mathcal F_{\xi(\mathcal T,\mathcal Y)} = \int_{\mathscr I^+}\D u\,\D^2 z\,F_{\xi(\mathcal T,\mathcal Y)}$ computed from \eqref{FT}--\eqref{FY} into pure hard and soft parts has been proposed to be
\begin{equation}
    \begin{split}
        & \mathcal F^{H}_{\xi(\mathcal T,0)} = -\frac{1}{16\pi G}\int_{\mathscr I^+}\D u\,\D^2 z\,\mathcal T\,\big[\tilde N_{zz}\tilde N_{\bar z\bar z}\big] - \int_{\mathscr I^+}\D u\,\D^2 z\,\mathcal T\,T^{m(2)}_{uu},\\
        & \mathcal F^{S}_{\xi(\mathcal T,0)} = \frac{1}{8\pi G}\int_\Sigma \D^2 z\,\mathcal T\,\big[\mathscr D_z^2 \mathcal N^{(0)}_{\bar z\bar z}\big], \\
        & \mathcal F_{\xi(0,\mathcal Y)}^{H} = \frac{1}{16\pi G}\int_{\mathscr I^+}\D u\,\D^2 z\,\mathcal Y^z\,\left[ \frac{3}{2}\tilde C_{zz}\partial_z \tilde N_{\bar z\bar z} + \frac{1}{2}\tilde N_{zz}\partial_z \tilde C_{zz} + \frac{u}{2}\partial_z\left(\tilde N_{zz}\tilde N_{\bar z\bar z}\right)\right] \\
        & \hspace{45pt} - \int_{\mathscr I^+}\D u\,\D^2 z\,\mathcal Y^z \left[ T^{m(2)}_{uz} + \frac{u}{2}\,\partial_z\mathcal Y^z\, T^{m(2)}_{uu}\right] + \text{c.c.}, \\
        & \mathcal F_{\xi(0,\mathcal Y)}^{S} = \frac{1}{16\pi G}\int_\Sigma \D^2 z\,\mathcal Y^z\,\left[-\mathscr D^3_z \mathcal N^{(1)}_{\bar z\bar z} + \frac{3}{2}C_{zz}^{(\circ)}\mathscr D \mathcal N_{\bar z\bar z}^{(0)} + \frac{1}{2}\mathcal N_{\bar z\bar z}^{(0)}\mathscr D_z C_{zz}^{(\circ)}\right]+\text{c.c.}
    \end{split} \label{FH FS gravity}
\end{equation}
where \eqref{electricity condition} was used to simplify the expressions of soft fluxes. Unlike the case of electromagnetism, the nonlinearity of Einstein theory implies that hard fluxes do include pure gravitational terms, notably the first term in $F^H_{\xi(1,0)}$, responsible for the Bondi mass loss. Inverting the symplectic structure yields the following Poisson brackets
\begin{align}
    &\big\{\tilde N_{z_1 z_1}(u_1,z_1,\bar z_1),\tilde C_{\bar z_2\bar z_2}(u_2,z_2,\bar z_2)\big\} = -16\pi G\,\delta(u_1-u_2)\,\delta^{(2)}(z_1-z_2),\\
    &\begin{aligned}
        &\big\{\mathcal N^{(0)}_{z_1 z_1}(z_1,\bar z_1),C^{(0)}_{\bar z_2\bar z_2}(z_2,\bar z_2)\big\} = -16\pi G\,\delta^{(2)}(z_1-z_2) ,\\
        &\big\{\Pi_{z_1 z_1}(z_1,\bar z_1) ,N^{vac}_{\bar z_2\bar z_2}(z_2,\bar z_2)\big\} = -16\pi G\,\delta^{(2)}(z_1-z_2) ,
    \end{aligned}
\end{align}
from which one can derive the very important property that the hard and soft parts of the fluxes act independently on \eqref{Gamma hard} and \eqref{Gamma soft} as
\begin{equation}
    \begin{split}
        &\big\{ \mathcal F^{H}_{\xi(\mathcal T,\mathcal Y)},\Gamma^{H} \big\} = \delta_{\xi(\mathcal T,\mathcal Y)}\Gamma^{H},\quad \big\{ \mathcal F^{H}_{\xi(\mathcal T,\mathcal Y)},\Gamma^{S} \big\} = 0,\\
        &\big\{ \mathcal F^{S}_{\xi(\mathcal T,\mathcal Y)},\Gamma^{S} \big\} = \delta_{\xi(\mathcal T,\mathcal Y)}\Gamma^{S} , \quad \big\{ \mathcal F^{S}_{\xi(\mathcal T,\mathcal Y)},\Gamma^{H} \big\} = 0 ,
    \end{split} \label{variations generated}
\end{equation}
and form separately two representations of the extended BMS algebra \eqref{commutation relations 1}--\eqref{commutation relations 2} \cite{Donnay:2021wrk,Donnay:2022hkf}.

As usual, a similar analysis can be performed at past null infinity $\mathscr I^-$ by trading the retarded coordinates for the advanced ones. At the vicinity of $\mathscr I^-$, the solution space is expanded as 
\begin{equation}
    \begin{split}
        ds^2 = \, &\left(\frac{2M^{(-)}}{r'}+\mathcal O(r'^{-2})\right) \D v^2 + 2 \left( 1+\mathcal O(r'^{-2})\right) \D v \D r'\\
        &+ \left(r'^2 \mathring{q}_{AB} + r'\, C_{AB}^{(-)} + \mathcal{O}(r'^{0})\right) \D x^A \D x^B \label{Bondi gauge metric PAST} \\
        &- \left(-\frac{1}{2}\partial^B C_{AB}^{(-)} - \frac{2}{3r'}(N_A^{(-)} + \frac{1}{4}C^{(-)}_{AB} \partial_C C^{BC}_{(-)}) + \mathcal O(r'^{-2}) \right)\D v \D x^A ,
    \end{split}
\end{equation}
still denoting $x^A = (z',\bar z') = (z,\bar z)$. The Bondi time evolution/constraint equations for the Bondi mass aspect $M^{(-)}(v,z,\bar z)$ and the angular momentum aspect $N_A^{(-)} (v,z,\bar z)$ at $\mathscr I^-$ read as
\begin{equation} 
    \begin{split}
        \partial_v M^{(-)} &= \frac{1}{8} N_{AB}^{(-)} N^{AB}_{(-)} + \frac{1}{4} \partial_A \partial_B N^{AB}_{(-)} + 4\pi G\, T^{m(2,-)}_{vv} \, , \\ 
        \partial_v N_A^{(-)} &= -D_A M^{(-)} + \frac{1}{16} \partial_A (N_{BC}^{(-)} C^{BC}_{(-)}) - \frac{1}{4} N^{BC}_{(-)} D_A C_{BC}^{(-)} - 8\pi G\, T^{m(2,-)}_{vA} \\
        &\quad  -\frac{1}{4} \partial_B (C^{BC}_{(-)} N_{AC}^{(-)} - N^{BC}_{(-)} C_{AC}^{(-)}) + \frac{1}{4} \partial_B \partial^B \partial^C C_{AC}^{(-)}- \frac{1}{4} \partial_B \partial_A \partial_C C^{BC}_{(-)} ,
    \end{split} \label{constraint bondi past}
\end{equation} 
with the (past) Bondi news tensor $N^{(-)}_{AB}=\partial_v C^{(-)}_{AB}$ encoding the incoming radiation. Under the BMS symmetries acting at $\mathscr I^-$
\begin{equation}
    \xi\Big|_{\mathscr I^-} = \left(\mathcal T(z,\bar z) + \frac{v}{2}(\partial_z\mathcal Y^z + \partial_{\bar z}\mathcal Y^{\bar z})\right)\partial_v + \mathcal Y^z(z)\partial_z + \mathcal Y^{\bar z}(\bar z)\partial_{\bar z},
\end{equation}
the variation of the shear is still given by \eqref{transfo CAB} with $u\mapsto v$ and a sign flip in $\delta_\xi^S C_{zz}$. The covariant phase space analysis and a similar split between hard and soft boundary degrees of freedom yield again \eqref{variations generated}. Around spatial infinity, the requirement \eqref{antipodal on fields} now implies $C_{zz}(u\to-\infty,z,\bar z) = -C_{zz}^{(-)}(v\to+\infty,z,\bar z)$, where the minus sign is due to $r'=-r$ and the convention \eqref{Boundary value PAST}. This agrees with the antipodal matching in the sense of \cite{Strominger:2013jfa,Kapec:2014opa}.

\section{Sourced quantum field theory}
\label{sec:Sourced quantum field theory}

As discussed in Section \ref{sec:Asymptotic symmetry analysis} the charges associated with the asymptotic symmetries are not conserved due to the radiation leaking through $\mathscr{I}$. We argued in \cite{Donnay:2022aba} (see also Section \ref{sec:Holographic conformal Carrollian field theory}) that the non-conservation of the bulk charges can be holographically interpreted as a coupling of the dual theory with some external sources. In this section, we discuss a general framework that allows to deal with symmetries in presence of external sources. The details on the newly introduced notions and the resulting construction will be presented in the upcoming paper \cite{Barnich:2022xx} (see also \cite{Troessaert:2015nia,Wieland:2020gno}). We argue that, at the classical level, the inclusion of external sources in field theories allows to derive a generalized version of Noether's theorem giving an interpretation of flux-balance laws for the Noether currents from symmetry principle. We then obtain the quantum analogue of this result by writing the sourced Ward identities using a path integral formulation. Finally, we exemplify these sourced Ward identities for a sourced theory exhibiting $U(1)$ symmetries and BMS symmetries.

\subsection{Generalized Noether symmetries}
\label{sec:Generalized Noether symmetries}

In this section, we briefly discuss a generalization of Noether's theorem that allows to write some flux-balance laws associated with (generalized) symmetries in presence of external sources. We consider a theory living on a $n$-dimensional manifold $\mathscr{M}$ with coordinates $x^a$ and boundary $\partial\mathscr M$. The dynamical fields are denoted by $\Psi^i(x)$ and the external sources by $\sigma^m(x)$. The latter are defined as local functions with no associated equations of motion, contrarily to $\Psi^i(x)$. The action reads as
\begin{equation}
S[\Psi|\sigma] = \int_{\mathscr M} \D^n x\, L[\Psi|\sigma] .
\end{equation}
To derive flux-balance laws for the Noether charges from first principles, it will be useful to allow variations of the sources on the phase space. 
The variation of the action reads as
\begin{equation}
    \delta S = \int_{\mathscr M} \D^n x\, \frac{\delta S}{\delta \Psi^i}\delta\Psi^i + \int_{\mathscr M} \D^n x\, \frac{\delta S}{\delta \sigma^m} \delta \sigma^m, \label{delta S}
\end{equation}
discarding the boundary terms if the fields and sources are sufficiently decaying while approaching $\partial\mathscr M$. For a fixed set of sources $\sigma^m(x)$, the action is stationary for arbitrary variations $\delta\Psi^i$ if and only if the equations of motion $\frac{\delta S}{\delta \Psi^i} = 0$ are obeyed by the dynamical fields $\Psi^i(x)$.

If the theory admits some Noether symmetries $\delta_K \Psi^i = K^i[\Psi]$ in absence of external sources, turning on the sources will usually break these symmetries. However, the Noetherian symmetries $\delta_K \Psi^i = K^i[\Psi]$ of the theory without source can be promoted to generalized symmetries of the sourced theory in the sense of \cite{Troessaert:2015nia} (notice that the present notion of generalized symmetries shall not be confused with the concept of higher-form symmetries discussed in quantum field theory and that goes under the same name, see \textit{e.g.} \cite{Gaiotto:2014kfa}). The detailed analysis of the implications of the existence of such symmetries requires some care in the definition of a generalized symmetry and is beyond the scope of this article. This analysis will be presented elsewhere  \cite{Barnich:2022xx}. For the purpose of the present work we will only rely on the following features, which we will take as a naive definition for a generalized symmetry: the joint action on both the fields and sources
\begin{equation}
\delta_K \Psi^i = K^i[\Psi|\sigma] , \quad \delta_K \sigma^m = K^m[\sigma] \label{transfo sources}
\end{equation}  
is then a symmetry in the sense that it is required to be a symmetry of the sourced equations of motion
\begin{equation}
  \frac{\delta S}{\delta \Psi^i}=0 \quad \Longrightarrow \quad \delta_K \left(\frac{\delta S}{\delta \Psi^i}\right) = 0
\end{equation}
but is not required to preserve the action
\begin{equation}
    \delta_K L = \partial_a B^a_K + V_K[\sigma],\quad V_K[\sigma=0]=0.
    \label{symmetries L}
\end{equation}
As the second equation of \eqref{transfo sources} expresses that the external sources transform among themselves, this translates into the fact that the generalized symmetries break the invariance of the action in \eqref{symmetries L} only by terms depending on the sources which vanish when $\sigma= 0$. Consequently, in the absence of source ($\sigma=0$), the symmetries are fully restored as rightful variational or Noether symmetries. Hence one can say that generalized symmetries are symmetries of the sourced equations of motion extending the Noetherian symmetries of the source-free action. In this framework, writing the usual Noether current as $\bm j_K = j^a_K\, (\D^{n-1}x)_a$, the following relation can be obtained
\begin{equation}
\partial_a j^a_K = K^i \frac{\delta S}{\delta \Psi^i} + F_K  , \label{modified noether off shell}
\end{equation} 
where $\bm F_K = F_K \, (\D^nx)$ is a flux term whose explicit form is given by \cite{Barnich:2022xx}
\begin{equation}
    F_K = K^m \frac{\delta L}{\delta \sigma^m } -V_K . \label{flux expression explicit}
\end{equation} On-shell, \eqref{modified noether off shell} leads to the flux-balance law:
\begin{equation}
\boxed{
\D\bm j_K = \bm F_K[\Psi|\sigma] .
} \label{modified noether}
\end{equation} 
The latter generalizes the first Noether theorem in presence of external sources: the conservation of the Noether current $\bm j_K$ associated with a symmetry of characteristics $K$ is broken by the flux term $\bm F_K$. If one turns off the sources, we have $F_K[\Psi|\sigma=0] = 0$ and one recovers the standard conservation law $\D\bm j_K = 0$.

\subsection{Sourced Ward identities}
\label{sec:Sourced Ward identities}
Let us now derive Ward identities associated with the generalized symmetries \eqref{transfo sources}. The derivation follows the usual steps by properly taking into account the presence of external sources (see \textit{e.g.} \cite{DiFrancesco:1997nk} for the standard derivation of Ward identities from the path integral). For a fixed source $\sigma$, the partition function reads as 

\begin{equation}
    \mathcal Z_\sigma[J_i] = \int \mathcal D[\Psi ]_\sigma \, \exp \frac{i}{\hbar} \big( S[\Psi |\sigma ] + J_A\Psi^A \big) , \label{Z sigma}
    \end{equation} where we allow the path integral measure $\mathcal{D}[\Psi ]_\sigma$ to depend upon the sources $\sigma^m$ and we use the convenient notation $J_A\Psi^A \equiv \int_{\mathscr M} \D^n x \, J_i(x)\Psi^i(x)$.  Here $J_i(x)$ are classical sources introduced in the definition of the partition function. They are not meant to be quantized and will be sent to zero when evaluating the correlation function. We stress that they should be distinguished from the sources $\sigma^m$ that have a physical meaning and break the conservation of Noether currents: the sources $\sigma^m$ will be integrated in the final path integral and will generate a flux.

     Then, we suggest to consider that the partition function of the full sourced quantum theory should be computed by integrating \eqref{Z sigma} over the sources as 
\begin{equation}
    \mathcal Z[J_i, J_m] = \int \mathcal D[\sigma]\, \mathcal Z_\sigma[J_i]\, \exp \frac{i}{\hbar} J_M\sigma^M  . \label{full Z}
\end{equation} 
We used again the shorthand $J_M\sigma^M \equiv \int_{\mathscr M} \D^n x \, J_m(x)\sigma^m(x)$ and introduced a second bunch of classical sources $J_m$ as part of the definition of the partition function. The sources $\sigma^m$ are therefore promoted as \textit{source operators} and will play an important role in the Carrollian holographic correspondence discussed in Section \ref{sec:Holographic correspondence}. A very similar procedure was adopted in \cite{Compere:2008us} in the AdS/CFT context where the path integral was performed over the boundary sources. We will further comment on this in the discussion closing the paper (Section \ref{sec:Discussion}). Notice that, even though the source operators $\sigma^m$ and field operators $\Psi^i$ seem on the same footing in \eqref{full Z}, they are distinguished by their role played in the generalized symmetries introduced in Section \ref{sec:Generalized Noether symmetries}. In the absence of source, the symmetries of the partition functions are restored. We now investigate the implications of generalized symmetries on the correlation functions when the sources are turned on. Inserting the off-shell relation \eqref{modified noether off shell} in the path integral, we get
\begin{align}
&\frac{\partial}{\partial x^a} \int \mathcal D[\sigma] \int \mathcal D[\Psi]_\sigma\, j^a_K(x) \, \exp \frac{i}{\hbar}\big(S[\Psi|\sigma]+J_A\Psi^A+J_M\sigma^M\big) \nonumber \\
&= \int \mathcal D[\sigma] \int\mathcal D[\Psi]_\sigma\, \left( K^i[\Psi(x)|\sigma(x)]\frac{\delta S}{\delta \Psi^i(x)} + F_K(x) \right) \exp \frac{i}{\hbar}\big(S[\Psi|\sigma]+J_A\Psi^A+J_M\sigma^M\big) .
\label{interm 7}
\end{align}
The first term in the right-hand side can be reworked by noticing that
\begin{equation}
\begin{split}
&\frac{\hbar }{i}\frac{\delta}{\delta\Psi^i(x)} \, \exp \frac{i}{\hbar}\big(S[\Psi|\sigma]+J_A\Psi^A+J_M\sigma^M\big) \\
&= \frac{\delta}{\delta\Psi^i(x)}\big(S[\Psi|\sigma]+J_A\Psi^A\big)\exp \frac{i}{\hbar}\big(S[\Psi|\sigma]+J_A\Psi^A+J_M\sigma^M\big) \\
&= \left( \frac{\delta S}{\delta \Psi^i(x)} + J_i(x)\right) \exp \frac{i}{\hbar}\big(S[\Psi|\sigma]+J_A\Psi^A+J_M\sigma^M\big),
\end{split}
\end{equation}
and assuming that $\delta(0) = 0$ and the path integral is invariant under field translation, \textit{i.e.} $\mathcal D[\Psi+\delta\Psi]_\sigma = \mathcal D[\Psi]_\sigma$ for any variation $\delta\Psi^i$: 
\begin{equation}
\begin{split}
&\int \mathcal D[\sigma]\int\mathcal D[\Psi]_\sigma \left( K^i[\Psi(x)|\sigma(x)]\frac{\delta S}{\delta \Psi^i(x)} \right) \exp \frac{i}{\hbar}\big(S[\Psi|\sigma]+J_A\Psi^A+J_M\sigma^M\big) \\
& = - \int \mathcal D[\sigma]\int\mathcal D[\Psi]_\sigma \, J_i(x) K^i[\Psi(x)|\sigma(x)]\, \exp \frac{i}{\hbar}\big(S[\Psi|\sigma]+J_A\Psi^A+J_M\sigma^M\big).
\end{split}
\end{equation}
As a conclusion, \eqref{interm 7} is simply
\begin{equation}
\begin{split}
&\frac{\partial}{\partial x^a} \int \mathcal D[\sigma]\int\mathcal D[\Psi]_\sigma \, j^a_K(x) \, \exp \frac{i}{\hbar}\big(S[\Psi|\sigma]+J_A\Psi^A+J_M\sigma^M\big) \\
&= \int \mathcal D[\sigma]\int\mathcal D[\Psi]_\sigma\, \Big[ - J_i(x)K^i[\Psi(x)|\sigma(x)] + F_K(x) \Big] \exp \frac{i}{\hbar}\big(S[\Psi|\sigma]+J_A\Psi^A+J_M\sigma^M\big) .
\end{split} \label{interm10}
\end{equation}
Acting on the left-hand side with $N$ successive derivatives with respect to $J_i(x)$ and setting $J_i(x) = 0 = J_m(x)$ afterwards gives 
\begin{equation}
    \boxed{
    \begin{aligned}
        \frac{\partial}{\partial x^a}\big\langle j^a_K(x)\,X_N^\Psi \big\rangle = \frac{\hbar}{i}\sum_{k=1}^N \delta^{(n)}(x-x_k)\,\delta_{K^{i_k}} \big\langle X_N^\Psi \big\rangle + \big\langle F_K(x)\,X_N^\Psi \big\rangle,
    \end{aligned}
    } \label{local infinitesimal Ward identity}
\end{equation}
where we have introduced
\begin{equation}
    X_N^\Psi \equiv \Psi^{i_1}(x_1)\dots \Psi^{i_N}(x_N), \quad \delta_{K^{i_k}}X_N^\Psi = \Psi^{i_1}(x_1)\dots K^{i_k}[ \Psi(x_k)] \dots  \Psi^{i_N}(x_N) \label{XNPsi}
\end{equation}
to denote a collection of $N$ quantum insertions and their transformation under the symmetry of characteristics $K$ and recalled that, by definition,
\begin{equation}
\big\langle X_N^\Psi \big\rangle \equiv \left(\frac{\hbar}{i}\right)^N \frac{\delta}{\delta J_{i_1}(x_1)}\cdots \frac{\delta}{\delta J_{i_N}(x_N)}\mathcal Z[J]\Big|_{J=0}.
\end{equation}
Now acting on the left-hand side of \eqref{interm10} with $N$ successive derivatives with respect to the other $J_m(x)$ yields
\begin{equation}
    \boxed{
    \begin{aligned}
        \frac{\partial}{\partial x^a}\big\langle j^a_K(x)\,X_N^\sigma \big\rangle = \big\langle F_K(x)\,X_N^\sigma \big\rangle,
    \end{aligned}
    } \label{local infinitesimal Ward identity SOURCES}
\end{equation}
where
\begin{equation}
    X_N^\sigma \equiv \sigma^{m_1}(x_1)\dots \sigma^{m_N}(x_N) \label{XNsource}
\end{equation}
denotes a collection of $N$ insertions of quantum sources. The equations \eqref{local infinitesimal Ward identity} and \eqref{local infinitesimal Ward identity SOURCES} are the local version of the infinitesimal Ward identities in presence of external sources. With no field insertion in the correlators, one finds 
\begin{equation}
    \partial_a \big\langle  j^a_K(x) \big\rangle
=  \big\langle F_K(x) \big\rangle ,
\end{equation}
which reproduces the classical flux-balance equation \eqref{modified noether}. Integrating \eqref{local infinitesimal Ward identity} on the whole manifold $\mathscr{M}$ with boundary $\partial \mathscr{M}$ gives
\begin{equation}
\boxed{
    \sum_{k=1}^N \delta_{K^{i_k}} \big\langle X_N^\Psi\big\rangle = \frac{i}{\hbar} \Big\langle \left(\int_{\mathscr M} \bm F_K - \int_{\partial \mathscr M}\bm j_K \right)\,X_N^\Psi\Big\rangle .
} \label{infinitesimal Ward identity integrated}
\end{equation}
The standard textbook result is recovered by setting the sources to zero and assuming that the Noether currents vanish at the boundary, so that the right-hand side of \eqref{infinitesimal Ward identity integrated} vanishes. Finally integrating \eqref{local infinitesimal Ward identity SOURCES} on $\mathscr M$ leads to
\begin{equation}
    \boxed{
    \Big\langle \left(\int_{\mathscr M} \bm F_K - \int_{\partial \mathscr M}\bm j_K \right)\,X_N^\sigma\Big\rangle = 0.
    } \label{infinitesimal Ward identity integrated SOURCES}
\end{equation}

\paragraph{Remark} At this stage, and since the partition function \eqref{full Z} can be rewritten as
\begin{equation}
    \mathcal Z[J_i, J_m] = \int \mathcal D[\sigma]\, \int \mathcal D[\Psi ]_\sigma \, \exp \frac{i}{\hbar} \big( S[\Psi |\sigma ] + J_A\Psi^A + J_M\sigma^M\big),
\end{equation} 
one might wonder what is responsible for the asymmetry between the Ward identities for the fields \eqref{local infinitesimal Ward identity} and the sources \eqref{local infinitesimal Ward identity SOURCES}. In the following lines, we wish to highlight that, besides a possible dependence of the measure $\mathcal D[\Psi ]_\sigma$  in the sources, this asymmetry can ultimately be traced back to the fact that the generalized symmetries treat fields and sources differently. To convince oneself of this point, it is instructive to consider the sub-case where the measures $\mathcal D[\sigma]$ and $\mathcal D[\Psi]$ are assumed to commute and use this property to rewrite the Ward identities in a form that puts sources and fields on an equal footing. To achieve this, we only need to reintroduce the explicit form of the flux \eqref{flux expression explicit}, which has not yet been used in the derivation of the local version of the Ward identities \eqref{local infinitesimal Ward identity} and \eqref{local infinitesimal Ward identity SOURCES}. Injecting \eqref{flux expression explicit} into \eqref{interm10}, we have
\begin{equation}
    \begin{split}
        &\frac{\partial}{\partial x^a} \int \mathcal D[\sigma] \int \mathcal D[\Psi]_\sigma\, j^a_K(x) \, \exp \frac{i}{\hbar}\big(S[\Psi|\sigma]+J_A\Psi^A+J_M\sigma^M\big)  \\
        &= \int \mathcal D[\sigma] \int\mathcal D[\Psi]_\sigma\, \left( -K^i[\Psi(x)|\sigma(x)] J_i(x)  +K^m[\sigma(x)] \frac{\delta S}{\delta \sigma^m(x) } -V_K[\sigma (x)] \right) \times \cdots \\
        &\qquad \times \exp \frac{i}{\hbar}\big(S[\Psi|\sigma]+J_A\Psi^A+J_M\sigma^M\big).
    \end{split}
\end{equation}
Assuming that one can commute the measures $\mathcal D[\sigma]$ and $\mathcal D[\Psi]$ and that the path integral for the sources is invariant under translation as well, \textit{i.e.} $\mathcal{D}[\sigma + \delta \sigma] = \mathcal{D}[\sigma]$, the second term in the right-hand side can be reworked as
\begin{equation}
    \begin{split}
        &\frac{\partial}{\partial x^a} \int \mathcal D[\sigma] \int \mathcal D[\Psi]_\sigma\, j^a_K(x) \, \exp \frac{i}{\hbar}\big(S[\Psi|\sigma]+J_A\Psi^A+J_M\sigma^M\big)  \\
        &= \int \mathcal D[\sigma] \int\mathcal D[\Psi]_\sigma\, \left( - K^i[\Psi(x)|\sigma(x)]J_i(x)  -K^m[\sigma(x)] J_m (x) -V_K[\sigma (x)] \right) \times\cdots \\
        &\qquad \times \exp \frac{i}{\hbar}\big(S[\Psi|\sigma]+J_A\Psi^A+J_M\sigma^M\big).
    \end{split}
\end{equation}
Deriving $N$ times with respect to $J_i(x)$ and then setting $J_m = 0=J_i$ gives
\begin{equation}
    \frac{\partial}{\partial x^a}\big\langle j^a_K(x)\,X_N^\Psi \big\rangle = \frac{\hbar}{i}\sum_{k=1}^N \delta^{(n)}(x-x_k)\,\delta_{K^{i_k}} \big\langle X_N^\Psi \big\rangle -\big\langle V_K(x)\,X_N^\Psi \big\rangle. \label{WI champs}
\end{equation} Similarly, deriving $N$ times with respect to $J_m(x)$ and then setting $J_m = 0=J_i$ gives 
\begin{equation}
    \frac{\partial}{\partial x^a}\big\langle j^a_K(x)\,X_N^\sigma \big\rangle = \frac{\hbar}{i}\sum_{k=1}^N \delta^{(n)}(x-x_k)\,\delta_{K^{m_k}} \big\langle X_N^\sigma \big\rangle - \big\langle V_K(x)\,X^\sigma_N \big\rangle . \label{WI sources}
\end{equation} 
Under the above assumptions, these expressions are equivalent to \eqref{local infinitesimal Ward identity} and \eqref{local infinitesimal Ward identity SOURCES} respectively. They highlight that the asymmetry between fields and sources in the treatment of generalized symmetries is always present as a result of the appearance of $V_K[\sigma(x)]$ in the right-hand sides. In particular, these Ward identities are distinctly different from the usual Ward identities of a Noetherian symmetry.

\subsection{Sourced \texorpdfstring{$U(1)$}{U(1) } Ward identities}
\label{sec:Sourced $U(1)$ Ward identities}

As a warm-up, we apply the above framework to the simple case of a field theory exhibiting a $U(1)$ symmetry. We assume that the fields $\Psi_Q$ inserted in the correlators transform as
\begin{equation}
    \delta_\lambda \Psi_Q = -i \lambda Q \Psi_Q
    \label{U(1) primary}
\end{equation} under the $U(1)$ symmetry parametrized by some arbitrary function $\lambda (x)$. The infinitesimal Ward identity \eqref{local infinitesimal Ward identity} in presence of external sources then simply reads as 
\begin{equation}
   \begin{split}
        \frac{\partial}{\partial x^a} \Big\langle  j^a_\lambda(x) \, \Psi^{i_1}_{Q_{i_1}}(x_1)\dots\Psi^{i_N}_{Q_{i_N}}(x_N) \Big\rangle  &- \hbar \sum_{k=1}^N \lambda (x_k ) \,Q_i\, \delta^{(n)}(x-x_k) \, \big\langle \Psi^{i_1}_{Q_{i_1}}(x_1)\dots  \Psi^{i_N}_{Q_{i_N}}(x_N) \big\rangle  \\
    &=  \Big\langle F_\lambda(x)\, \Psi^{i_1}_{Q_{i_1}}(x_1)\dots\Psi^{i_N}_{Q_{i_N}}(x_N) \Big\rangle .
   \end{split} \label{infinitesimal ward QED}
\end{equation} While this example is quite trivial, it will be relevant when discussing electrodynamics from the Carrollian holographic perspective.

\subsection{Sourced conformal Carrollian Ward identities}
\label{sec:Sourced conformal Carrollian Ward identities}
We now make the same exercise for the case of $3d$ conformal Carrollian field theory. After reviewing conformal Carrollian symmetries and introducing associated primary fields, we write the sourced Ward identities involving the Carrollian momenta. Notice importantly that we make here a slight abuse of terminology, as the very meaning of what exactly \textit{is} a quantum Carrollian field theory is still largely unknown. For us, this term will refer to a $3d$ theory which enjoys conformal Carrollian symmetries.

\subsubsection{Conformal Carrollian symmetries}
\label{sec:Conformal Carrollian symmetries}

A Carrollian structure \cite{1977asst.conf....1G,Henneaux:1979vn, Ashtekar:2014zsa,Duval:2014uva,Duval:2014lpa,Bekaert:2015xua,Morand:2018tke,Ciambelli:2019lap,Figueroa-OFarrill:2019sex, Herfray:2020rvq,Herfray:2021xyp, Herfray:2021qmp,Henneaux:2021yzg,Bekaert:2022ipg} on a $3d$ manifold $\mathscr{I}$ is a couple $(q_{ab}, n^a)$, where $q_{ab}$ is a degenerate metric with signature $(0,+,+)$ and $n^a$ is a vector field in the kernel of the metric, \textit{i.e.} $q_{ab} n^a = 0$. For convenience, we choose coordinates $(u,z,\bar{z})$ on $\mathscr{I}$ such that $q_{ab} dx^a dx^b = 0\, du^2 + 2 d z d\bar{z}$ and $n^a \partial_a = \partial_u$. We still denote $x^A = (z,\bar z)$, $A = 1,2$.  The conformal Carrollian symmetries are generated by vector fields $\bar\xi = \bar \xi^a \partial_a$ satisfying
\begin{equation}
    \mathscr{L}_{\bar{\xi}} q_{ab} = 2 \alpha q_{ab}, \qquad \mathscr{L}_{\bar\xi} n^a = - \alpha n^a,
    \label{conformal Carroll symmetries}
\end{equation} where $\alpha$ is a function on $\mathscr I$. The solution $\bar{\xi}$ of \eqref{conformal Carroll symmetries} is given by  
\begin{equation}
    \bar{\xi} = \left[\mathcal{T} + \frac{u}{2} ( \partial_z \mathcal Y^z + \partial_{\bar z} \mathcal Y^{\bar z}) \right]\partial_u + \mathcal{Y}^z \partial_z + \mathcal{Y}^{\bar z} \partial_{\bar{z}} ,
   \label{conformal Carroll symmetries here}
\end{equation}
with $\mathcal T = \mathcal T(z,\bar z)$, $\mathcal Y^z = \mathcal Y^z(z)$ and $\mathcal Y^{\bar z} = \mathcal Y^{\bar z}(\bar z)$. These vector fields are called conformal Carrollian Killing vectors and $\alpha = \frac{1}{2}( \partial_z \mathcal Y^z + \partial_{\bar z} \mathcal Y^{\bar z})$. They precisely coincide with the restriction to $\mathscr{I}$ of the asymptotic Killing vector fields \eqref{AKV Bondi}. Furthermore, their standard Lie bracket on $\mathscr{I}$, $[\bar{\xi}(\mathcal{T}_1,\mathcal Y^z_1,\mathcal Y^{\bar z}_1),\bar{\xi}(\mathcal{T}_2,\mathcal Y^z_2,\mathcal Y^{\bar z}_2) ] = \bar{\xi}(\mathcal{T}_{12},\mathcal Y^z_{12},\mathcal Y^{\bar z}_{12})$, reproduces \eqref{commutation relations 2}. This shows that the conformal Carroll algebra in $3d$ is isomorphic to the BMS algebra in $4d$ \cite{Duval:2014uva,Duval:2014lpa}.

We will call the global conformal Carrollian algebra the subalgebra generated by 
\begin{enumerate}
	\item Carrollian translations, $P_a=\partial_a$ or $P_0 \equiv \partial_u$, $P_1 = \partial_z$, $P_2 = \partial_{\bar z}$.
	\item Carrollian rotation, $J \equiv x_1\partial_2 - x_2\partial_1 = -z\partial_z+ \bar z \partial_{\bar z}$.
	\item Carrollian boosts, $B_A \equiv x_A\partial_u$ or $B_1 = \bar z\partial_u$, $B_2 = z\partial_u$.
	\item Carrollian dilatation, $D \equiv x^a\partial_a$ or $D = u\partial_u + z\partial_z+\bar z\partial_{\bar z}$.
	\item Carrollian special conformal transformations, $K_0 \equiv -x^Ax_A \partial_u = -2z\bar z\partial_u$, and $K_A \equiv 2 x_A D - x^Bx_B \partial_A$, or $K_1 = 2 u \bar z\partial_u + 2 \bar z^2\partial_{\bar z}$ and $K_2 = 2 u z \partial_u + 2 z^2\partial_z$.
\end{enumerate} 
Here, $x_A = q_{AB} x^B$. In this basis, the commutation relations \eqref{commutation relations 2} for the global conformal Carrollian subalgebra \cite{Bagchi:2016bcd,Chen:2021xkw} can be split as follows: first, one easily checks that $B_1$, $B_2$ $P_0$ and $K_0$ form an abelian subalgebra
\begin{equation}
[B_A,B_B] = 0,\quad [B_A,P_0] = [B_A,K_0] = 0,\quad [K_0,P_0]=0. \label{global conformal Carrollian
subalgebra ABELIAN}
\end{equation}
The six remaining generators form a Lorentz subalgebra, however in an unusual basis:
\begin{equation}
\begin{split}
&[D,P_{0,A}] = -P_{0,A} , \quad [D,K_{0,A}] = K_{0,A} , \quad [D,B_A] = [D,J] = 0 , \label{global conformal Carrollian
subalgebra} \\
&[J,G_A] = q_{2A}G_1- q_{1A}G_2 , \ G \in \{P,B,K\} , \quad [J,P_0] = [J,K_0] = 0 , \\
&[B_A,P_B] = - q_{AB}P_0 , \quad [B_A,K_B] = - q_{AB}K_0 ,  \\
&[K_0,P_A] = 2B_A , \quad [K_A,P_0] = -2B_A ,\quad [K_A,P_B] = -2 q_{AB}D - 2J\delta_{1,[A}\delta_{B],2} .     
\end{split}
\end{equation}
We present in Appendix \ref{sec:isomorphism Poincare Carroll} the concrete relations defining the isomorphism between the global conformal Carrollian algebra in $3d$ and the Poincaré algebra in $4d$.

In conformal field theory, primary fields are required to transform consistently under the action of the infinite-dimensional Witt algebra $\mathfrak{Witt}$. Quasi-primaries on the other hand only need to behave well under the action of the finite-dimensional subalgebra of M\"obius transformations $\mathfrak{sl}(2,\mathbb{C})$ (the ``global'' conformal algebra). The BMS algebra $\mathfrak{bms}_4$ should play in the Carrollian context a role similar to the Witt algebra of conformal field theory with the Poincar\'e algebra $\mathfrak{iso}(3,1)$ (the ``global'' conformal Carrollian algebra in $3d$) playing the role of M\"obius transformations. Crucially, the embeddings of the Poincar\'e (resp. M\"obius) algebra inside the BMS  (resp. Witt) algebra are not unique and correspond to an extra piece of geometry. In the asymptotically flat case, these are the gravity vacua discussed in \cite{Ashtekar:2014zsa,Compere:2016jwb}. These embeddings can be locally realized by a choice of Poincar\'e operators \cite{Herfray:2020rvq,Herfray:2021qmp} (generalizing M\"obius operators \cite{calderbank_mobius_2006} of conformal geometry) with vanishing ``curvature'' (corresponding to the curvature of a Cartan connection). From these considerations, we will say that a field is a conformal Carrollian primary (also referred to as Carrollian tensor \cite{Ciambelli:2018xat,Ciambelli:2018wre,Ciambelli:2019lap,Freidel:2021qpz}) if it transforms infinitesimally as
\begin{equation}
    \delta_{\bar{\xi}} \Phi_{(k,\bar{k})} = \left[\left(\mathcal{T} + \frac{u}{2}(\partial_z\mathcal{Y}^z + \partial_{\bar z} \mathcal{Y}^{\bar z})\right)\partial_u + \mathcal Y^z \partial_z + \mathcal Y^z \partial_{\bar z} +   k\, \partial_z \mathcal Y^z + \bar{  k}\, \partial_{\bar z} \mathcal Y^{\bar z}\right] \Phi_{(  k,\bar{  k})}\label{Carrollian tensor def}
\end{equation}
under full conformal Carroll symmetries \eqref{conformal Carroll symmetries}. Here the Carrollian weights $(k,\bar{k})$ are some integers or half-integers. Quasi-conformal Carrollian primary fields are only required to transform properly as \eqref{Carrollian tensor def} under the global subalgebra displayed above. Importantly, recalling that $\partial_u \bar\xi^u = \frac{1}{2}(\partial_z \mathcal Y^z+\partial_{\bar z}\mathcal Y^{\bar z})$ and $[\delta_{\bar\xi},d] = 0$, it can be deduced from \eqref{Carrollian tensor def} that $\partial_u \Phi_{(k,\bar k)}(u,z,\bar z)$ is also transforming as a conformal Carrollian primary with weights $(k+\frac{1}{2},\bar k+\frac{1}{2})$.

\subsubsection{Classical flux-balance laws}

Let us now specify the flux-balance law \eqref{modified noether} for a $3d$ conformal Carrollian field theory. We assume that Noether currents associated with the global conformal Carrollian symmetries \eqref{conformal Carroll symmetries here} take the following Brown-York expression \cite{Brown:1992br, Chandrasekaran:2021hxc}
\begin{equation}
    j^a_{\bar \xi} = {\mathcal{C}^a}_{b} \bar{\xi}^b 
    \label{Carrollian Noether currents}
\end{equation} where ${\mathcal{C}^a}_b$ is the Carrollian stress tensor. Its components are called the Carrollian momenta \cite{Hartong:2015usd,deBoer:2017ing,Ciambelli:2018xat, Ciambelli:2018wre, Ciambelli:2018ojf, Donnay:2019jiz,deBoer:2021jej,Chandrasekaran:2021hxc,Freidel:2021qpz,Freidel:2021dfs} and denoted as
\begin{align}
    {\mathcal{C}^a}_{b} = \left[
        \begin{array}{cc}
            \mathcal{M} &\mathcal{N}_B \\
            \mathcal{B}^A &{\mathcal{A}^A}_{B}
        \end{array} 
    \right]. \label{Carrollian momenta def}
\end{align}
If the conformal Carrollian field theory under consideration is sourced, the currents \eqref{Carrollian Noether currents} are no longer meant to be conserved but still obey some flux-balance laws as in \eqref{modified noether} or in coordinates
\begin{equation}
    \partial_a j^a_{\bar \xi} = F_a[\sigma]\bar\xi^a \label{carroll flux balance}
\end{equation}
where $\sigma$ denotes again the external sources coupled to the theory. Here we assumed that the flux can be written as $F_{\bar{\xi}} = F_a \bar\xi^a$, which is sufficient for the holographic purposes discussed in this paper. The flux-balance law \eqref{carroll flux balance} is obeyed for the generators of the global Carroll subalgebra (isomorphic to Poincaré algebra) if and only if ${\mathcal{C}^a}_b$ satisfies the following constraints: 
\begin{equation}
    \renewcommand{\arraystretch}{1.25}
    \begin{tabular}{c|c|c}
        & \text{Generator} & \text{Constraint} \\ 
        \hline
        \text{Carrollian translations} & $\partial_a$ & $\partial_a {\mathcal C^a}_b = F_b$ \\
        \hline
        \text{Carrollian rotation} & $-z\partial_z+\bar z \partial_{\bar z}$ & ${\mathcal C^z}_z - {\mathcal C^{\bar z}}_{\bar{z}} = 0$ \\
        \hline
        \text{Carrollian boosts} & $x^A\partial_u$ & ${\mathcal C^A}_u = 0$ \\
        \hline
        \text{Carrollian dilatation} & $x^a\partial_a$ & ${\mathcal C^a}_a = 0$
    \end{tabular}
\label{classical constraints on the C}
\end{equation}
In Appendix \ref{sec:Constrains on the Carrollian stress tensor}, we show that the Carrollian special conformal transformations $K_0,K_A$ do not impose further constraints. Furthermore, the above global conformal Carrollian symmetries are enough to completely constrain ${\mathcal{C}^a}_b$, \textit{i.e.} \eqref{carroll flux balance} is automatically satisfied by the supertranslation (and superrotation) currents provided \eqref{classical constraints on the C} holds. Notice also that the term ${\mathcal C^a}_b\partial_a\bar \xi^b$ does not contribute to the left-hand side of \eqref{carroll flux balance} as a consequence of \eqref{conformal Carroll symmetries here} and \eqref{classical constraints on the C}: this is in line with the hypothesis of flux being linear in the symmetry parameters. 

In terms of the Carrollian momenta \eqref{Carrollian momenta def}, the constraints \eqref{classical constraints on the C} read as
\begin{equation}
\boxed{
\begin{array}{rl}
    \partial_u \mathcal{\mathcal{M}} = F_u , &\quad \mathcal{B}^A =0 , \\
    \partial_u \mathcal{N}_z - \frac{1}{2}\partial_z \mathcal{M} + \partial_{\bar z} {\mathcal{A}^{\bar{z}}}_z =  F_z , &\quad {\mathcal{A}^z}_z + \frac{1}{2} \mathcal{M} = 0 , \\
    \partial_u \mathcal{N}_{\bar{z}} -\frac{1}{2}\partial_{\bar z} \mathcal M + \partial_z {{\mathcal A}^z}_{\bar{z}}  =   F_{\bar{z}} , &\quad {\mathcal{A}^{\bar{z}}}_{\bar{z}} +  \frac{1}{2} \mathcal{M} = 0 .
\end{array}
}
\label{flux balance carrollian momenta}
\end{equation} 
In the left column, the constraints take the form of flux-balance equations for the Carrollian momenta $\mathcal{M}$ and $\mathcal{N}_A$ that are sourced by the fluxes $F_a$. In the right column, the constraints completely fix the Carrollian momenta $\mathcal{B}^A$, ${\mathcal{A}^z}_z$ and ${\mathcal{A}^{\bar{z}}}_{\bar{z}}$. Notice that ${\mathcal{A}^z}_{\bar{z}}$ and ${\mathcal{A}^{\bar{z}}}_{z}$ are not fixed by the symmetries. 

\subsubsection{Sourced Ward identities}

At the quantum level, the analog of the constraints \eqref{classical constraints on the C} will be provided by the sourced infinitesimal Ward identities \eqref{local infinitesimal Ward identity} for the specific case of a $3d$ conformal Carrollian field theory. Similarly to the $2d$ CFT case (see for instance \cite{DiFrancesco:1997nk}), the Ward identities of a $3d$ conformal Carrollian field theory can be rewritten very simply in terms of ${\mathcal{C}^a}_b$. We assume that the operators $\Psi^i(x)$ inserted in the correlators are (quasi-)conformal Carrollian primary fields transforming as \eqref{Carrollian tensor def}. The sourced Ward identities \eqref{local infinitesimal Ward identity} for the (global) conformal Carrollian symmetries imply
\begin{equation}
   \begin{split}
   &\partial_a \big\langle {\mathcal{C}^a}_b \, X_N^\Psi \big\rangle + \frac{\hbar}{i}\sum_{i}\delta^{(3)}(x-x_i)\frac{\partial}{\partial {x^b_i}}\big\langle X_N^\Psi \big\rangle = \big\langle F_b \, X_N^\Psi \big\rangle , \\
   &\big\langle({\mathcal{C}^z}_z - {\mathcal{C}^{\bar z}}_{\bar{z}} )X_N^\Psi\big\rangle + \frac{\hbar}{i}\sum_i \delta^{(3)}(x-x_i)\, (k_i - \bar k_i)\big\langle X_N^\Psi \big\rangle  = 0 ,\\ 
   &\big\langle {\mathcal{C}^A}_u\, X_N^\Psi\big\rangle = 0 , \\
   &\big\langle {\mathcal{C}^a}_a\, X_N^\Psi \big\rangle + \frac{\hbar}{i}\sum_i \delta^{(3)}(x-x_i)\, (k_i+\bar k_i)\big\langle X_N^\Psi \big\rangle = 0    
   \end{split}
   \label{constraints quantum}
\end{equation}
where $X_N^\Psi$ is defined in \eqref{XNPsi}. The derivation of \eqref{constraints quantum} is pretty similar to the classical case \eqref{classical constraints on the C} (see Appendix \ref{sec:Constrains on the Carrollian stress tensor}) and will not be repeated. Now, taking linear combinations of \eqref{constraints quantum}, we express the Ward identities in terms of the Carrollian momenta: 
\begin{equation}
\boxed{
\begin{aligned}
        &\partial_u \big\langle\mathcal M\, X_N^\Psi \big\rangle + \frac{\hbar}{i}\sum_i \delta^{(3)}(x-x_i)\partial _{u_i} \big\langle X_N^\Psi \big\rangle = \big\langle F_u\, X_N^\Psi \big\rangle ,  \\[1em]
        &\partial_u \big\langle\mathcal N_z\, X_N^\Psi\big\rangle - \frac{1}{2}\partial_z \big\langle\mathcal M\, X_N^\Psi\big\rangle + \partial_{\bar z} \big\langle{\mathcal A^{\bar z}}_z\, X_N^\Psi\big\rangle\\
        &\quad + \frac{\hbar}{i}\sum_i \left[ \delta^{(3)}(x-x_i) \partial_{z_i} \big\langle X_N^\Psi\big\rangle - \partial_z \left(\delta^{(3)}(x-x_i)\, k_i\, \big\langle X_N^\Psi\big\rangle\right)\right] = \big\langle F_z\, X_N^\Psi \big\rangle ,\\[.5em]
        &\partial_u \big\langle \mathcal N_{\bar z}\, X_N^\Psi \big\rangle - \frac{1}{2}\partial_{\bar z} \big\langle\mathcal M\, X_N^\Psi \big\rangle + \partial_z \big\langle {\mathcal A^z}_{\bar z}\, X_N^\Psi\big\rangle\\
        &\quad + \frac{\hbar}{i}\sum_i \left[ \delta^{(3)}(x-x_i) \partial_{\bar z_i} \big\langle X_N^\Psi \big\rangle - \partial_{\bar z} \left(\delta^{(3)}(x-x_i)\, \bar k_i\, \big\langle X_N^\Psi \big\rangle\right)\right] = \big\langle F_{\bar z}\, X_N^\Psi \big\rangle ,\\[1em]
        &\big\langle \mathcal{B}^A\,X_N^\Psi \big\rangle = 0  ,\\[1em]
        &\big\langle ({\mathcal A^z}_z+ \frac{1}{2}\mathcal M)X_N^\Psi \big\rangle + \frac{\hbar}{i}\sum_i \delta^{(3)}(x-x_i)\,k_i\,\big\langle X_N^\Psi \big\rangle = 0 ,\\[1em]
        &\big\langle ({\mathcal A^{\bar{z}}}_{\bar{z}}+ \frac{1}{2}\mathcal M)X_N^\Psi \big\rangle + \frac{\hbar}{i}\sum_i \delta^{(3)}(x-x_i)\,\bar k_i\,\big\langle X_N^\Psi \big\rangle = 0 . 
\end{aligned}
}
\label{Ward identities Carrollian momenta}
\end{equation}
With no field insertion in the correlators, the expectation values of the operators reproduce the classical relations \eqref{flux balance carrollian momenta}. 

\section{Holographic conformal Carrollian field theory}
\label{sec:Holographic conformal Carrollian field theory}

In this section, we discuss the main ingredients needed for a holographic description of asymptotically flat spacetime in terms of a dual sourced conformal Carrollian field theory. First, we argue that the dual theory lives on $\hat{\mathscr{I}} = \mathscr{I}^- \sqcup \mathscr{I}^+$ where the gluing between $\mathscr{I}^-$ and $\mathscr{I}^+$ is obtained by identifying antipodally $\mathscr{I}^+_-$ and $\mathscr{I}^-_+$. We will make this gluing precise by introducing the geometry of time-ordered conformal Carrollian manifolds. We then propose an identification between scattering elements in position space and Carrollian correlation functions by relating the bulk quantities introduced in Section \ref{sec:Asymptotic symmetry analysis} and the boundary objects discussed in Section \ref{sec:Sourced quantum field theory}. Eventually, we deduce the explicit form of low-point correlation functions of a Carrollian CFT. In particular, we find a new branch of solutions of the two-point function and argue that this is the appropriate one for holographic purposes.

\subsection{Time-ordered conformal Carrollian manifolds}
\label{sec:Time-ordered conformal Carrollian manifolds}

In this section, we introduce some geometrical description of the manifold on which the dual sourced Carrollian CFT is living. As reviewed in Section \ref{sec:Conformal Carrollian symmetries}, a Carrollian structure on a manifold $\mathscr{M}$ is made of a pair $(q_{ab}, n^a)$ where $q_{ab} n^b =0$ and $\mathscr L_n q_{ab} \propto q_{ab}$. A conformal Carrollian structure is then defined as an equivalence class $[q_{ab},n^a]$ for the equivalence relation $(q_{ab},n^a) \sim (\Omega^2q_{ab},\Omega^{-1}n^a)$ where $\Omega$ is a nowhere vanishing function on $\mathscr M$. Equipped with such a structure, $(\mathscr{M}, [q_{ab},n^a])$ is called a conformal Carrollian manifold.

To obtain the universal structure \cite{1977asst.conf....1G,Ashtekar:2014zsa} at, say, future null infinity, we would need to add two hypotheses to our definition of conformal Carrollian manifold. First, fix the topology of $\mathscr M$ as $\mathbb R\times \Sigma$ where $\mathbb R$ is spanned by the flow of $n^a$ and $\Sigma$ is the space of null generators (for our purpose, we choose $\Sigma$ to be the one-puncture complex plane, see \textit{e.g.} \cite{Barnich:2021dta,Donnay:2021wrk}). Second, require that the vector field $n^a$ is complete and nowhere vanishing. However, as we will argue later, the Carrollian CFT is not living at $\mathscr{I}^+$ or $\mathscr{I}^-$ separately, but should really be seen as living on the whole conformal boundary of asymptotically flat spacetimes. Therefore, we would like to give a geometrical notion of ``past'' and ``future'' null infinity after gluing $\mathscr I^-$ with $\mathscr I^+$. The idea is to single out the separating surface $\Sigma_0\simeq \Sigma$ as the unique locus where the Carrollian vector $n^a$ vanishes. This provides a definition of ``time-ordered'' conformal Carrollian manifold, which is a conformal Carrollian manifold $(\mathscr M,[q_{ab},n^a])$ satisfying all the hypotheses above except that $n^a$ now vanishes on a co-dimension one surface $\Sigma_0$. For a chosen connection $\nabla_a$ defined on $\mathscr M$ (nothing will in fact depend on this choice), we demand
\begin{equation}
    \nabla_a n^b \big|_{\Sigma_0} = 0 ,\quad \nabla_a \nabla_b n^c \big|_{\Sigma_0} \neq 0. \label{def sigma 0}
\end{equation}
Now, around $\Sigma_0$, we can always choose local coordinates $(s,x^A)$ such that
\begin{equation}
    s\big|_{\Sigma_0} = 0, \quad n^a \partial_a = \tilde f(s,x^A) \partial_s
\end{equation}
for some function $\tilde f$. Since $n^a$ vanishes on $\Sigma_0$, the constraints \eqref{def sigma 0} only involves partial derivatives in the coordinates and are satisfied if $\partial_s \tilde f \to 0$ and $\partial_s^2\tilde f\to 2 f(x^A)$ as $s\to 0$ for some other function $f$ on $\Sigma$. The latter condition is solved by $\tilde f(s,x^A) = s^2 f(x^A)+\mathcal O(s^3)$ and we have
\begin{equation}
    n^a\partial_a = \left(s^2 f+\mathcal O(s^3)\right)\partial_s .
\end{equation}
Notice that the orientation of $n^a$ does not change across $\Sigma_0$. This provides a notion of global ordering of the points in $\mathscr M$: points are ``in the past'' of $\mathscr M$ if the flow generated by $n^a$ takes them towards the separating surface $\Sigma_0$ and ``in the future'' if the flow takes them away from it. Points of $\Sigma_0$ are fixed points of the flow of $n^a$, they are neither in the past nor in the future of $\mathscr M$.

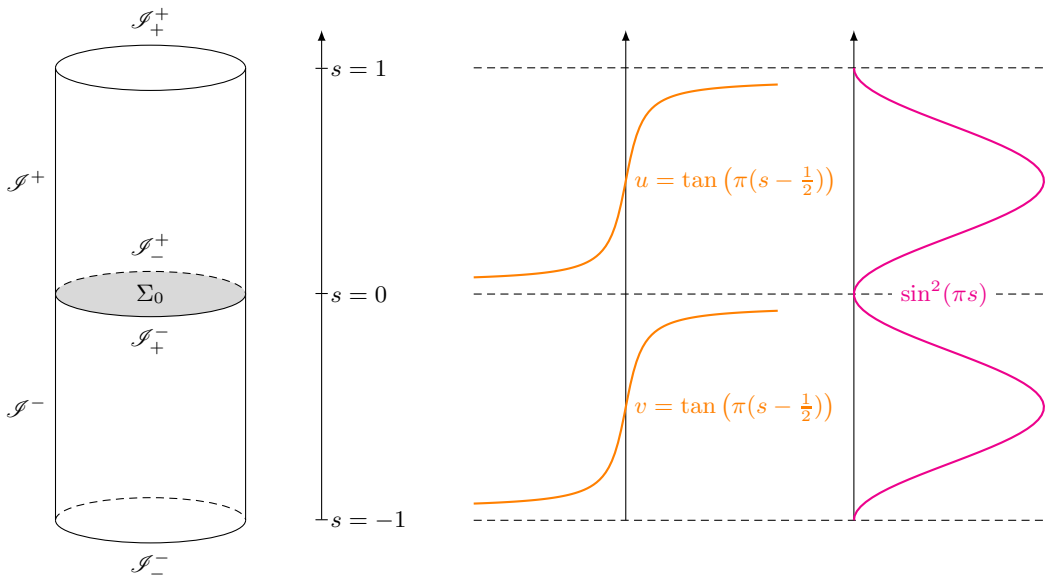
\begin{figure}[ht!]
    \centering
    \begin{tikzpicture}
    	\def\cx{2.5}; \def\cy{3}; \def\arq{0.3};
    	\coordinate (smL) at (0,0);
    	\coordinate (smR) at (\cx,0);
    	\coordinate (szL) at (0,\cy);
    	\coordinate (szR) at (\cx,\cy);
    	\coordinate (spL) at (0,2*\cy);
    	\coordinate (spR) at (\cx,2*\cy);
    	\draw[] (smL) -- (spL);
    	\draw[] (smR) -- (spR);
    	\draw[densely dashed] (smL) arc[x radius=\cx/2, y radius=\arq, start angle=180, end angle=0];
    	\draw[] (smL) arc[x radius=\cx/2, y radius=\arq, start angle=-180, end angle=0];
    	\fill[black!15] (szR) arc[x radius=\cx/2, y radius=\arq, start angle=0, end angle=360];
    	\draw[densely dashed] (szL) arc[x radius=\cx/2, y radius=\arq, start angle=180, end angle=0];
    	\draw[] (szL) arc[x radius=\cx/2, y radius=\arq, start angle=-180, end angle=0];
    	\draw[] (spL) arc[x radius=\cx/2, y radius=\arq, start angle=180, end angle=0];
    	\draw[] (spL) arc[x radius=\cx/2, y radius=\arq, start angle=-180, end angle=0];
    	\node[left] at ($(smL)!0.5!(szL)$) {$\mathscr I^-$};
    	\node[left] at ($(szL)!0.5!(spL)$) {$\mathscr I^+$};
    	\node[above] at ($(spL)!0.5!(spR)+(0,\arq)$){$\mathscr I^+_+$};
    	\node[below] at ($(smL)!0.5!(smR)-(0,\arq)$){$\mathscr I^-_-$};
    	\node[] at ($(szL)!0.5!(szR)$){$\Sigma_0$};
    	\node[above] at ($(szL)!0.5!(szR)+(0,\arq)$){$\mathscr I^+_-$};
    	\node[below] at ($(szL)!0.5!(szR)-(0,\arq)$){$\mathscr I^-_+$};
    	\def\decal{1}; 
    	\coordinate (arm) at ($(smR)+(\decal,0)$);
    	\coordinate (arz) at ($(szR)+(\decal,0)$);
    	\coordinate (arp) at ($(spR)+(\decal,0)$);
    	\draw[|-] (arm)node[right]{$s=-1$} -- (arz)node[right]{$s=0$};
    	\draw[|-|] (arz) -- (arp)node[right]{$s=1$};
    	\draw[-latex] (arp) -- ($(arp)+(0,0.5)$);
    	\coordinate (limm) at ($(arm)+(2*\decal,0)$);
    	\coordinate (limz) at ($(arz)+(2*\decal,0)$);
    	\coordinate (limp) at ($(arp)+(2*\decal,0)$);
    	\def\gasize{4}; \def\gbsize{3.5};
    	\draw[densely dashed] (limm) -- ($(limm)+(\gasize+\gbsize,0)$);
    	\draw[densely dashed] (limz) -- ($(limz)+(\gasize+\gbsize,0)$); 
    	\draw[densely dashed] (limp) -- ($(limp)+(\gasize+\gbsize,0)$);
    	\draw[-latex] ($(limm)+(\gasize/2,0)$) -- ($(limp)+(\gasize/2,0.5)$);
    	\draw[-latex] ($(limm)+(\gasize+\decal,0)$) -- ($(limp)+(\gasize+\decal,0.5)$);
    	\def\posx{\cx+3*\decal+\gasize/2};
    	\def\posyp{3*\cy/2};
    	\def\posym{\cy/2};
    	\draw[orange,thick,domain=-\gasize/2:\gasize/2,samples=100] plot (\x+\posx,{0.015*atan(6*(\x))+\posyp});
    	\draw[orange,thick,domain=-\gasize/2:\gasize/2,samples=100] plot (\x+\posx,{0.015*atan(6*(\x))+\posym});
    	\draw[orange] (\posx,\posyp)node[anchor=west]{\footnotesize $u = \tan\left(\pi(s-\tfrac{1}{2})\right)$};
    	\draw[orange] (\posx,\posym)node[anchor=west]{\footnotesize $v = \tan\left(\pi(s-\tfrac{1}{2})\right)$};
    	\def\posxx{\cx+4*\decal+\gasize};
    	\draw[magenta,thick,domain=-\cy:\cy,samples=100] plot ({2.5*(sin(180*\x/\cy))^2+\posxx},\x+\cy);
    	\node[magenta,fill=white,right] at ($(arz)+(3*\decal+\gasize+0.5,0)$) {\footnotesize $\sin^2(\pi s)$};
    \end{tikzpicture}
    \caption{Time-ordered conformal boundary $\hat{\mathscr I}$ of asymptotically flat spacetimes.}
    \label{fig:coord on whole scri}
\end{figure}

Going back to the conformal boundary of asymptotically flat spacetimes, we are led to consider
\begin{equation}
    \hat{\mathscr I} \equiv \mathscr I^- \sqcup \mathscr I^+
\end{equation}
as conformal Carrollian manifold $\mathscr M$, where $\sqcup$ represents a union by gluing of the two conformal boundaries. According to the above discussion, we will take $\hat{\mathscr I}$ to be a three-dimensional time-ordered conformal Carrollian manifold. The separating surface along which the gluing is performed is
\begin{equation}
    \Sigma_0 \simeq \mathscr I^+_- \simeq \mathscr I^-_+ ,
\end{equation}
\textit{i.e.} $\mathscr I^+$ is glued to $\mathscr I^-$ by identifying continuously the past surface $\mathscr I^+_-$ of $\mathscr I^+$ and the future surface $\mathscr I^-_+$ of $\mathscr I^-$ around spatial infinity. On $\mathscr I^+$, we can always choose local coordinates $(u,x^A)$ such that $u$ is the retarded time and $n^a  \partial_a =\partial_u$. Analogously on $\mathscr I^-$, there exists a local coordinate system $(v,x^A)$ such that $v$ is the advanced time and $n^a  \partial_a=\partial_v$. Let us recall that our coordinate choices are such that the angular coordinates $x^A = (z,\bar z)$ are antipodally identified between $\mathscr I^+$ and $\mathscr I^-$ (see appendix \ref{app:Bondi adv and ret}). Since $n$ vanishes on $\Sigma_0$ (defined as the transverse $\Sigma$ in the limit $u\to-\infty$ or $v\to+\infty$), both coordinate systems cannot be extended on the whole $\hat{\mathscr I}$. On both sides, we parametrize
\begin{equation}
    \left\{
    \begin{aligned}
        u &= \tan \left(\pi(s-\tfrac{1}{2})\right),\, s>0 \quad\text{on}\quad \mathscr I^+,\\
        v &= \tan \left(\pi(s-\tfrac{1}{2})\right),\, s<0\quad\text{on}\quad \mathscr I^-.
    \end{aligned}
    \right. \label{s for uv}
\end{equation}
The Figure \ref{fig:coord on whole scri} depicts the situation.

The time reparametrizations \eqref{s for uv} define a global ``time'' coordinate
\begin{equation}
    s : \hat{\mathscr I}\to \,\ ]\text{$-1$},+1[
\end{equation}
where the separating surface is located at $\Sigma_0 = \{s=0\}\subset\hat{\mathscr I}$. We have $s = -\pi^{-1}\text{arccot}\, u$ (and similarly for $v$), hence the Carrollian vector reads now as
\begin{equation}
    n^a \partial_a = \frac{1}{\pi}\sin^2(\pi s)\partial_s .
    \label{n on whole scri}
\end{equation}
In this coordinate system, it is manifest that $n \equiv 0$ if and only if $s=0$ (\textit{i.e.} on the separating surface) on $\hat{\mathscr I}$. On $\mathscr I^+$ and $\mathscr I^-$, conformal Carrollian automorphisms act infinitesimally as \eqref{conformal Carroll symmetries here} or
\begin{equation}
    \left\{
    \begin{aligned}
        \bar\xi^+(u,x^A) &= \left(\mathcal T^+ + \tfrac{u}{2}\partial_B \mathcal Y^B_+\right)\partial_u + \mathcal Y^A_+\partial_A,\quad\text{on}\quad \mathscr I^+ ,\\
        \bar\xi^-(v,x^A) &= \left(\mathcal T^- + \tfrac{v}{2}\partial_B \mathcal Y^B_-\right)\partial_v + \mathcal Y^A_-\partial_A,\quad\text{on}\quad \mathscr I^-  .\\
    \end{aligned}
    \right. 
\end{equation}
Using \eqref{s for uv} and \eqref{n on whole scri}, we can define the unique global smooth infinitesimal automorphism of the time-ordered geometry $\hat{\mathscr I}$ as
\begin{equation}
    \bar\xi(s,x^A) = \frac{1}{\pi}\left(\mathcal T-\frac{1}{2}\cot\left(\pi s\right)\partial_B \mathcal Y^B \right)\sin^2(\pi s)\partial_s + \mathcal Y^A\partial_A \label{xi on whole scri}
\end{equation}
where $\mathcal T(x^A)$, $\mathcal Y^A(x^B)$ are smooth functions on the 2-sphere such that
\begin{equation}\label{diagonal BMS}
    \mathcal T(x^A) = \mathcal T^+(x^A) = \mathcal T^-(x^A),\quad \mathcal Y^A_+(x^B) = \mathcal Y^A(x^B) = \mathcal Y^A_-(x^B) .
\end{equation}
In this picture, the antipodal matching for BMS generators advocated in \cite{Strominger:2013jfa,He:2014laa, Kapec:2014opa} and confirmed later in \cite{Troessaert:2017jcm,Compere:2017knf,Henneaux:2018cst, Henneaux:2018hdj, Prabhu:2019fsp,Prabhu:2021cgk,Mohamed:2021rfg,Capone:2022gme} by a phase space analysis at spacelike infinity, corresponds to the requirement that $\bar\xi$ given by \eqref{xi on whole scri} is smooth at $\Sigma_0$. A final remark is that the diagonal BMS group selected in this way preserves the location of the separating surface since
\begin{equation}
    \bar\xi\big|_{\Sigma_0} = \bar\xi(s=0,x^A) = \mathcal Y^A\partial_A \in \mathfrak{diff}(\Sigma_0)  .
\end{equation}
For the sake of simplicity, in what follows, we will keep the coordinate $u$ (resp. $v$) to denote the Carrollian time on $\mathscr I^+$ (resp. $\mathscr I^-$). It will be important to remember that combinations such as $u-v$ are in fact invariant under the action of the diagonal BMS group \eqref{diagonal BMS}.

\subsection{Carrollian holographic correspondence}
\label{sec:Holographic correspondence}
Having discussed the geometry of the conformal boundary, we are now ready to develop our proposal on how a putative conformal Carrollian theory living on $\hat{\mathscr I}$ can encode a scattering of massless particles.

\subsubsection{Boundary operators as conformal Carrollian fields}

As already anticipated in Section \ref{sec:The three bases for scattering amplitudes in flat spacetime}, we will argue that $\mathcal{S}$-matrix elements written in position space can naturally be interpreted as correlation functions of operators sourcing the Carrollian CFT living on $\hat{\mathscr I}$. The latter will be taken as the source operators $\sigma^m(x)$ introduced in Section \ref{sec:Sourced Ward identities}, assumed to transform as conformal Carrollian primaries \eqref{Carrollian tensor def}, namely
\begin{equation}
    \delta_{\bar{\xi}} \sigma_{(k,\bar{k})}(x) = \left[\left(\mathcal{T} + \frac{u}{2}(\partial_z\mathcal{Y}^z + \partial_{\bar z} \mathcal{Y}^{\bar z})\right)\partial_u + \mathcal Y^z \partial_z + \mathcal Y^z \partial_{\bar z} +   k\, \partial_z \mathcal Y^z + \bar{  k}\, \partial_{\bar z} \mathcal Y^{\bar z}\right] \sigma_{(  k,\bar{  k})}(x)
    \label{SOURCE Carroll}
\end{equation}
For instance, it is suggestive to write boundary correlators in the form
\begin{equation}
    \big\langle \sigma_{(k_1,\bar k_1)}^{\text{out}}(x_1) \dots \sigma_{(k_n,\bar k_n)}^{\text{out}}(x_n)  \sigma_{(k_{n+1},\bar k_{n+1})}^{\text{in}}(x_{n+1}) \dots \sigma_{(k_N,\bar k_N)}^{\text{in}}(x_N)  \big\rangle
\label{correlator source carr}
\end{equation} 
where $\sigma_{(k_1,\bar k_1)}^{\text{out}}(x_1)$ $\dots$ $\sigma_{(k_n,\bar k_n)}^{\text{out}}(x_n)$ are $n$ insertions of source operators at $\mathscr{I}^+$ and $\sigma_{(k_{n+1},\bar k_{n+1})}^{\text{in}}(x_{n+1})$ $\dots$ $\sigma_{(k_N,\bar k_N)}^{\text{in}}(x_N)$ are $N-n$ insertions at $\mathscr{I}^-$. For outgoing insertions, $x_i = (u_i,z_i,\bar z_i)$ while for incoming insertions, $x_j = (v_j,z_j,\bar z_j)$. The boundary values \eqref{outgoing spin s} and \eqref{incoming spin s} of the operators in position space introduced in Section \ref{sec:Massless scattering in flat spacetime} are natural candidates for conformal Carrollian primaries sourcing the dual theory (for $s=2$, these are quasi-primaries), \textit{i.e.}
\begin{equation}
    \sigma_{(k,\bar k)}^{\text{out}}(u,z,\bar z) \eqholo \lim_{r\to+\infty} \left(r^{1-s}\phi^{(s)\,\text{out}}_{z\dots z}(u,r,z,\bar z)\right) = \bar\phi^{(s)\,\text{out}}_{z\dots z}(u,z,\bar z), \label{Carroll from bulk to bnd}
\end{equation}
for ``out'' insertions in retarded Bondi coordinates and the similar identification
\begin{equation}
    \sigma_{(k,\bar k)}^{\text{in}}(v,z,\bar z) \eqholo \lim_{r\to-\infty} \left(r^{1-s}\phi^{(s)\,\text{in}}_{z\dots z}(v,r,z,\bar z)^\dagger\right) =  \bar\phi^{(s)\,\text{in}}_{z\dots z}(v,z,\bar z)^\dagger \label{Carroll from bulk to bnd INCOMING}
\end{equation}
for ``in'' insertions in advanced Bondi coordinates and the hermitian conjugated contributions for reversed helicities. From \eqref{Poincare on boundary value}, the conformal Carrollian weights $k,\bar k$ in \eqref{SOURCE Carroll} are fixed in terms of the helicity $J$ as
\begin{equation}
    k = \frac{1\pm J}{2},\quad \bar k = \frac{1\mp J}{2}, \label{k and kbar in terms of J}
\end{equation}
the upper sign for outgoing fields and the lower sign for incoming fields. 

Aside of the source operators, the Carrollian CFT is populated by fields $\Psi^i(x)$. Among them, one finds the Carrollian stress tensor ${\mathcal C^a}_b$ that gathers the Carrollian momenta \eqref{Carrollian momenta def} and will be identified with the gravitational momenta in the bulk, irrespective of the presence of radiation. Furthermore, the propagating massless fields are not only described by their boundary values but also by a tower of subleading pieces such as the components $F_{ur}^{(2)}$ of the Faraday tensor for spin-$1$ field or the subleading tensors $D_{AB},E_{AB},\dots$ in the expansion of the angular part of the Bondi metric. These are all natural candidates for the fields of the Carrollian CFT whose evolution is modified by the presence of source operators that holographically encode the radiative degrees of freedom. 

\paragraph{Remark} The fact that the algebraic constraints \eqref{k and kbar in terms of J} hold, \textit{i.e.} that the conformal Carrollian operators constructed from the boundary value of bulk quantum spin-$s$ fields can be labeled only with one quantum number $J$, comes from the conformal compactification process and its impact on the fall-offs near $\mathscr I^+$ imposed for radiative fields. Indeed, under the combined action of Weyl rescalings and BMS symmetries, the boundary value of a spin-$s$ field with helicity $J= \pm s$ and Weyl weight $W$ will transform infinitesimally as
\begin{equation}
\begin{split}
    &\delta_{\bar{\xi},\Omega} \bar\phi^W_{z\dots z} = \big[\mathcal{Y}^z \partial_z + \mathcal Y^{\bar z} \partial_{\bar{z}} + s\, \partial_z \mathcal{Y}^z - W \Omega \big] \bar\phi^W_{{z\dots z}} ,\\
    &\delta_{\bar{\xi},\Omega} \bar\phi^W_{\bar{z}\dots \bar{z}} = \big[\mathcal{Y}^z \partial_z + \mathcal Y^{\bar z} \partial_{\bar{z}} + s\, \partial_{\bar{z}} \mathcal{Y}^{\bar{z}} - W \Omega \big] \bar\phi^W_{\bar{z}\dots \bar{z}} .
\end{split}
\end{equation} 
Fixing the representative of the conformal boundary metric to be the flat metric implies $\delta_{\bar{\xi},\Omega} \mathring{q}_{z\bar{z}} = \partial_z \mathcal{Y}^z + \partial_{\bar{z}}\mathcal{Y}^{\bar{z}} - 2 \Omega = 0$, so that  
\begin{equation}
\begin{split}
    &\delta_{\bar{\xi},\Omega = \alpha} \bar\phi^W_{{z\dots z}} = \big[\mathcal{Y}^z \partial_z + \mathcal Y^{\bar z} \partial_{\bar{z}} + ( s - \tfrac{1}{2}W) \partial_z \mathcal{Y}^z - \tfrac{1}{2} W \partial_{\bar{z}} \mathcal{Y}^{\bar{z}} \big] \bar\phi^W_{{z\dots z}} ,\\
    &\delta_{\bar{\xi},\Omega = \alpha} \bar\phi^W_{{\bar{z}\dots \bar{z}}} = \big[\mathcal{Y}^z \partial_z    + \mathcal Y^{\bar z} \partial_{\bar{z}} - \tfrac{1}{2} W  \partial_z \mathcal{Y}^z  + (s - \tfrac{1}{2} W) \partial_{\bar{z}} \mathcal{Y}^{\bar{z}} \big] \bar\phi^W_{\bar{z}\dots\bar{z}} .
\end{split}
\end{equation} 
Using \eqref{Carroll from bulk to bnd} and by identification with the transformation \eqref{SOURCE Carroll}, we find
\begin{equation}
\begin{split}
    &\bar\phi^W_{{z\dots z}}\mapsto \sigma_{(k,\bar k)} \quad\text{with}\quad k = s - \tfrac{1}{2} W, \quad \bar{k} = - \tfrac{1}{2} W,\\
    &\bar\phi^W_{{\bar z\dots \bar z}}\mapsto \sigma_{(k,\bar k)} \quad\text{with}\quad k = -\tfrac{1}{2}W , \quad \bar k = s - \tfrac{1}{2}W  . 
\end{split}
\end{equation} 
Finally, the radiative falloff \eqref{Phi s final} fixes the Weyl weight $W$ of the boundary value of the field because, under a boundary Weyl rescaling induced by $r\to \Omega^{-1}r$, the boundary spin-$s$ field scales as $\bar\phi^W_{{z\dots z}}\to \Omega^{1-s}\bar\phi^W_{{z\dots z}}$ so $W=s-1$, which gives again \eqref{k and kbar in terms of J} with the upper choice of sign. The reasoning can also be applied for incoming fields defined as \eqref{Carroll from bulk to bnd INCOMING}, the hermitian conjugation being responsible of the sign flip in \eqref{k and kbar in terms of J}.

Owing to \eqref{Carroll from bulk to bnd}--\eqref{Carroll from bulk to bnd INCOMING}, we propose the following holographic identification between conformal Carrollian correlators \eqref{correlator source carr} and $\mathcal S$-matrix elements written in position space \eqref{Carrollian S-matrix}: 
\begin{equation}
    \boxed{
    \begin{gathered}
        \big\langle \sigma_{(k_1,\bar k_1)}^{\text{out}}(x_1) \dots \sigma_{(k_n,\bar k_n)}^{\text{out}}(x_n)  \sigma_{(k_{n+1},\bar k_{n+1})}^{\text{in}}(x_{n+1}) \dots \sigma_{(k_N,\bar k_N)}^{\text{in}}(x_N) \big\rangle \\[5pt]
        \eqholo \\[5pt]
        \frac{1}{(2\pi)^N}\prod_{k=1}^n \int_0^{+\infty} \D\omega_k \, e^{-i\omega_k u_k}  \prod_{\ell=n+1}^{N} \int_0^{+\infty} \D\omega_{\ell} \, e^{i\omega_{\ell} v_\ell}  \mathcal A_N ( p_1;\dots; p_N),
    \end{gathered}
    } \label{carrollmap}
\end{equation}
or equivalently
\begin{equation}
    \begin{gathered}
        \big\langle \sigma_{(k_1,\bar k_1)}^{\text{out}}(x_1) \dots \sigma_{(k_n,\bar k_n)}^{\text{out}}(x_n)  \sigma_{(k_{n+1},\bar k_{n+1})}^{\text{in}}(x_{n+1}) \dots \sigma_{(k_N,\bar k_N)}^{\text{in}}(x_N) \big\rangle \\[5pt]
        \eqholo \\[5pt]
        \langle 0 | \bar\phi_{I_1}^{(s)}(x_1)^{\text{out}}\dots \bar\phi_{I_n}^{(s)}(x_n)^{\text{out}} \bar\phi^{(s)}_{I_{n+1}}(x_{n+1})^{\text{in}\,\dagger}\dots \bar\phi^{(s)}_{I_N}(x_N)^{\text{in}\,\dagger} | 0 \rangle,
    \end{gathered} \label{correlators boundary values}
\end{equation}
where $I_i=z\dots z$ if the helicity is positive ($J_i=+s$), and $I_i = \bar z\dots\bar z$ otherwise, and Carrollian weights $(k_i,\bar k_i)$ are fixed by bulk helicities according to \eqref{k and kbar in terms of J}. 

Particularizing \eqref{correlators boundary values} to the insertion of two fields only, we obtain, by virtue of the regulated expression \eqref{Wuup}, that the Carrollian two-point function is given by
\begin{equation}
    \begin{split}
        &\big\langle \sigma^{\text{out}}_{(k_1,\bar k_1)}(u,z_1,\bar z_1)\sigma^{\text{in}}_{(k_2,\bar k_2)}(v,z_2,\bar z_2) \big\rangle \\
        &= \frac{K_{(s)}^2}{4\pi} \left[ \frac{1}{\beta} - \left(\gamma+\ln|u-v|+\frac{i\pi}{2}\text{sign}(u-v)\right) \right]\delta^{(2)}(z_1-z_2)\,\delta_{k^+_{12},2}\,\delta_{k^-_{12},0}
    \end{split} \label{Carroll 2pt holo}
\end{equation}
denoting $k^\pm_{12}\equiv \sum_i(k_i\pm \bar k_i)$. There is correlation if $k^+_{12}$ and $k^-_{12}$ obey two algebraic conditions: the first condition $k_{12}^+ = 2$ is imposed by \eqref{k and kbar in terms of J}, \textit{i.e.} the fact that the inserted fields are identified with boundary values of bulk radiative fields, while the second implies the conservation of helicity $J_1 = J_2$. As above, $\beta\in\mathbb R^+_0$ is an infrared regulator than can be arbitrarily close to zero, encoding the fact that position space amplitudes are divergent in the low-energy regime. Finally, from the identification \eqref{Carroll from bulk to bnd} and recalling \eqref{commu boundary fields}, one gets that the source operators \eqref{Carroll from bulk to bnd} obey the following canonical commutation relations
\begin{equation}
    \big[{\sigma}^{\text{out}}_{(k_1,\bar k_1)}(u_1,z_1,\bar z_1),{\sigma}^{\text{out}}_{(k_2,\bar k_2)} (u_2,z_2,\bar z_2)\big] 
    = -\frac{i}{2} K_{(s)}^2\,\text{sign}(u_1 - u_2) \delta^{(2)}(z_1-z_2)\,\delta_{k^+_{12},2}\,\delta_{k^-_{12},0}.
    \label{commu carroll}
\end{equation}
Analogous considerations hold for ``in'' insertions.

\subsubsection{Electrodynamics}
\label{sec:electrodynamics holographic correspondence}

We now detail some holographic aspects of electrodynamics reviewed in Section \ref{sec:electro} in terms of the dual sourced Carrollian CFT. For the sake of conciseness, we give all definitions for $\mathscr I^+$ and drop the ``out'' subscripts to lighten the notations when the coordinate dependence is clear. The analogous relations can be easily deduced for $\mathscr I^-$.

The dual conformal Carrollian theory is invariant under $U(1)$ transformations. The source operators that are considered in this section are denoted as $\sigma_{(k,\bar k),Q}$ and bear conformal Carrollian weights $(k,\bar k)$ describing their transformations under boundary diffeomorphisms \eqref{conformal Carroll symmetries here} and a ``weight'' $Q$ under the internal $U(1)$ symmetry, which is interpreted as the electric charge of the bulk field. Following \eqref{Carroll from bulk to bnd}, we introduce
\begin{equation}
    \begin{split}
        \sigma_z^{(A)}(u,z,\bar z) &\equiv \sigma_{(1,0),0}(u,z,\bar z) = A_z^{(0)}(u,z,\bar z), \quad \sigma_{\bar z}^{(A)}(u,z,\bar z) = \big(\sigma_z^{(A)}\big)^\dagger, \\
        \sigma^{(\phi)}(u,z,\bar z) &\equiv \sigma_{(\frac{1}{2},\frac{1}{2}),Q}(u,z,\bar z) = \phi^{(0)}(u,z,\bar z) = \lim_{r\to\infty} r\,\phi(u,r,z,\bar z),
    \end{split}
    \label{new}
\end{equation}
coinciding with the boundary values of the bulk gauge field $A_\mu$ and some massless scalar matter field $\phi$ at $\mathscr I^+$. These fields are conformal Carrollian primaries with respective weights $(1,0)$ and $(\frac{1}{2},\frac{1}{2})$ by virtue of \eqref{k and kbar in terms of J}, on which the global $U(1)$ symmetry acts as
\begin{equation}
    \delta_\lambda \sigma_z^{(A)} (u,z,\bar z) =  \partial_z\lambda^{(0)}(z,\bar z), \quad \delta_\lambda \sigma^{(\phi)} (u,z,\bar z) = -ieQ\lambda^{(0)}(z,\bar z)\sigma^{(\phi)} (u,z,\bar z). \label{gauge tr carroll}
\end{equation}
The homogeneous transformation determines the charge under the $U(1)$ transformation (which is proportional to the electric charge of the field) and the inhomogeneous transformation vanishes for parameters $\lambda^{(0)}\in\mathbb R$. The Noether current for the $U(1)$ symmetry is given by
\begin{equation}
\begin{split}
     \big\langle j^u_\lambda \big\rangle \equiv -\frac{1}{e^2}\lambda^{(0)} F^{(2)}_{ru} , \quad  \big\langle j^{z}_\lambda \big\rangle  \equiv 0 \equiv  \big\langle j^{\bar{z}}_\lambda \big\rangle
\end{split} \label{identification j QED}
\end{equation}
and the corresponding fluxes are
\begin{equation}
    F_\lambda = \lambda^{(0)}\,\modj^{(2)}_u\big(\sigma^{(\phi)}\big) -\frac{1}{e^2}\Big(\partial_u \sigma_z^{(A)}\partial_{\bar z}\lambda^{(0)}+ \partial_u \sigma_{\bar z}^{(A)} \partial_{z}\lambda^{(0)} \Big). \label{identification F QED}
\end{equation}
The factors are adjusted to precisely match with the charges \eqref{charge maxwell future} obtained by the bulk computation. With the identifications \eqref{identification j QED} and \eqref{identification F QED}, one can then check explicitly that the time evolution equations in the sourced Ward identities \eqref{infinitesimal ward QED} reproduce the asymptotic Maxwell equation \eqref{maxwell on the bndry} when there is no insertion in the correlators. Notice finally that the fluxes \eqref{identification F QED} vanish identically if $\tilde A_z^{(0)} = 0 = \tilde A_{\bar z}^{(0)}$ and $\modj^{(2)}_u = 0$. 

\subsubsection{Gravity}
We now argue that a quantum conformal Carrollian field theory coupled with external sources is a viable candidate to describe holographically gravity in $4d$ asymptotically flat spacetimes reviewed in Section \ref{sec:gravity asymptotics}. We propose the following correspondence between Carrollian momenta \eqref{Carrollian momenta def} (left-hand side) and gravitational data \eqref{BMS momenta} (right-hand side) at $\mathscr{I}^+$:
\begin{equation}
    \big\langle {\mathcal C^u}_u \big\rangle \eqholo \frac{\bar M}{4\pi G},\quad \big\langle {\mathcal C^u}_A \big\rangle \eqholo \frac{1}{8\pi G}\big(\bar N_A + u\,\partial_A\bar M\big),\quad \big\langle {\mathcal C^A}_B\big\rangle + \frac{1}{2}{\delta^A}_B\big\langle {\mathcal C^u}_u \big\rangle \eqholo 0. \label{holographic correspondence grav}
\end{equation}
The factors are fixed by demanding that the gravitational charges \eqref{grav charge} correspond to the Noether currents \eqref{Carrollian Noether currents} of the conformal Carrollian field theory integrated on a section $u=\text{const}$. The correspondence \eqref{holographic correspondence grav} is inspired by the AdS/CFT dictionary where the holographic stress-energy tensor of the CFT is identified with some subleading order pieces in the expansion of the bulk metric \cite{Balasubramanian:1999re , deHaro:2000vlm}. Indeed, recall that the Carrollian momenta are nothing but the components of an ultra-relativistic stress tensor living at null infinity.

Following the identification \eqref{Carroll from bulk to bnd}, we identify the source operators in the Carrollian CFT as the asymptotic shear
\begin{equation}
    \begin{split}
        \sigma^{(g)}_{zz} (u,z,\bar z) &\equiv \sigma_{(\frac{3}{2},-\frac{1}{2})}(u,z,\bar z) = C_{zz}(u,z,\bar z), \\
        \sigma^{(g)}_{\bar z\bar z}(u,z,\bar z) &\equiv \sigma_{(-\frac{1}{2},
     \frac{3}{2})}(u,z,\bar z) = C_{\bar z\bar z}(u,z,\bar z),
    \end{split}
\end{equation}
and similarly at $\mathscr I^-$. They are quasi-conformal Carrollian primary fields because of \eqref{transfo CAB}. The homogeneous part of the transformation determines the Carrollian weights of the field to be $k = \frac{3}{2}$ and $\bar k = -\frac{1}{2}$. In the presence of matter, boundary values of radiative null fields will also be considered as source operators $\sigma^{(\phi)}$ with weights $(k,\bar k)$ fixed by the helicity of the bulk field as \eqref{k and kbar in terms of J}. All these sources are responsible for the dissipation in the conformal Carrollian field theory through the fluxes $F_{\bar \xi} = F_a\bar\xi^a$ with 
\begin{equation}
    \begin{split}
        F_u &= \frac{1}{16\pi G}\Big[ \partial_z^2\partial_u \sigma^{(g)}_{\bar z\bar z} + \frac{1}{2}\sigma^{(g)}_{\bar z\bar z}\partial_u^2\sigma^{(g)}_{zz} + \text{h.c.} \Big] - T_{uu}^{m(2)}\big(\sigma^{(\phi)}\big) ,\\
        F_z &= \frac{1}{16\pi G}\Big[ -u\partial_z^3\partial_u \sigma_{\bar z\bar z}^{(g)} + \sigma_{zz}^{(g)}\partial_z\partial_u\sigma_{\bar z\bar z}^{(g)} - \frac{u}{2}(\partial_z\sigma_{zz}^{(g)}\partial_u^2\sigma_{\bar z\bar z}^{(g)} + \sigma_{zz}^{(g)}\partial_z\partial_u^2 \sigma^{(g)}_{\bar z\bar z}) \Big]\\
        &\quad - T_{uz}^{m(2)}\big(\sigma^{(\phi)}\big) + \frac{u}{2}\partial_z T_{uu}^{m(2)}\big(\sigma^{(\phi)}\big)  , \qquad F_{\bar z} = (F_z)^\dagger .
    \end{split} \label{flux expressions sources}
\end{equation} 
Taking the identifications \eqref{holographic correspondence grav} and \eqref{flux expressions sources} into account, one can then check explicitly that the time evolution equations in the sourced Ward identities \eqref{Ward identities Carrollian momenta} reproduce the gravitational retarded time evolution equations \eqref{EOM1} when there is no insertion in the correlators. Notice that imposing the BMS-invariant conditions $\tilde C_{zz} = 0 = \tilde C_{\bar z\bar z}$ and 
$T_{uu}^{m(2)} = 0 = T_{uz}^{m(2)} = T_{u\bar z}^{m(2)}$ makes the flux vanish identically. 

\subsection{Holographic Ward identities for a massless scattering}
\label{sec:Holographic Ward identities for a massless scattering}

In this section, we specify the holographic sourced Ward identities discussed in Sections \ref{sec:Sourced $U(1)$ Ward identities} and \ref{sec:Sourced conformal Carrollian Ward identities} to the case where the bulk process consists of a massless scattering.

Let us consider a scattering of null particles in some asymptotically flat spacetime at null infinity. Through the holographic identification \eqref{carrollmap}, the scattering amplitudes are encoded by correlation functions $\big\langle X_N^\sigma \big\rangle$ of source operators in a putative dual conformal Carrollian theory living on $\hat{\mathscr I}$. The integrated version of the sourced Ward identity \eqref{infinitesimal Ward identity integrated SOURCES} is specified as
\begin{equation}
    \Big\langle \left( \int_{\hat{\mathscr{I}}}  {\bm F}_{\bar{\xi}} - \int_{\mathscr{I}^+_+}  {\bm j}_{\bar{\xi}} + \int_{\mathscr{I}^-_-}  {\bm j}_{\bar{\xi}} \right) \, X_N^\sigma \Big\rangle = 0, \label{stage Ward}
\end{equation}
for any conformal Carrollian vector $\bar\xi$, where $\bm j_{\bar\xi}|_{\mathscr I^-_-}$, $\bm j_{\bar\xi}|_{\mathscr I^+_+}$ are sensitive to motions of massive particles and the flux $\bm F_{\bar\xi}$ is given by \eqref{flux expressions sources}. Hypothesizing that the process under consideration only encompasses massless fields and nothing arrives at past and future timelike infinities, the Noether current $\bm j_{\bar\xi}$ vanishes at $\mathscr{I}^-_-$ and $\mathscr{I}^+_+$. Hence \eqref{stage Ward} becomes
\begin{equation}
    \boxed{\big\langle \mathcal F_{\bar\xi}\,X_N^\sigma\big\rangle = 0,} \quad\text{where}\quad \mathcal F_{\bar\xi} = \int_{\hat{\mathscr{I}}}  {\bm F}_{\bar{\xi}}. \label{interm11}
\end{equation}
Taking into account that the insertion of the flux operator $\mathcal F_{\bar\xi}$ generates the transformation of the source operators in $X^\sigma_N$, we have
\begin{equation}
    \boxed{\delta_{\bar{\xi}} \big\langle X_N^\sigma  \big\rangle  = 0 ,} \label{carroll invariance}
\end{equation} 
hence recovering the invariance of the correlators under conformal Carrollian symmetries. Notice that with no source inserted ($N=0$), we obtain the natural result $\big\langle \mathcal F_{\bar\xi}\big\rangle = 0$. The consequence of the relation \eqref{carroll invariance} has been studied \textit{e.g.} in \cite{Chen:2021xkw,Bagchi:2009ca,Bagchi:2017cpu}. In the next section, we will revisit how to deduce the generic form of correlators from these symmetry constraints.

Prior to that, let us mention that the derivation can also be made for electrodynamics. In that case, the conformal Carrollian correlators obey
\begin{equation}
    \big\langle \mathcal F_\lambda\, X^\sigma_N \big\rangle = 0 \quad\Leftrightarrow\quad \delta_{\lambda} \big\langle X_N^\sigma  \big\rangle  = 0  \label{electro invariance}
\end{equation}
for any gauge parameter $\lambda$. In particular, this constrains  the two-point function to satisfy
\begin{equation}
    \delta_{\lambda} \big\langle X_2^\sigma \big\rangle = 0 \quad\Rightarrow\quad \left(\lambda^{(0)}(z_1,\bar z_1)Q_1-\lambda^{(0)}(z_2,\bar z_2)Q_2\right) \big\langle X_2^\sigma \big\rangle = 0  \label{delta X2 QED} 
\end{equation}
using \eqref{Carroll from bulk to bnd}--\eqref{Carroll from bulk to bnd INCOMING} and \eqref{gauge tr carroll}. For the particular transformation $\lambda^{(0)} = c\in\mathbb R_0$, this imposes the algebraic constraint $Q_1=Q_2$,  which is nothing but the statement of conservation of electric charge. For a generic function $\lambda^{(0)}(z,\bar z)$, this further imposes $\big\langle X_2^\sigma \big\rangle \propto \delta^{(2)}(z_1-z_2)\,\delta_{Q_1,Q_2}$, which is consistent with conformal Carrollian invariance \eqref{carroll invariance} as we will show in the next section.

\subsection{Conformal Carrollian invariant correlation functions}
\label{sec:Conformal Carrollian invariant correlation functions}

In this section, we deduce the explicit form of the two- and three-point correlation functions in Carrollian CFT from the Ward identities \eqref{carroll invariance}. The computation of higher-point functions is left to future endeavor.

Let $\Phi_{(  k_1,\bar{ {k}}_1)}(x_1)$ and $\Phi_{(  k_2,\bar{ {k}}_2)}(x_2)$ be two quasi-conformal Carrollian primary operators. We want to study the constrains on the two-point function $\big\langle X_2 \big\rangle \equiv \big\langle \Phi_{(  k_1,\bar{ {k}}_1)}(x_1)\Phi_{(  k_2,\bar{ {k}}_2)}(x_2)\big\rangle$ implied by $\delta_{\bar\xi}\big\langle X_2 \big\rangle=0$, where $\bar\xi$ denotes an element of the global part of the conformal Carroll algebra (or equivalently Poincaré algebra), which transforms the inserted fields as \eqref{Carrollian tensor def}. Invariance under Carrollian translations generated by $P_a = \partial_a$ gives
\begin{equation}
       \frac{\partial}{\partial x^a_1}\big\langle X_2 \big\rangle + \frac{\partial}{\partial x^a_2}\big\langle X_2 \big\rangle = 0 \quad\Rightarrow\quad \big\langle X_2 \big\rangle = \big\langle X_2 \big\rangle (u_{12},z_{12},\bar z_{12})
\end{equation}
where $u_{12} \equiv u_1-u_2$ and $z_{12}\equiv z_1-z_2$. Invariance under Carrollian boosts $B_A = x_A\partial_u$ gives
\begin{equation}
    x_A^{12}\partial_{u_{12}} \big\langle X_2 \big\rangle = 0 \quad \Rightarrow\quad \big\langle X_2 \big\rangle = f_{(2)}(z_{12},\bar z_{12}) + g_{(2)}(u_{12})  \delta^{(2)}(z_{12})  .
\end{equation}
The general solution thus involves two distinct branches \cite{Bagchi:2022emh,Chen:2021xkw}. The time-independent one $\big\langle X_2 \big\rangle^{ti}$ is meant to be invariant under the stabilizer group of $u=\text{const.}$ cuts of the conformal Carrollian manifold (\textit{i.e.} the conformal group in two dimensions), while the other branch $\big\langle X_2 \big\rangle^{td}$ involves explicitly the time direction but at the price to reduce the angular dependence to contact terms, which can be expected for an ``ultralocal'' theory of fields.

\subsubsection{Time-independent branch} Selecting first the time-independent branch by setting $g_{(2)} \equiv 0$, the invariance under the Carrollian dilatation $D = x^a\partial_a =u\partial_u +z\partial_z+\bar z\partial_{\bar z}$ imposes
\begin{equation}
        z_{12}\partial_{z_{12}} f_{(2)}+\bar z_{12}\partial_{\bar z_{12}} f_{(2)}+k^+_{12} f_{(2)} = 0\quad\Rightarrow \quad f_{(2)} = \frac{\tilde c_{(2)}}{z_{12}^a\bar z_{12}^b},\ a+b=k^+_{12} .
\end{equation}
Invariance under Carrollian rotation $J= -z\partial_z+\bar z\partial_{\bar z}$ also imposes
\begin{equation}
    z_{12}\partial_{z_{12}}f_{(2)}-\bar z_{12}\partial_{\bar z_{12}}f_{(2)}+k^-_{12}f_{(2)} = 0 \quad \Rightarrow \quad a-b = k^-_{12},
\end{equation}
which then implies that $a =\sum_i k_i$ and $b = \sum_i \bar k_i$. Because of time-independence, $K_0$ brings no additional constraint. $K_1$ and $K_2$ respectively impose that $k_1=k_2$ and $\bar k_1 = \bar k_2$, which allows to conclude that \cite{Chen:2021xkw,Bagchi:2022emh}
\begin{equation}
    \big\langle X_2 \big\rangle^{ti} = f_{(2)} = \frac{\tilde c_{(2)}}{(z_1-z_2)^{k_1+k_2}(\bar z_1-\bar z_2)^{\bar k_1+\bar k_2}}\delta_{k_1,k_2}\delta_{\bar k_1,\bar k_2}  .
\end{equation}
This is exactly the standard two-point function for a 2$d$ CFT. However, although this branch is allowed from a symmetry analysis, it is not related to dynamical bulk events such as scattering processes since it has no time dependence. 

\subsubsection{Time-dependent branch} We are thus rather interested in the time-dependent branch $\big\langle X_2 \big\rangle^{td}$ where we set $f_{(2)} \equiv 0$. The Ward identity encoding the invariance under Carrollian rotation $J$ simply adds one algebraic constraint as
\begin{equation}
        g_{(2)}\left(z_{12}\partial_{z_{12}} -\bar z_{12}\partial_{\bar z_{12}} + k^-_{12}\right)\delta^{(2)}(z_{12}) = k^-_{12}g_{(2)}\delta^{(2)}(z_{12}) = 0\quad\Rightarrow\quad k^-_{12} = 0 .
\end{equation} 
Now, invariance under Carrollian dilatation $D$ gives
\begin{equation}
        u_{12}\partial_{u_{12}} \big\langle X_2 \big\rangle^{td} + z_{12}\partial_{z_{12}}\big\langle X_2 \big\rangle^{td} + \bar z_{12}\partial_{\bar z_{12}} \big\langle X_2 \big\rangle^{td} + k^+_{12}\big\langle X_2 \big\rangle^{td} = 0 ,
\end{equation}
which, using the fact that $x\delta'(x) \simeq -\delta(x)$ in the sense of distributions, becomes
\begin{equation}
        u_{12}g_{(2)}'(u_{12}) + (k^+_{12} -2) g_{(2)}(u_{12}) = 0  . \label{eq for g(u12)}
\end{equation}
When \eqref{eq for g(u12)} is satisfied, we can check that the Ward identities for $K_0$, $K_1$ and $K_2$ do not bring additional constraints. Indeed, the proof for $K_0$ is immediate, because
\begin{equation}
    \left( |z_1|^2\partial_{u_1} + |z_2|^2\partial_{u_2} \right)\big\langle X_2 \big\rangle^{td} \propto \left( |z_1|^2-|z_2|^2 \right)\delta^{(2)}(z_1-z_2) = 0 .
\end{equation}
The Ward identity for $K_1$ is equivalent to the following differential constraint:
\begin{equation}
    \begin{split}
        & u_{12}z_1\partial_{u_{12}}\big\langle X_2 \big\rangle^{td} + (z_1^2-z_2^2)\partial_{z_{12}}\big\langle X_2 \big\rangle^{td} + 2 z_1 \big\langle X_2 \big\rangle^{td} \textstyle\sum_i k_i = 0 \\
        &\Rightarrow\ (2-k_{12}^-)z_1 \big\langle X_2 \big\rangle^{td} + (z_1+z_2)z_{12}\partial_{z_{12}}\big\langle X_2 \big\rangle^{td} + 2 z_1\big\langle X_2 \big\rangle^{td} \textstyle\sum_i k_i = 0 \\
        &\Rightarrow\ \left(2 - \textstyle\sum_i k_i - \textstyle\sum_i \bar k_i - 2 +2 \textstyle\sum_i k_i\right)z_1\big\langle X_2 \big\rangle^{td} = \left( \textstyle\sum_i k_i - \textstyle\sum_i \bar k_i \right)z_1\big\langle X_2 \big\rangle^{td} = 0 ,
    \end{split}
\end{equation}
using successively \eqref{eq for g(u12)} and the constraint $k^-_{12} = 0$. This concludes the derivation since the invariance under the last special conformal transformation $K_2$ is proven in an analogous way. The deal now consists in solving carefully the master equation \eqref{eq for g(u12)}, which offers a few surprises. 

\paragraph{Continuous set of functional solutions.} 
For generic values of $k^+_{12}\neq 2$, the general functional solution of this equation is
    \begin{equation}
    g_{(2)}(u_{12}) \propto \frac{1}{u_{12}^{k^+_{12} - 2}} ,
    \end{equation}
    leading to the following correlator
    \begin{equation}
    \boxed{
        \big\langle X_2 \big\rangle^{td,f} = \frac{c_{f,(2)}}{(u_1-u_2)^{k^+_{12} - 2}}\, \delta^{(2)}(z_1-z_2)\, \delta_{k^-_{12},0}.
    }
    \label{G2 td f}
    \end{equation}
    This class of solutions is consistent with previous analyses \cite{Banerjee:2018gce,Bagchi:2022emh,Chen:2021xkw}.

    When $k^+_{12}=2$, the above solution seems to be time-independent again. But if the insertions represent boundary values of bulk scattering fields, this algebraic constraint is precisely implied by \eqref{k and kbar in terms of J}. The only hope to keep a time-dependent solution from \eqref{G2 td f} that can match \textit{e.g.} with \eqref{Carroll 2pt holo} in the limit $k_{12}^+\to 2$ is to have a particular dependency in $k^+_{12}$ in the overall constant, left unfixed by the symmetries. Going to Fourier space, $g_{(2)}(u_{12}) = \int_{-\infty}^{+\infty}\D \omega\,G(\omega)\,e^{-i\omega u_{12}}$ and posing $\beta\equiv k^+_{12} -2$, \eqref{eq for g(u12)} becomes
    \begin{equation}
        \int_{-\infty}^{+\infty}\D\omega \left[-\omega G'(\omega) + (\beta-1)G(\omega)\right] e^{-i\omega u_{12}} = 0 . \label{eq for G in fourier}
    \end{equation}
    Assuming that $\beta>0$, the distribution
    \begin{equation}
        G_\beta(\omega) = 2\pi c\,\omega^{\beta-1}\,\Theta(\omega)
    \end{equation}
    is solution of \eqref{eq for G in fourier}, where $c$ is a constant. Indeed, enforcing that $\omega \geq 0$ to interpret it afterwards as the (light-cone) energy of particles,
    \begin{equation}
        \int_{-\infty}^{+\infty}\D\omega \left[-\omega G'_\beta(\omega) + (\beta-1)G_\beta(\omega)\right] e^{-i\omega u_{12}} = -2\pi c \int_{-\infty}^{+\infty}\D\omega\,\omega^{\beta}\,\delta(\omega)\, e^{-i\omega u_{12}} = 0 .
    \end{equation}  
    Inverting the Fourier transform for $\beta>0$ gives
    \begin{equation}
        g_{(2)}(u_{12}) = \frac{1}{2\pi}\int_{-\infty}^{+\infty}\D \omega \, G_\beta(\omega)\,e^{-i\omega u_{12}} = c\int_0^{+\infty} \D\omega\,\omega^{\beta-1}\,e^{-i\omega u_{12}},
    \end{equation}
    which involves the integral \eqref{I beta}. This fixes the dependency of $c_{f,(2)}$ as
    \begin{equation}
        g_{(2)}(u_{12}) = \frac{c\,\Gamma[k^+_{12}-2]}{u_{12}^ {k^+_{12}-2}} \xrightarrow[]{\quad k^+_{12}\to 2\quad } c \left[\frac{1}{k^+_{12}-2} -\big(\gamma + \ln|u_{12}| \big) \right] + \mathcal O(k^+_{12}-2) 
    \label{log funct}
    \end{equation}
    only focusing on the functional branch by fixing $c\in\mathbb R$. As we discussed in Section \ref{sec:asymptotic states in direct basis}, the pole is related to an infrared divergence, which seems entangled with the large-$r$ limit of the bulk fields inducing the algebraic constraints \eqref{k and kbar in terms of J}. As we shall now see, this is however not the unique time dependent solution for $k^+_{12}=2$.
    
\paragraph{Discrete set of distributional solutions.} When $k^+_{12} = 2+n$ for $n\in\mathbb N$, the solution of \eqref{eq for g(u12)} is enriched by a discrete set of distributional solutions. For $n=0$, using the fact that $x\delta(x) \simeq 0$ in the sense of distributions, we see that $g_{(2)}(u_{12}) \propto f_{(0)}(u_{12}) \equiv \text{sign}(u_{12})$ is a solution. We thus have
    \begin{equation}
        \big\langle X_2 \big\rangle^{td,d} = c_{d,(2)}\,\text{sign}(u_1-u_2)\, \delta^{(2)}(z_1-z_2)\, \delta_{k^-_{12},0}\,\delta_{k^+_{12},2}  , \label{two point carroll sgn}
    \end{equation}
    among the possible solutions of the Carrollian Ward identity for $k^+_{12}=2$. For any $n\in\mathbb N_0$, it can be shown by recurrence that \eqref{eq for g(u12)} is solved by the successive distributional derivatives of the sign function, denoted by $f_{(n)}(u_{12})\equiv \frac{\D^n}{\D u_{12}^n}f_{(0)}(u_{12})$. Indeed, if $f_{(n)}(u_{12})$ is solution of \eqref{eq for g(u12)} for $k_+^{12}-2=n$, \textit{i.e.} $u_{12} f'_{(n)}(u_{12}) + n\,f_{(n)}(u_{12})=0$, then
    \begin{equation}
            u_{12}f'_{(n+1)}(u_{12}) + (n+1)f_{(n+1)}(u_{12}) = \left(u_{12} f'_{(n)}(u_{12})\right)' + n\,f'_{(n)}(u_{12}) = 0
    \end{equation}
    So in general, we can write
    \begin{equation}
        \boxed{
        \big\langle X_2 \big\rangle_{(n)}^{td,d} = c_{d,(2)}\,\frac{\D^n}{\D u_1^n}\text{sign}(u_1-u_2)\, \delta^{(2)}(z_1-z_2)\, \delta_{k^-_{12},0}\,\delta_{k^+_{12},2+n} ,\quad\forall n\in\mathbb N .
        } \label{two point carroll distri}
    \end{equation}
    To the best of our knowledge, this particular solution has not been derived in the previous Carrollian literature. Together with \eqref{log funct}, this proves the Carrollian invariance of the boundary two-point function \eqref{Carroll 2pt holo} and the commutator \eqref{commu carroll}.  More precisely, the holographic dictionary \eqref{Carroll from bulk to bnd}, \eqref{correlators boundary values} and the considerations of Section \ref{sec:Low-point amplitudes} favors the following particular linear combination of \eqref{log funct} and \eqref{two point carroll sgn}:
   \begin{equation}
    \boxed{
        \big\langle X_2 \big\rangle^{td} \propto \left[\frac{1}{k_{12}^+-2}-\left(\gamma+\ln|u_1-u_2| + \frac{i\pi}{2}\text{sign}(u_1-u_2)\right)\right]\,\delta^{(2)}(z_1-z_2)\,\delta_{k^+_{12},2}\,\delta_{k^-_{12},0} .
    } \label{carroll 2pt holo fixed}
    \end{equation}
    We thus regard the above expression as the form of the two-point function for source operators in any holographic Carrollian CFT. The expression \eqref{Wuup} of the two-point amplitude in position space fixes the overall constant in \eqref{carroll 2pt holo fixed} to $(4\pi)^{-1} K^2_{(s)}$. 

\paragraph{Remark} Acting with $n$ time derivatives on the correlators increments $k_{12}^+ \to k_{12}^++n$ (because each time derivative increases Carrollian weights by $\frac{1}{2}$), starting from the correlators of boundary value fields with $k_{12}^+ =2$. For each $n\in\mathbb N_0$, we thus have two branches to consider. For instance,
\begin{equation}
\begin{split}
    &\big\langle\partial_{u_1}\Phi_{z_1z_1}(u_1,z_1,\bar z_1) \Phi_{\bar z_2\bar z_2}(u_2,z_2,\bar z_2)\big\rangle \propto \frac{\delta^{(2)}(z_1-z_2)}{u_1-u_2} \quad\text{or}\quad \delta(u_1-u_2)\,\delta^{(2)}(z_1-z_2) ,\\
    &\big\langle\partial_{u_1}\Phi_{z_1z_1}(u_1,z_1,\bar z_1) \partial_{u_2}\Phi_{\bar z_2\bar z_2}(u_2,z_2,\bar z_2)\big\rangle \propto \frac{\delta^{(2)}(z_1-z_2)}{(u_1-u_2)^2} \quad\text{or}\quad \delta'(u_1-u_2)\,\delta^{(2)}(z_1-z_2) ,
\end{split}
\end{equation}
for $k_{12}^+ = 3$ and $4$. Both choices lead to time-dependent correlation functions which are possibly well-behaved for the $\mathcal B$-transform. However, the compatibility with the holographic dictionary imposes to choose the distributional branch for (expectation values of) commutators and the functional branch for the correlation functions. As a curiosity, we remark that the coexistence of inverse power law in time and distributional dependencies is reminiscent of what happens in $2d$ CFT, where the analogous branches can be mapped onto each other by means of shadow transforms.

Let us conclude this section by giving some comments on the three-point function $\big\langle X_3 \big\rangle$. It has already been pointed out that the time-dependent three-point function $\big\langle X_3 \big\rangle^{td}$ is identically zero in a conformal Carrollian quantum theory \cite{Chen:2021xkw,Bagchi:2022emh,Banerjee:2018gce}. This can be seen as a consequence of the invariance under Carrollian time translation $P_0$, boosts $B_A$ and special conformal transformation $K_0$, or, in other words, invariance under Poincaré translations (see Appendix \ref{sec:isomorphism Poincare Carroll} for the dictionary). One has the following algebraic constraints
\begin{equation}
    \begin{split}
        \sum_{i=1}^3 \partial_{u_i}\big\langle X_3 \big\rangle^{td} =  \sum_{i=1}^3 z_i\, \partial_{u_i}\big\langle X_3 \big\rangle^{td} =  \sum_{i=1}^3 \bar z_i\, \partial_{u_i}\big\langle X_3 \big\rangle^{td} =  \sum_{i=1}^3 |z_i|^2\, \partial_{u_i}\big\langle X_3 \big\rangle^{td} = 0 
    \end{split} \label{3pt constraints}
\end{equation}
solved by $\partial_{u_i}\big\langle X_3 \big\rangle^{td} = 0$, for all $i=1,2,3$ and the time-dependent three-point function is identically zero. Finally, by similar arguments, one can easily show that this is also the case for the one-point function, namely $\big\langle X_1 \big\rangle = \big\langle \Phi_{(k,\bar k)}\big\rangle = 0$, except for the identity operator.

\section{Relation with celestial holography}
\label{sec:Relation with celestial holography}

In Section \ref{sec:Holographic conformal Carrollian field theory}, we discussed the Carrollian holography proposal by providing some kinematical properties of the dual Carrollian CFT and its relation with gravity in the bulk. The goal of this section is to make contact between the Carrollian approach and the celestial holography paradigm. In order to do so, we start by recalling  some of the well-established symmetry constraints for celestial CFT induced from the bulk analysis, namely the BMS Ward identities encoding bulk soft theorems. We then relate Carrollian CFT and celestial CFT by mapping the Carrollian source operators, their correlation functions and associated Ward identities to those of the CCFT.

\subsection{Ward identities of \texorpdfstring{$2d$}{2d } celestial CFT currents}

As already reviewed in Section \ref{sec:asymptotic states in Mellin basis}, for a massless scattering, one can express the $\mathcal{S}$-matrix elements in a boost-eigenstate basis made of the conformal primary wave functions by performing a Mellin transform on the amplitudes in energy eigenstates, see \eqref{Celestial S-matrix}. The key ingredient of celestial holography is to identify the $\mathcal{S}$-matrix elements written in this conformal basis with correlation functions of $2d$ celestial CFT, namely
\begin{equation}
\begin{split}
    &\big\langle \mathcal{O}^{\text{out}}_{(\Delta_1, J_1)}(z_1, \bar{z}_1) \dots \mathcal{O}^{\text{out}}_{(\Delta_n, J_n)}(z_n, \bar{z}_n) \mathcal{O}^{\text{in}}_{(\Delta_{n+1}, J_{n+1})}(z_{n+1}, \bar{z}_{n+1}) \dots \mathcal{O}^{\text{in}}_{(\Delta_{N}, J_{N})}(z_{N}, \bar{z}_{N}) \big\rangle_{\text{CCFT}} \\
    &\eqholo \langle \text{out} | \text{in} \rangle_{\text{boost}} =\int_0^{+\infty} \D\omega_1 \, \omega_1^{\Delta_1-1} \int_0^{+\infty} \D\omega_2 \, \omega_2^{\Delta_2-1} \dots \int_0^{+\infty} \D\omega_N \, \omega_N^{\Delta_N-1} \mathcal A_N(p_1;\dots; p_N),
\end{split} 
\label{identification bulk boundary celestial}
\end{equation} where we recall that $\mathcal A_N$ denotes the amplitude of $N$-particle scattering ($n$ of which are outgoing) in the usual energy-eigenstate basis. CCFT operators are characterized by a pair of numbers: $\Delta$ is the conformal dimension corresponding to the boost eigenvalue in the bulk and $J$ is the $2d$ spin identified with the bulk helicity. Very often, one trades $(\Delta, J)$ for the conformal weights $(h,\bar{h})$ defined as
\begin{equation}
    h=\frac{\Delta+J}{2} , \quad \bar{h}=\frac{\Delta-J}{2} . \label{h and hbar in terms of Delta J}
\end{equation}
CCFT operators can also wear a number associated with additional global symmetries, \textit{e.g.} the electric charge $Q$ for the $U(1)$ symmetry, which is sometimes dropped to simplify the notation. In the following, we will use the shorthand notation
\begin{equation}
    \mathcal X_N = \mathcal{O}^{\text{out}}_{(\Delta_1, J_1)}(z_1, \bar{z}_1) \dots \mathcal{O}^{\text{out}}_{(\Delta_n, J_n)}(z_n, \bar{z}_n) \mathcal{O}^{\text{in}}_{(\Delta_{n+1}, J_{n+1})}(z_{n+1}, \bar{z}_{n+1}) \dots \mathcal{O}^{\text{in}}_{(\Delta_{N}, J_{N})}(z_{N}, \bar{z}_{N})
\end{equation}
for $N$ insertions in CCFT correlators.

Using the identification \eqref{identification bulk boundary celestial}, soft theorems in the bulk can be rewritten as Ward identities associated to ``conformally soft'' currents (\textit{i.e.} with $\Delta \in \mathbb Z$ operators) in CCFT \cite{Donnay:2018neh}. The soft photon theorem can be rewritten as a $U(1)$ Kac-Moody Ward identity \cite{He:2014cra,Cheung:2016iub,Donnay:2018neh} 
\begin{equation}
    \big\langle J(z)\, \mathcal X_N \big\rangle = \hbar \sum_{q=1}^N \frac{e \eta_q Q_q}{z-z_q} \big\langle \mathcal X_N \big\rangle , \label{SOFT PHOTON}
\end{equation}
where $J(z)$ is a $U(1)$ Kac-Moody current of conformal weights $(1,0)$ and $\eta_q = \pm 1$ for outgoing/incoming fields. Note that this sign disappears in the all out convention. Similarly, the leading soft graviton theorem can be encoded in the supertranslation Ward identities of the CCFT \cite{Strominger:2013jfa,He:2014laa,Donnay:2018neh}
\begin{equation}
\big\langle P(z,\bar{z}) \,\mathcal X_N  \big\rangle + \hbar \sum_{q=1}^{N} \frac{\eta_q}{z-z_q}  \hat\partial_{\Delta_q} \big\langle \mathcal X_N \big\rangle = 0,
\label{leading soft}
\end{equation} where $P(z, \bar{z})$ is the supertranslation current with conformal weights $(\frac{3}{2},\frac{1}{2})$ and $\hat\partial_\Delta$ shifts the conformal dimension $\Delta$ by unit increment. The subleading soft graviton theorem is described by the CCFT Ward identities of superrotations
\cite{Kapec:2016jld,Cheung:2016iub,Fotopoulos:2019tpe,Fotopoulos:2019vac}:
\begin{equation}
\big\langle T(z)\,\mathcal X_N \big\rangle \label{subleading soft} +{\hbar}\sum_{q=1}^N \Big[ \frac{\partial_q}{z-z_q}  + \frac{h_q}{(z-z_q)^2} \Big] \, \big\langle \mathcal X_N \big\rangle=0, 
\end{equation} 
where $T(z)$ is the holomorphic celestial stress tensor with conformal weights $(2,0)$ (similar results hold for the anti-holomorphic stress tensor).

The low-point correlation functions in the CCFT can be deduced from the bulk amplitudes using \eqref{identification bulk boundary celestial}. Alternatively, they can be deduced by studying the constraints implied by \eqref{leading soft} and \eqref{subleading soft} \cite{Stieberger:2018onx, Law:2019glh}. The two-point function reads explicitly as
\begin{equation}
\begin{split}
\mathcal M(\Delta_1,z_1,\bar z_1;\Delta_2,z_2,\bar z_2) &\eqholo \big\langle \mathcal{O}^{\text{out}}_{(\Delta_1, J_1)} (z_1, \bar{z}_1) \mathcal{O}^{\text{in}}_{(\Delta_2, J_2)}(z_2, \bar{z}_2)  \big\rangle \\
&= (2\pi)^4\,\mathscr C_{(2)}\,\delta(\nu_1+\nu_2)\,\delta^{(2)}(z_1-z_2)\, \delta_{J_1,J_2},  
\end{split} \label{2pt celestial final}
\end{equation} 
with $\Delta_q = c+i\nu_q$, while the $3$-point correlation function vanishes in Lorentzian signature for the bulk spacetime. The latter can be made non-vanishing by formulating the CCFT correlators with complexified $(z,\bar z)$, \textit{i.e.} $\bar z\neq z^*$, which amounts to consider holographic duals of bulk amplitudes written in the split metric signature $(-,+,-,+)$ \cite{Pasterski:2017ylz}. 

Another important information in the CCFT that one can deduce is the form of the OPEs. The latter can be obtained from the bulk amplitudes using \eqref{identification bulk boundary celestial} and taking the collinear limit for the particles, which amounts to take the limit $(z_1, \bar{z}_1) \to (z_2, \bar{z}_2)$. The knowledge of the OPEs allows to deduce new symmetries for scattering amplitudes, which includes the $w_{1+\infty}$ algebra \cite{Strominger:2021lvk,Strominger:2021mtt,Ball:2021tmb}. An interesting observation for the current discussion is that the OPEs between the CCFT currents $P(z, \bar{z})$ and $T(z)$ can be deduced from the BMS charge algebra \cite{Donnay:2021wrk}, the latter being interpreted as an algebra for Noether currents in the Carrollian CFT at null infinity. This constitutes a first important insight suggesting that the Carrollian CFT and the CCFT can be related to one another. We explore this idea in further details in the following section.

\subsection{From Carrollian to celestial holography}

In this section, we show that the celestial Ward identities associated to large gauge and BMS symmetries can be recovered from Carrollian correlation functions involving (quasi-)conformal Carrollian primary source operators of specific weights.

\subsubsection{Soft photon theorem}

We consider again here the holographic description of massless scalar electrodynamics and we use the objects defined in Section \ref{sec:electrodynamics holographic correspondence}. The goal now is to deduce the celestial Ward identity \eqref{SOFT PHOTON} from the $U(1)$ invariance of Carrollian correlators \eqref{electro invariance} after some $\mathcal B$-transforms. We are giving more details about the simpler $U(1)$ case, as the gravity case proceeds similarly. 

First, it is useful to express the soft photon current $J(z)$ in terms of conformal Carrollian fields. 
Recalling \eqref{new}, we write $\dot{\sigma}{}_z^{(A)} \equiv  \partial_u \sigma_z^{(A)}$, which carries Carrollian weights $(\frac{3}{2},\frac{1}{2})$ and $\dot{\sigma}{}_{\bar z}^{(A)} = (\dot{\sigma}{}_z^{(A)})^\dagger$. The soft flux \eqref{flux soft maxwell} then reads
\begin{equation}
    \mathcal F^S_\lambda = \int_{\hat{\mathscr I}}\D^3 x\, F_\lambda^S(s,z,\bar z) = \frac{1}{e^2} \int_\Sigma \D^2 z\, \lambda^{(0)}\, \left(\int_{-\infty}^{+\infty}\D u\, \partial_{\bar z} \dot{\sigma}{}_z^{(A)} - \int_{-\infty}^{+\infty}\D v\, \partial_{\bar z} \dot{\sigma}{}_z^{(A)} \right) + \text{h.c.},
\end{equation}
where all objects have been promoted to quantum operators. Now using the electricity condition \eqref{electricity condition for QED} after performing the time integrals explicitly,
\begin{equation}
\begin{split}
    \mathcal F^S_\lambda &= \frac{2}{e^2} \int_\Sigma \D^2 z\, \lambda^{(0)}\, \left(\int_{-\infty}^{+\infty}\D u\, \partial_{\bar z} \dot{\sigma}{}_z^{(A)} - \int_{-\infty}^{+\infty}\D v\, \partial_{\bar z} \dot{\sigma}{}_z^{(A)} \right) \\
    &= -\frac{2}{e^2} \int_\Sigma \D^2 z\, \partial_{\bar z}\lambda^{(0)}\, \left(\int_{-\infty}^{+\infty}\D u\, \dot{\sigma}{}_z^{(A)} - \int_{-\infty}^{+\infty}\D v\, \dot{\sigma}{}_z^{(A)} \right) = \frac{1}{2\pi} \int_\Sigma \D^2 z\, \partial_{\bar z}\lambda^{(0)}\, J(z) ,
\end{split} \label{FSlambda}
\end{equation}
using the expression of the soft photon current \cite{He:2014cra,Lysov:2014csa}
\begin{equation}
   J(z) \equiv -\frac{4\pi}{e^2} \left(\int_{-\infty}^{+\infty}\D u \, \dot{\sigma}{}_z^{(A)}- \int_{-\infty}^{+\infty}\D v\, \dot{\sigma}{}_z^{(A)} \right).
\end{equation}
This operator inserts a soft photon of positive helicity. Notice that by using the electricity condition to trade $\partial_{\bar z}\dot{\sigma}{}_z^{(A)}$ in favor of $\partial_z \dot{\sigma}{}_{\bar z}^{(A)}$, one can insert a soft photon of negative helicity instead by means of the hermitian conjugated operator $\bar J(\bar z)$. Starting from the Ward identity \eqref{electro invariance}, splitting the flux $\mathcal F_\lambda = \mathcal F_\lambda^H + \mathcal F_\lambda^S$ as in \eqref{flux hard maxwell}--\eqref{flux soft maxwell} and using $\big\langle\mathcal F_\lambda^H\, X_N^\sigma\big\rangle = i\hbar\,\delta_\lambda^H \big\langle X_N^\sigma \big\rangle$ as a consequence of \eqref{symmetry generated by fluxes maxwell}, we find
\begin{equation}
     \frac{1}{i\hbar}  \big\langle \mathcal F_\lambda^S\, X_N^\sigma \big\rangle + \delta_\lambda^H \big\langle X_N^\sigma \big\rangle = \frac{1}{2\pi i\hbar} \int_\Sigma \D^2 z\, \partial_{\bar z}\lambda^{(0)} \big\langle J(z)\, X_N^\sigma\big\rangle + \delta_\lambda^H \big\langle X_N^\sigma\big\rangle = 0,
\end{equation}
owing to \eqref{FSlambda}. Here $\delta_\lambda^H \big\langle X_N^\sigma \big\rangle$ represents the homogeneous part of the $U(1)$ transformation of the source operators, \textit{i.e.}
\begin{equation}
    \delta_\lambda^H \sigma^{\text{out/in}}_{(k_j,\bar k_j),Q_j}(u_j/v_j,z_j,\bar z_j) = \mp i e Q_j \lambda^{(0)}(z_j,\bar z_j) \sigma^{\text{out/in}}_{(k_j,\bar k_j),Q_j}(u_j/v_j,z_j,\bar z_j),
\end{equation}
obtained explicitly from \eqref{U(1) primary} and \eqref{Carroll from bulk to bnd}--\eqref{Carroll from bulk to bnd INCOMING}. Particularizing for $\lambda^{(0)}(z,\bar z) = \frac{1}{z-w}$ and using the property
\begin{equation}
    \delta^{(2)}(z-w) = \frac{1}{2\pi} \partial_{\bar z}\left(\frac{1}{z-w}\right), \label{delta on z bar z}
\end{equation}
we have
\begin{equation}
    \frac{1}{i\hbar}\big\langle J(w)\,X_N^\sigma \big\rangle + ie\sum_{q=1}^N \frac{\eta_q\,Q_q}{w-z_q} 
\big\langle X_N^\sigma \big\rangle = 0, \label{QED soft thm carroll}
\end{equation}
where $\eta_q=\pm 1$ for incoming/outgoing insertions. The last step needed to translate this result into the celestial picture amounts to relating Carrollian outgoing and incoming source operators to the celestial operators by means of the $\mathcal B$-transform as in \eqref{a(Delta) from Phi(u)}, \textit{i.e.}
\begin{equation}
\boxed{
\begin{aligned}
     \mathcal{O}^{\text{out}}_{(\Delta_i,J_i),Q_i} (z_i, \bar{z}_i)  &= \kappa_\Delta^+ \,\lim_{\epsilon\to 0^+}\int^{+\infty}_{-\infty} \frac{\D u_i}{(u_i+i\epsilon)^{\Delta_i}}\, \sigma^{\text{out}}_{(k_i,\bar k_i),Q_i}(u_i,z_i,\bar z_i) , \\
     \mathcal{O}^{\text{in}}_{(\Delta_j,J_j),Q_j} (z_j, \bar{z}_j)  &= \kappa_\Delta^- \,\lim_{\epsilon\to 0^+}\int^{+\infty}_{-\infty} \frac{\D v_j}{(v_j-i\epsilon)^{\Delta_j}}\, \sigma^{\text{in}}_{(k_j,\bar k_j),Q_j}(v_j,z_j,\bar z_j)\,,
     \label{integral transforms QED}
\end{aligned}
}
\end{equation}
which is also consistent with the extrapolate-style dictionary of \cite{Pasterski:2021dqe}. Importantly, let us recall that, in our picture, the Carrollian sources that are $\mathcal B$-transformed in \eqref{integral transforms QED} are identified with the boundary values of bulk massless fields, namely implying that the $2d$ spins are implicitly fixed by Carrollian weights as \eqref{k and kbar in terms of J}. Denoting as before the set of celestial insertions by $\mathcal X_N$, one finally checks that \eqref{QED soft thm carroll} becomes \eqref{SOFT PHOTON}, which is nothing but the celestial encoding of Weinberg's soft photon theorem.

\paragraph{Remark} Let us stress that, in this last step to relate Carrollian and celestial results, our proposal to exchange time and conformal dimension by means of \eqref{integral transforms QED} differs from the proposal of \cite{Bagchi:2022emh} to use the ``modified Mellin transform'' introduced in \cite{Banerjee:2018gce}. By construction, the $\mathcal B$-transform \eqref{Btransform}, defined in Section \ref{sec:From null infinity to the celestial sphere} as the combination of Fourier and Mellin transforms acting on ladder operators, maps boundary values of bulk scattering fields onto celestial operators and vice-versa thanks to the inversion formula \eqref{full inverse B transform}. On the other hand, the modified Mellin transform maps functions in Fourier space onto functions depending both on (retarded) time and conformal dimension, which thus cannot be interpreted as boundary values of scattering fields. Nevertheless, since both integral transforms provide amplitudes which solve conformal Carrollian Ward identities, the link between the $\mathcal B$-transform and the modified Mellin transform would be worth exploring.

\subsubsection{Soft graviton theorems} 

Now we turn our interest to scattering processes involving gravitons in asymptotically flat spacetime. Let $\big\langle X_N^\sigma \big\rangle$ be a conformal Carrollian correlator with $N$ insertions, among which can be found either boundary values for the gravitational field (\textit{i.e.} quasi-conformal Carrollian primary source operators $\sigma^{(g)}_{zz}$ of weights $(\frac{3}{2},-\frac{1}{2})$) or null matter fields (\textit{i.e.} conformal Carrollian primary source operators $\sigma^{(\phi)}$). 

From the considerations of Section \ref{sec:Holographic Ward identities for a massless scattering}, we impose \eqref{interm11} for each conformal Carrollian transformation \eqref{conformal Carroll symmetries here}. The splitting in hard and soft variables induces a corresponding separation in the integrated fluxes as in \eqref{FH FS gravity}. Considering first a supertranslation (by setting $\mathcal Y^z = 0 = \mathcal Y^{\bar z}$) and defining $\dot\sigma{}_{zz}^{(g)} \equiv \partial_u \sigma^{(g)}_{(\frac{3}{2},-\frac{1}{2})}$ and $\dot\sigma{}_{\bar z\bar z}^{(g)} = (\dot\sigma{}_{zz}^{(g)})^\dagger$, we promote the soft flux to the following quantum operator 
\begin{equation}
    \begin{split}
        \mathcal F_{\bar\xi(\mathcal T,0)}^S &= \int_{\hat{\mathscr I}}\D^3 x\, F_{(\mathcal{T},0)}^S(s,z,\bar z)  \\
        &= \frac{1}{16\pi G}\int \D^2 z\, \mathcal T \left(\int_{-\infty}^{+\infty}\D u\, \mathscr D^2_{\bar z}\dot\sigma{}_{zz}^{(g)} + \int_{-\infty}^{+\infty}\D v\, \mathscr D^2_{\bar z}\dot\sigma{}_{zz}^{(g)} \right) + \text{h.c.} \\
        &= \frac{1}{8\pi G}\int_\Sigma \D^2 z\, \mathcal T \left(\int_{-\infty}^{+\infty}\D u\, \mathscr D^2_{\bar z}\dot\sigma{}_{zz}^{(g)} + \int_{-\infty}^{+\infty}\D v\, \mathscr D^2_{\bar z}\dot\sigma{}_{zz}^{(g)} \right) = -\frac{1}{2\pi}\int d^2 z \,\partial_{\bar z}\mathcal T\, P(z,\bar z) . 
    \end{split}
\end{equation}
The second equality involves the electricity condition \eqref{electricity condition} after performing the time integrals, the third equality uses an integration by parts on the angles and the following definition the supertranslation current \cite{He:2014laa,Strominger:2013jfa}
\begin{equation}
    P(z,\bar z) \equiv  \frac{1}{4G}\left(\int_{-\infty}^{+\infty}\D u \, \mathscr D_{\bar z}\dot\sigma{}_{zz}^{(g)} + \int_{-\infty}^{+\infty}\D v \, \mathscr D_{\bar z}\dot\sigma{}_{zz}^{(g)} \right),
\end{equation}
which inserts of a soft graviton of positive helicity. Notice again that by using the electricity condition to remove $\mathscr D_{\bar z}\dot\sigma{}_{zz}^{(g)}$ in favor of $\mathscr D_z \dot\sigma{}_{\bar z\bar z}^{(g)}$, we would insert a soft graviton of negative helicity instead. With this definition, \eqref{interm11} becomes
\begin{equation}
    \frac{1}{i\hbar}\big\langle \mathcal F_{\bar\xi(\mathcal T,0)}\,X_N^\sigma \big\rangle + \delta^H_{\bar\xi} \big\langle X_N^\sigma\big\rangle = -\frac{1}{2\pi i\hbar} \int_\Sigma \D^2 z \,\partial_{\bar z}\mathcal T\, \big\langle P(z,\bar z)\,X_N^\sigma\big\rangle + \delta^H_{\bar\xi}\big\langle X_N^\sigma\big\rangle=0,
\end{equation}
where we used the factorization property \eqref{variations generated} and $\delta^H_{\bar\xi}$ reproduces the homogeneous transformation \eqref{SOURCE Carroll}. We have
\begin{equation}
    -\frac{1}{2\pi i\hbar} \int_\Sigma \D^2 z \,\partial_{\bar z}\mathcal T\, \big\langle P(z,\bar z)\,X_N^\sigma \big\rangle + \left(\sum_{i=1}^n \mathcal T(z_i,\bar z_i)\partial_{u_i} + \sum_{j=n+1}^N \mathcal T(z_j,\bar z_j)\partial_{v_j}\right)\big\langle X_N^\sigma \big\rangle=0, \label{id temp 1}
\end{equation}
assuming that the first $n$ fields are holographically identified with outgoing radiative modes. We now perform the $\mathcal B$-transforms
\begin{equation}
\begin{split}
    \mathcal{O}^{\text{out}}_{(\Delta_i,J_i)} (z_i, \bar{z}_i)  &= \kappa^+_\Delta \,\lim_{\epsilon\to 0^+}\int^{+\infty}_{-\infty} \frac{\D u_i}{(u_i+i\epsilon)^{\Delta_i}} \, \sigma^{\text{out}}_{(k_i,\bar k_i)}(u_i,z_i,\bar z_i) , \\
     \mathcal{O}^{\text{in}}_{(\Delta_j,J_j)} (z_j, \bar{z}_j) &= \kappa^-_\Delta \,\lim_{\epsilon\to 0^+}\int^{+\infty}_{-\infty} \frac{\D v_j}{(v_j-i\epsilon)^{\Delta_j}}\, \sigma^{\text{in}}_{(k_j,\bar k_j)}(v_j,z_j,\bar z_j) ,
     \label{integral transforms gravity}
\end{split}
\end{equation}
such that
\begin{equation}
    \delta_{\bar\xi(\mathcal T, 0)}^H \mathcal O^{\text{out/in}}_{(\Delta_j,J_j)}(z_j,\bar z_j) = \mp i\mathcal T(z_j,\bar z_j)\hat\partial_{\Delta_j}\mathcal O^{\text{out/in}}_{(\Delta_j,J_j)}(z_j,\bar z_j),
\end{equation}
in accordance with \textit{e.g.} \eqref{a delta transfo poincare inf} and the holographic map \eqref{Carroll from bulk to bnd}--\eqref{Carroll from bulk to bnd INCOMING}. Notice that one also has
\begin{equation}
    \delta_{\bar\xi(0,\mathcal Y)}^H \mathcal O^{\text{out/in}}_{(\Delta_j,J_j)}(z_j,\bar z_j) = \left(\mathcal Y^{z_j}(z_j)\partial_{z_j} + \bar{\mathcal Y}^{\bar z_j}(\bar z_j)\partial_{\bar z_j} + h_j\,\partial_{z_j}\mathcal Y^{z_j} + \bar h_j\,\partial_{\bar z_j}\bar{\mathcal Y}^{\bar z_j} \right) \mathcal O^{\text{out/in}}_{(\Delta_j,J_j)}(z_j,\bar z_j) \label{delta Y celestial final}
\end{equation}
where $h_j,\bar h_j$ are fixed as \eqref{h and hbar in terms of Delta J}. Then particularizing \eqref{id temp 1} for $\mathcal T(z,\bar z) = \frac{1}{z-w}$, trading $X_N^\sigma$ for $\mathcal X_N$ through \eqref{integral transforms gravity} and using \eqref{delta on z bar z}, we recover \eqref{leading soft}, namely the celestial encoding of the (leading) soft graviton theorem.

The case of a holomorphic superrotation (setting $\mathcal T = 0 = \mathcal Y^{\bar z}$, the anti-holomorphic case being analogous) taken as $\mathcal Y^z(z) = \frac{1}{z-w}$, can be considered in a similar fashion. The related soft flux is promoted as the following quantum operator 
\begin{equation}
\begin{split}
    &\mathcal F^S_{\bar\xi(0,\mathcal Y)} = \int_{\hat{\mathscr I}}\D^3 x\, F^S_{\bar\xi(0,\mathcal Y)}(s,z,\bar z) = -i\,T(w)
\end{split}
\end{equation}
for the holomorphic stress tensor $T(z)$ \cite{Donnay:2021wrk,Donnay:2022hkf}
\begin{equation}
    T(z) \equiv \frac{i}{16\pi G}\int \frac{\D^2 w}{z-w}  \left[\int_{-\infty}^{+\infty}\D u\left( u\, \mathscr D^3_w \dot\sigma{}_{\bar w\bar w}^{(g)} + \frac{3}{2}\sigma^{(g,0)}_{ww}\mathscr D_w \dot\sigma{}_{\bar w\bar w}^{(g)} + \frac{1}{2}\dot\sigma{}_{\bar w\bar w}^{(g)}\mathscr D_w \sigma^{(g,0)}_{ww}\right) + (u\mapsto v)\right]
\end{equation}
where $\sigma^{(g,0)}_{ww}(w,\bar w)$ denotes the quantum operator representing the soft variable $C^{(0)}_{ww}$ defined by \eqref{split CAB hard soft}. Then proceeding the same way as before and recalling \eqref{delta Y celestial final}, one recovers the $2d$ CFT Ward identity \eqref{subleading soft} from \eqref{carroll invariance} after performing the integral transforms \eqref{integral transforms gravity}. 

\section{Discussion}
\label{sec:Discussion}

In this paper, we have provided more details about the Carrollian approach to flat space holography. Let us summarize the steps of this proposal. First, consider gravity in $4d$ asymptotically flat spacetimes without radiation. In this case, the putative dual theory is an honest Carrollian CFT without external source. The insertions of operators in the correlators of this theory, denoted by $\Psi^i$, are typically components of the Carrollian momenta. This situation is similar to what is usually considered in AdS/CFT with Dirichlet boundary conditions, where the correlation functions in the CFT involve the holographic stress tensor. It is also very reminiscent of the situation arising in $3d$ asymptotically flat spacetimes where the bulk theory is topological. 

The second step consists in introducing the radiation in $4d$ asymptotically flat spacetimes. In this case, a first observation is that the BMS charges are no longer conserved due to the radiation reaching null infinity. Therefore, if one identifies the BMS charges in the bulk theory with the Noether charges of the putative dual Carrollian CFT, something has to spoil the global symmetries in the dual theory to yield the non-conservation. In \cite{Donnay:2022aba} and in the present paper, we have argued that the right set-up to spoil the symmetries and encode the radiation at null infinity is to consider a sourced Carrollian CFT. The situation would be very similar to the case of AdS/CFT if one considered leaky boundary conditions instead of conservative boundary conditions such as the standard Dirichlet boundary conditions. In that case, the boundary metric, which plays the role of source, is allowed to fluctuate on the phase space and becomes a field of the dual theory. In the flat case, for pure gravity, the source operators $\sigma^m$ correspond to the asymptotic shear and encode insertions of gravitons at null infinity. The correlators of source operators are therefore identified with $\mathcal{S}$-matrix elements in the bulk. In this sence, from the Carrollian perspective, the $\mathcal{S}$-matrix is described by the source sector of the theory. Obviously, this sector does not exist in the $3d$ case since there is no propagating degree of freedom and no scattering process occurring. 

In the last part of the paper, we have then shown that the source sector of the Carrollian CFT could be related to the celestial CFT. More precisely, the source operators in the Carrollian CFT are mapped on the operators of the CCFT.

We end up this manuscript by providing future potentially interesting directions for this work.
\begin{itemize}[label=$\rhd$]
    \item The formalism of sourced quantum field theory introduced in Section \ref{sec:Sourced quantum field theory} was applied in this work to flat space holography and its associated asymptotic dynamics of the bulk spacetime. We believe that this set-up describing a  sourced system could be applied to a much broader scope than the one presented here. Indeed, any gravitational system with leaky boundary conditions could in principle be holographically described using this formalism. For instance, it would be worth applying this framework in the case of hypersurfaces at finite distance (see \textit{e.g.} \cite{Donnay:2015abr,Donnay:2016ejv,Hopfmuller:2018fni,Chandrasekaran:2018aop,Donnay:2019jiz,Adami:2020amw, Adami:2021nnf,Chandrasekaran:2021hxc}), such as black hole horizons, to deduce some insights on holography for finite spacetime regions. 
    \item As mentioned in the introduction, most of the results obtained in Carrollian holography are deduced from AdS/CFT by taking a flat limit in the bulk, leading to an ultra-relativistic limit at the boundary. However, as highlighted in \cite{Compere:2019bua,Compere:2020lrt,Ruzziconi:2020cjt,Fiorucci:2020xto,Fiorucci:2021pha,Geiller:2022vto}, one has to start from leaky boundary conditions in $4d$ AdS if one wants to obtain radiative spacetimes in the limit. This contrasts with the standard Dirichlet boundary conditions that are usually considered in AdS/CFT. Therefore, the first step would be to obtain a holographic description of AdS spacetimes with leaky boundary conditions by coupling the dual CFT with some external sources and using the formalism of Section \ref{sec:Sourced quantum field theory}. Then, taking a flat limit in the bulk will imply an ultra-relativistic limit at the boundary. The hope is that one might get an explicit realization of the sourced Carrollian CFT using this procedure. It would also be interesting to revisit in this context the flat limit procedure of scattering amplitudes in AdS in the spirit of the works done in \cite{Gary:2009mi,Penedones:2010ue,Fitzpatrick:2011hu,Hijano:2020szl} and relate it to our set-up.

    \item The newly uncovered $w_{1+\infty}$ symmetries in the CCFT \cite{Strominger:2021lvk} have not yet been given
a clear interpretation in the Carrollian CFT. As suggested by the recent analysis of \cite{Freidel:2021ytz}, the information on these symmetries might be encoded in the subleading orders of the bulk metric. In our analysis, we have only considered Carrollian stress tensor or source operators. However, nothing would prevent us to consider Carrollian fields that are holographically identified with subleading orders in the expansion of the bulk metric. It might be instructive to revisit these $w_{1+\infty}$ symmetries in those terms and provide an interpretation at null infinity. 
    \item Finally, it would be a great progress if one could provide an explicit example of Carrollian CFT living at null infinity that would holographically capture some features of gravity in the bulk (for the celestial approach, see \textit{e.g} \cite{Costello:2022jpg,Miller:2022fvc,Kar:2022vqy,Rosso:2022tsv} for recent top-down models). A good starting point is the BMS geometric action constructed in \cite{Barnich:2022bni} which, together with \cite{Nguyen:2020hot}, furnishes an effective description of the dual Carrollian CFT for non-radiative spacetimes. The source sector of the Carrollian CFT is not yet known but we believe that the analysis provided in the present paper imposes strong constraints on it. For instance, the Carrollian weights of the source operators are completely fixed via \eqref{k and kbar in terms of J}. Moreover, the $2$-point correlation function for source operators is identified with the bulk $2$-point amplitude as in \eqref{Carroll 2pt holo}. In particular, this tells us that the propagator of source operators is $u$-dependent, which suggests that the source sector of the Carrollian CFT is a timelike electric-type of Carrollian theory \cite{Duval:2014uoa, Henneaux:2021yzg, deBoer:2021jej,Hansen:2021fxi,Baiguera:2022lsw}. 
\end{itemize}

\paragraph{Acknowledgements} We would like to thank Arjun Bagchi, Glenn Barnich, Nicolas Boulanger, Andrea Campoleoni, Geoffrey Comp\`ere, Florian Ecker, Erfan Esmaeili, Laurent Freidel, Gaston Giribet, Daniel Grumiller, Carlo Heissenberg, K\'evin Nguyen, Sabrina Pasterski, Marios Petropoulos, Stefan Prohazka, Andrea Puhm and Jakob Salzer, for useful conversations. We are also grateful to all the participants of the Carroll Workshops in Vienna and Mons for inspiring and stimulating discussions. LD is partially supported by INFN Iniziativa Specifica ST\&FI. AF and RR are supported by Austrian Science Fund (FWF), project P 32581-N. LD, AF and RR also acknowledge support from the FWF START project Y 1447-N.

\appendix

\section*{Appendices}
\addcontentsline{toc}{section}{Appendices}
\renewcommand{\thesubsection}{\Alph{subsection}}

\numberwithin{equation}{subsection}

\subsection{Bondi coordinates for Minkowski spacetime} 
\label{app:Bondi adv and ret}

In the main text, we make an extensive use of the parametrization of Minkowski space by Bondi coordinates. We found convenient to work with a flat representative of the conformal boundary metric, which amounts to perform a boundary Weyl rescaling with respect to the usual choice of round boundary representative. The aim of this appendix is to install our conventions and notations regarding this choice of coordinates.

\paragraph{Bondi coordinates with round boundary representative} In retarded Bondi coordinates $\{u_\circ,r_\circ,z_\circ,\bar z_\circ\}$ ($u_\circ \in\mathbb R$, $r_\circ\in\mathbb R^+$, $z_\circ\in\mathbb C$), the Minkowski line elements reads as
\begin{equation}
    \D s^2 = -\D u_\circ^2 - 2\D u_\circ \D r_\circ + \frac{4\,r_\circ^2}{(1+z_\circ \bar z_\circ)^2}\D z_\circ\D \bar z_\circ.
\end{equation}
Cuts of constant $u_\circ$ of future null infinity $\mathscr I^+ = \{r_\circ\to+\infty\}$ are spheres on which the line element is the unit round-sphere metric in stereographic coordinates $z_\circ = e^{i\varphi}\cot\frac{\theta}{2}$. The change of coordinates from the Cartesian chart $X^\mu =\{t,\vec x\}$ is given by
\begin{equation}
    X^\mu = u_\circ \, \delta_0^\mu + r_\circ\,\frac{\sqrt{2}}{1+z_\circ\bar z_\circ}\,q^\mu(z_\circ,\bar z_\circ)
\end{equation}
where
\begin{equation}
    q^\mu(z,\bar z) \equiv \frac{1}{\sqrt{2}} \Big(1+z\bar z,z+\bar z,-i(z-\bar z),1-z\bar z\Big) \label{q in terms of z bar z future}
\end{equation}
is the standard parametrization of a null direction pointing toward the direction $(z_\circ,\bar z_\circ)$ on the celestial sphere. In a similar way, one defines advanced Bondi coordinates $\{v_\circ,r_\circ',z_\circ',\bar z_\circ'\}$ ($v_\circ \in\mathbb R$, $r_\circ'\in\mathbb R^+$, $z_\circ'\in \mathbb{C}$) in which the Minkowski line element is
\begin{equation}
    \D s^2 = -\D v_\circ^2 + 2\D v_\circ \D r'_\circ + \frac{4\,r_\circ'^2}{(1+z'_\circ \bar z'_\circ)^2}\D z'_\circ\D \bar z'_\circ.
\end{equation}
In terms of the Cartesian coordinates, we have
\begin{equation}
    X^\mu = v_\circ \, \delta_0^\mu - r'_\circ\,\frac{\sqrt{2}}{1+z'_\circ\bar z'_\circ}\,q^\mu(z'_\circ,\bar z'_\circ) .
\end{equation}
For any finite value of $r_\circ,r'_\circ$, both Bondi coordinate systems are related as
\begin{equation}
    v_\circ = u_\circ + 2r_\circ ,\quad r'_\circ = r_\circ ,\quad z'_\circ = -\frac{1}{\bar z_\circ} , \label{uo to vo}
\end{equation}
where in particular the holomorphic coordinates are related by the antipodal map $(\theta,\varphi)\mapsto(\pi-\theta,\varphi+\pi)$, or $z_\circ\mapsto -\frac{1}{\bar z_\circ}$. This particularly convenient choice \cite{Strominger:2013jfa} allows to label a light ray crossing the spacetime with the same value of the angular coordinates in both coordinate systems. In other words, a light ray originating from a point $z'_\circ = a$ on the celestial sphere at past null infinity will pierce again the celestial sphere at future null infinity at the antipodal point identified by $z_\circ = a$.

\paragraph{Bondi coordinates with flat boundary representative} We now introduce the coordinate system $\{u,r,z,\bar z\}$ by trading the unit round-sphere metric on the boundary for the flat complex plane metric. The diffeomorphism that implements the boundary Weyl rescaling has been worked out \textit{e.g.} in \cite{Compere:2016hzt} and reads as
\begin{equation}
    r = \frac{\sqrt{2}}{1+z_\circ\bar z_\circ}r_\circ +\frac{u_\circ}{\sqrt{2}}\,,\quad u = \frac{1+z_\circ \bar z_\circ}{\sqrt{2}}u_\circ - \frac{z_\circ\bar z_\circ u_\circ^2}{2r}\,,\quad z = z_\circ - \frac{z_\circ u_\circ}{\sqrt{2}r} \label{flatten future}
\end{equation}
for retarded coordinates. Here $u,r\in\mathbb R$ and $z\in\mathbb C$. The relation with Cartesian coordinates is now given by
\begin{equation}
    X^\mu = u\,\partial_z\partial_{\bar z}q^\mu(z,\bar z) + r\, q^\mu(z,\bar z) \label{Retarded flat BMS coordinates (appendix)}
\end{equation}
and the Minkowski line element reads as
\begin{equation}
    \D s^2 = -2\D u\D r+2r^2\D z\D \bar z  .
\end{equation}
These coordinates are such that $\mathscr{I}^+$ and $\mathscr{I}^-$ are respectively obtained by taking the limits $r \to +\infty$ and $r\to -\infty$. Cuts of $\mathscr{I}^+$ are now complex planes endowed with a flat metric. Lines obtained by keeping $(u,z,\bar z)$ fixed form a null geodesic congruence extending from past to future null infinity. The matching between past and future null infinities is therefore here immediate.

Similarly, we can also flatten the boundary representative in advanced Bondi coordinates to reach the coordinates $\{v,r',z',\bar z'\}$. The diffeomorphism implementing the boundary Weyl rescaling is given explicitly by
\begin{equation}
    r' = \frac{\sqrt{2}}{1+z_\circ\bar z_\circ}r_\circ'-\frac{v_\circ}{\sqrt{2}}\,,\quad v = \frac{1+z'_\circ \bar z'_\circ}{\sqrt{2}}v_\circ + \frac{z'_\circ\bar z'_\circ v_\circ^2}{2r'}\,,\quad z' = z'_\circ + \frac{z'_\circ v_\circ}{\sqrt{2}r'} \label{flatten past}
\end{equation}
In these coordinates, related to the Cartesian coordinates via
\begin{equation}
    X^\mu = v\,\partial_{z'}\partial_{\bar z'}q^\mu(z',\bar z') - r'\, q^\mu(z',\bar z') , \label{Advanced flat BMS coordinates (appendix)}
\end{equation}
the Minkowski line element is
\begin{equation}
    \D s^2 = 2\D v\D r'+2 r'^2\D z'\D\bar z' .
\end{equation}
Here $\mathscr I^+$ is reached in the limit $r'\to -\infty$ while $\mathscr I^-$ is obtained by taking $r'\to+\infty$. Inverting \eqref{flatten future} recalling that $r_\circ\in\mathbb R^+$ and taking into account the diffeomorphism \eqref{uo to vo}, one can show that \eqref{flatten past} yields
\begin{equation}
    v = u ,\quad r' = -r ,\quad z' = z . \label{u to v}
\end{equation}
The Bondi coordinates with flat boundary representative are asymptotically related to those with round boundary representative as follows. In the limit where $r_\circ$ goes to infinity in the retarded Bondi coordinates, we deduce from \eqref{flatten future} that
\begin{equation}
    u = u_\circ\,\frac{1+z_\circ \bar z_\circ}{\sqrt{2}}+\mathcal O(r_\circ^{-1}) ,\quad r = r_\circ \,\frac{\sqrt{2}}{1+z_\circ\bar z_\circ} + \mathcal O(r_\circ^{0}) ,\quad z = z_\circ + \mathcal O(r_\circ^{-1}) ,
\end{equation}
while, in the limit where $r_\circ' \equiv r_\circ$ goes to infinity in the advanced Bondi coordinates, \eqref{flatten past} together with the matching \eqref{u to v} imply
\begin{equation}
    v = v_\circ\,\frac{1+z'_\circ \bar z'_\circ}{\sqrt{2}}+\mathcal O(r_\circ^{-1}) ,\quad r = -r_\circ \,\frac{\sqrt{2}}{1+z'_\circ\bar z'_\circ} + \mathcal O(r_\circ^{0}) ,\quad z = z'_\circ + \mathcal O(r_\circ^{-1}) .
\end{equation}
In this sense, the coordinate system $\{u,r,z,\bar z\}$ with flat boundary representative interpolates between the advanced and retarded coordinate systems with boundary round representatives at large distances. In particular points situated at $z_\circ$ on the future celestial sphere are identified with points at $z'_\circ = -\frac{1}{\bar z_\circ}$ on the past celestial sphere through a null ray defined with constant $(u,z,\bar z)$ such that $z \equiv z_\circ$ at future null infinity. Similar considerations apply for the advanced coordinate system up to performing the simple change of coordinates \eqref{u to v}.

\subsection{Isomorphism between global conformal Carrollian and Poincaré algebras}
\label{sec:isomorphism Poincare Carroll}

In Cartesian coordinates $X^\mu = \{t,x^i\}$, the Poincaré generators on Minkowski spacetime are
\begin{itemize}[label=$\rhd$]
\item Translations: $\texttt P_0 = \partial_t$, $\texttt P_i = \partial_{x^i}$, $i=1,2,3$ ;
\item Rotations: $\texttt R_i = x^j\partial_{x^k} - x^k\partial_{x^j}$, $(i,j,k) = (1,2,3)$, $(2,1,3)$, $(3,1,2)$ ;
\item Special Lorentz transformations: $\texttt B_i = x^i\partial_t + t\partial_{x^i}$, $i=1,2,3$.
\end{itemize}
They satisfy the well-known $\mathfrak{iso}(3,1)$ algebra 
\begin{equation}
    \begin{split}
        [\texttt{P}_i,\texttt{P}_j] &= 0 = [\texttt{P}_0,\texttt{P}_i], \quad [\texttt{R}_i,\texttt{P}_j] = -\varepsilon_{ijk} \texttt{P}_k, \quad [\texttt{R}_i,\texttt{P}_0] = 0,\quad [\texttt{B}_i,\texttt{P}_j] = -\delta_{ij}\texttt{P}_0, \\
        [\texttt{B}_i,\texttt{P}_0] &= -\texttt{P}_i,\quad [\texttt{R}_i,\texttt{R}_j] = -\varepsilon_{ijk}\texttt{R}_k, \quad [\texttt{R}_i,\texttt{B}_j] = -\varepsilon_{ijk}\texttt{B}_k, \quad [\texttt{B}_i,\texttt{B}_j] = \varepsilon_{ijk} \texttt{R}_k.
    \end{split} \label{poincare habituel}
\end{equation}
Performing the change of coordinates to retarded Bondi gauge with flat conformal frame at the boundary, one finds that the Poincaré generators can be expressed as \eqref{conformal Carroll symmetries here} on $\mathscr I^+$ with particular functions $\mathcal T,\mathcal Y^z$ and $\mathcal Y^{\bar z}$ given in Table \ref{tab:Poincare}. 

\begin{table}[ht!]
\centering
\renewcommand{\arraystretch}{1.3}
\begin{tabular}{c|c|c|c}
Generator & $\mathcal T(z,\bar z)$ & $\mathcal Y^z(z)$ & ${\mathcal Y}^{\bar z}(\bar z)$ \\ \hline
$\texttt P_0$ & $\tfrac{1}{\sqrt{2}}(1+z\bar z)$ & $0$ & $0$ \\ \hline
$\texttt P_1$ & $-\tfrac{1}{\sqrt{2}}(z+\bar z)$ & $0$ & $0$ \\ \hline
$\texttt P_2$ & $\tfrac{1}{\sqrt{2}}i(z-\bar z)$ & $0$ & $0$ \\ \hline
$\texttt P_3$ & $\tfrac{1}{\sqrt{2}}(1-z\bar z)$ & $0$ & $0$ \\ \hline
$\texttt R_1$ & $0$ & $\frac{1}{2}i(1-z^2)$ & $-\frac{1}{2}i(1-\bar z^2)$ \\ \hline
$\texttt R_2$ & $0$ & $\frac{1}{2}(1+z^2)$ & $\frac{1}{2}(1+\bar z^2)$ \\ \hline
$\texttt R_3$ & $0$ & $iz$ & $-i\bar z$ \\ \hline
$\texttt B_1$ & $0$ & $\frac{1}{2}(1-z^2)$ & $\frac{1}{2}(1-\bar z^2)$ \\ \hline
$\texttt B_2$ & $0$ & $\frac{1}{2}i(1+z^2)$ & $-\frac{1}{2}i(1+\bar z^2)$ \\ \hline
$\texttt B_3$ & $0$ & $z$ & $\bar z$
\end{tabular}
\caption{Poincaré generators in retarded Bondi gauge.}
\label{tab:Poincare}
\end{table}

From the intrinsic point of view, these generators do not necessarily seem natural but can be related to the standard generators of a Carrollian conformal symmetry algebra (see Section \ref{sec:Conformal Carrollian symmetries}) thanks to the following isomorphism:
\begin{equation}
\mathfrak{CCarr}_3 \simeq \mathfrak{Conf}_{2} \loplus \mathbb R^{4} \simeq \mathfrak{so}(3,1)\loplus \mathbb R^{4} \equiv \mathfrak{iso}(3,1).
\end{equation}
The dictionary is as follows:
\begin{equation}
\begin{array}{lll}
P_0 = \frac{1}{\sqrt{2}}(\bar{\texttt P}_0 + \bar{\texttt P}_3), \qquad & P_1 = -\frac{i}{2}(\bar{\texttt R}_1 + i \bar{\texttt R}_2 + i \bar{\texttt B}_1 + \bar{\texttt B}_2), \qquad & P_2 = \frac{i}{2}(\bar{\texttt R}_1 - i \bar{\texttt R}_2 - i \bar{\texttt B}_1 + \bar{\texttt B}_2),\\
J = i \bar{\texttt R}_3,\quad D = \bar{\texttt B}_3, \qquad & B_1 = -\frac{1}{\sqrt 2}(\bar{\texttt P}_1 - i \bar{\texttt P}_2),\qquad & B_2 = -\frac{1}{\sqrt 2}(\bar{\texttt P}_1 + i \bar{\texttt P}_2),\\
K_0 = -\sqrt{2}(\bar{\texttt P}_0-\bar{\texttt P}_3), \qquad & K_1 = -i(\bar{\texttt R}_1+i\bar{\texttt R}_2-i\bar{\texttt B}_1-\bar{\texttt B}_2),\qquad & K_2 = i(\bar{\texttt R}_1-i\bar{\texttt R}_2+i\bar{\texttt B}_1-\bar{\texttt B}_2).
\end{array} \label{poincare dictionary}
\end{equation}
The bar over the Poincaré generators means again their restriction to future null infinity. Applying the redefinitions \eqref{poincare dictionary} on \eqref{poincare habituel} gives the algebra \eqref{global conformal Carrollian
subalgebra ABELIAN}--\eqref{global conformal Carrollian
subalgebra}.

\subsection{Constraints on the Carrollian stress tensor}
\label{sec:Constrains on the Carrollian stress tensor}

In this appendix, we detail the proof of the identities \eqref{classical constraints on the C} obeyed by the Carrollian stress tensor ${\mathcal C^a}_b$ at the classical level. All of them stem from the flux-balance law \eqref{carroll flux balance}.

For Carrollian translations, $\bar\xi^a = \delta^a_b$, hence $\partial_a {\mathcal{C}^a}_{b} = F_b$ is immediate from \eqref{carroll flux balance}. Invariance under Carrollian rotation imposes
\begin{equation}
    0 = \partial_a\left(-{\mathcal C^a}_z z + {\mathcal C^a}_{\bar z}\bar z\right) + z F_z - \bar z F_{\bar z} = -z(\partial_a {\mathcal C^a}_z-F_z) +\bar z(\partial_a {\mathcal C^a}_{\bar z}-F_{\bar z}) - {\mathcal C^z}_z + {\mathcal C^{\bar z}}_{\bar z},
\end{equation}
which gives ${\mathcal C^z}_z = {\mathcal C^{\bar z}}_{\bar z}$ using the invariance by translation, \textit{i.e.} the first condition in \eqref{classical constraints on the C}.
For the Carrollian boosts, 
\begin{equation}
    0 = \partial_a\left({\mathcal C^a}_u x^A\right) - x^A F_u = \left(\partial_a{\mathcal C^a}_u - F_u\right)x^A + {\mathcal C^A}_u \quad\Rightarrow\quad {\mathcal C^A}_u = 0.
\end{equation}
The invariance under Carrollian dilatations finally gives
\begin{equation}
    0 = \partial_a ({\mathcal C^a}_bx^b) - F_b x^b =  \left(\partial_a{\mathcal C^a}_b - F_b\right)x^b + {\mathcal C^a}_a \quad\Rightarrow\quad {\mathcal C^a}_a = 0,
\end{equation}
which concludes the demonstration of \eqref{classical constraints on the C}.

It remains to show that the Carrollian special conformal transformations $K_0 = -2z\bar z\partial_u$, $K_1 = 2 u \bar z\partial_u + 2 \bar z^2 \partial_{\bar z}$ and $K_2 = 2 u z \partial_u + 2 z^2\partial_z$ do not impose further constraints. Let us prove this statement for $K_2$ (the proof for $K_0$ and $K_1$ is similar). The flux-balance law is particularized as
\begin{equation}
\begin{split}
    0 &= \partial_a ({\mathcal C^a}_u uz + {\mathcal C^a}_z z^2) - F_u uz - F_z z^2 \\
    &= (\partial_a{\mathcal C^a}_u - F_u)uz + (\partial_a{\mathcal C^a}_z - F_z)z^2 + {\mathcal C^u}_u z + {\mathcal C^z}_u u + 2 {\mathcal C^z}_z z \\
    &= \left({\mathcal C^u}_u + {\mathcal C^z}_z + {\mathcal C^{\bar z}}_{\bar z}\right) z = 0.
\end{split}
\end{equation}
The second equality holds by virtue of the invariance by translation and boosts, the third one holds thanks to the invariance by rotation while the last one uses the invariance by dilatation.

Let us finally prove that the global conformal Carrollian symmetries are enough to completely constrain ${\mathcal{C}^a}_b$, \textit{i.e.} \eqref{carroll flux balance} is automatically satisfied by the pure supertranslation and superrotation currents provided \eqref{classical constraints on the C} holds. Considering first a generic supertranslation $\bar\xi^u = \mathcal T(z,\bar z)$, $\bar\xi^A = 0$, we have
\begin{equation}
    0 = \partial_a({\mathcal C^a}_u \mathcal T) - F_u\mathcal T = (\partial_a{\mathcal C^a}_u-F_u)\mathcal T + {\mathcal C^A}_u \partial_A \mathcal T = 0
\end{equation}
using successively the invariance by Carrollian translations and boosts. For superrotations $\bar \xi^u = \frac{u}{2}\partial_A\mathcal Y^A$, $\bar\xi^z = \mathcal Y^z(z)$ and $\bar\xi^{\bar z} = \mathcal Y^{\bar z}(\bar z)$, we have
\begin{align}
        &\partial_a \left({\mathcal C^a}_u \tfrac{u}{2}\partial_A\mathcal Y^A+{\mathcal C^a}_A\mathcal Y^A\right) - F_u \tfrac{u}{2}\partial_A\mathcal Y^A - F_A \mathcal Y^A \nonumber \\
        &= \tfrac{u}{2}\partial_A\mathcal Y^A(\partial_a {\mathcal C^a}_u - F_u) + \mathcal Y^A (\partial_a {\mathcal C^a}_A - F_A)+ \tfrac{1}{2}{\mathcal C^u}_u \partial_A\mathcal Y^A + \tfrac{u}{2}{\mathcal C^B}_u \partial_B\partial_A\mathcal Y^A + {\mathcal C^B}_A \partial_B \mathcal Y^A \nonumber \\
        &= \tfrac{1}{2}{\mathcal C^u}_u (\partial_z \mathcal Y^z+\partial_{\bar z}{\mathcal Y}^{\bar z}) + {\mathcal C^z}_z\partial_z \mathcal Y^z + {\mathcal C^{\bar z}}_{\bar z}\partial_{\bar z} {\mathcal Y}^{\bar z} \\
        &= \tfrac{1}{2}{\mathcal C^u}_u (\partial_z\mathcal Y^z+\partial_{\bar z}{\mathcal Y}^{\bar z}) + \tfrac{1}{2}({\mathcal C^z}_z+{\mathcal C^{\bar z}}_{\bar z})\partial_z \mathcal Y^z + \tfrac{1}{2}({\mathcal C^z}_z+{\mathcal C^{\bar z}}_{\bar z}) \partial_{\bar z} {\mathcal Y}^{\bar z} \nonumber \\
        &= \tfrac{1}{2}({\mathcal C^u}_u+{\mathcal C^z}_z+{\mathcal C^{\bar z}}_{\bar z})(\partial_z\mathcal Y^z+\partial_{\bar z}{\mathcal Y}^{\bar z}) = 0. \nonumber
\end{align}
The third equality uses the invariance by translations and boosts. It also uses the fact that superrotations are holomorphic. The fourth equality uses the invariance by rotation while the last one invokes the invariance by dilatation. This concludes the proof of \eqref{classical constraints on the C}.

\hypersetup{urlcolor=blue}
\addcontentsline{toc}{section}{References}
\providecommand{\href}[2]{#2}\begingroup\raggedright\endgroup

\end{document}